\documentclass[final,5p,times,twocolumn,sort&compress,number]{elsarticle}
\usepackage[hidelinks]{hyperref}
\usepackage{lineno}
\usepackage{xurl}
\usepackage{xcolor}
\graphicspath{ {./graphics/} }
\usepackage[page]{appendix}
\usepackage{bibtopicprefix}
\usepackage{multirow}
\usepackage{lscape}
\usepackage{enumitem}
\usepackage{pbox}

\usepackage{floatrow}
\usepackage{supertabular,booktabs}

\usepackage{footmisc}
\usepackage{textgreek}
\usepackage{tcolorbox}
\usepackage{adjustbox}
\usepackage{changepage}
\usepackage{rotating}
\usepackage{calc}
\usepackage{tablefootnote}
\usepackage{natbib}
\usepackage{nicefrac}
\usepackage{booktabs}
\usepackage{float}
\usepackage[graphicx]{realboxes}
\usepackage{booktabs}
\usepackage{amssymb}% http://ctan.org/pkg/amssymb
\usepackage{pifont}% http://ctan.org/pkg/pifont
\usepackage{blindtext}
\usepackage[normalem]{ulem}
\useunder{\uline}{\ul}{}
\usepackage[rightcaption]{sidecap}
\usepackage{amsmath}
\usepackage{mathabx}
\usepackage{chapterbib}

\usepackage{morefloats}
\renewcommand{\citet}[1]{[W\citealp{#1}]}
\renewcommand{\tiny}{\fontsize{7.282pt}{9.282pt}\selectfont}
\definecolor{tumblue}{RGB}{0,101,189}
\definecolor{tumgray}{RGB}{255,255,255}
\definecolor{tumblack}{RGB}{0,0,0}
\usepackage{mdwlist}
\usepackage{supertabular,array}
\usepackage{ltablex,ragged2e}
\newcolumntype{L}{>{\RaggedRight}X}
\usepackage{eurosym}

\usepackage{xcolor}

\newif\ifdraft
\drafttrue  %switch to <<draftfalse>> to remove all comments
\ifdraft
\newcommand{\alnote}[1]{ {\textcolor{magenta} {**Andre: #1}}}
\newcommand{\jsnote}[1]{ {\textcolor{cyan} {**Johannes: #1}}}
\newcommand{\gm}[1]{ {\textcolor{red} {**Gonzalo: #1}}}

\fi

\newif\ifappendix
\appendixtrue % switch to \appendixfalse for removing appendix and literature

\journal{Journal of Network and Computer Applications}

\begin{document}

\begin{frontmatter}

\title{Revealing the Landscape of Privacy-Enhancing Technologies in the Context of Data Markets for the IoT: A Systematic Literature Review}

\author[mymainaddress]{Gonzalo Munilla Garrido\corref{mycorrespondingauthor1}}\cortext[mycorrespondingauthor1]{Corresponding author: Gonzalo Monilla Garrido\\\textit{Email address:} \href{mailto:gonzalo.munilla-garrido@tum.de}{gonzalo.munilla-garrido@tum.de}\\ \textit{Address:} Boltzmannstrasse 3, 85748, Garching}

\author[mysecondaddress]{Johannes Sedlmeir}

\author[mymainaddress]{\"Omer Uluda\u{g}}

\author[mymainaddress]{Ilias Soto Alaoui}

\author[mythirdaddress]{Andre Luckow} 

\author[mymainaddress]{Florian Matthes}

\address[mymainaddress]{Technical University of Munich, Department of Informatics, Munich, Germany}
%\address[mysecondaryaddress]{The BMW Group, Group IT, Munich, Germany}
\address[mysecondaddress]{Project Group Business \& Information Systems Engineering of the Fraunhofer FIT, Bayreuth, Germany}
\address[mythirdaddress]{Ludwig Maximilian University of Munich, Department of Computer Science, Munich, Germany}

\begin{abstract}
IoT data markets in public and private institutions have become increasingly relevant in recent years because of their potential to improve data availability and unlock new business models.
However, exchanging data in markets bears considerable challenges related to disclosing sensitive information.
Despite considerable research focused on different aspects of privacy-enhancing data markets for the IoT, none of the solutions proposed so far seems to find a practical adoption.
Thus, this study aims to organize the state-of-the-art solutions, analyze and scope the technologies that have been suggested in this context, and structure the remaining challenges to determine areas where future research is required.
To accomplish this goal, we conducted a systematic literature review on privacy enhancement in data markets for the IoT, covering $50$ publications dated up to July 2020, and provided updates with $24$ publications dated up to May 2022.
Our results indicate that most research in this area has emerged only recently, and no IoT data market architecture has established itself as canonical. Existing solutions frequently lack the required combination of anonymization and secure computation technologies.
Furthermore, there is no consensus on the appropriate use of blockchain technology for IoT data markets and a low degree of leveraging existing libraries or reusing generic data market architectures.
We also identified significant challenges remaining, such as the copy problem and the recursive enforcement problem that -- while solutions have been suggested to some extent -- are often not sufficiently addressed in proposed designs.
We conclude that privacy-enhancing technologies need further improvements to positively impact data markets so that, ultimately, the value of data is preserved through data scarcity and users' privacy and businesses-critical information are protected. 
% PERHAPS ADD IF NO LIMITS
\end{abstract}

% Choose keywords that are not included in the title to optimize the visibility in searches

\begin{comment}
\alnote{maybe structure that similar to the document. Separate functional purpose of a building block and technical implementation?}
\sep Systematic mapping study 
\end{comment}

\begin{comment}
Context
Objective
Method
Results
Conclusions (lengthier)
\end{comment}

\end{frontmatter}
% ^
%\linenumbers
\nolinenumbers

\section{Introduction}
\label{sec:introduction}

%P1 Short sentence about context. What is IoT from a global perspective? why it is used, its impact, benefits, but there is a warning.

\noindent IoT devices have been improved, mass-produced, and deployed in the past few decades through steady progress in information and communication technologies (ICTs) and motivated by a trend of data-driven decision-making, automation, and the opportunity for new business models.
IoT devices' primary collective purpose is to interact with the physical world and enable the measurement and collection of events and interactions~\citep{P23}.
These characteristics apply to IoT devices deployed in, for example, a factory or a powerline network and many devices employed by people, such as cell phones, laptops, or wearables. 
The \textit{volume}, \textit{velocity}, and \textit{variety} of the information generated by the IoT is immense, which drove practitioners to coin the term \textit{big data} and develop tools for their analysis~\citep{P4}.
Public and private institutions use \textit{big data} to promote the public good, innovations, and improve products and services.
Big data has become the foundation of the emerging data economy, which in Europe was worth nearly $2$\,\% of its GDP in 2016, close to $300$ billion Euros~\cite{European_Data_Market}.
However, the generation, collection, storage, processing, distribution, and analysis of \textit{big data} to realize such economic potential also come with challenges for enterprises and responsibilities towards society.

%P2 Problems in terms of accessibility and privacy. to solve the first one we have the emergence of data markets, and for the second one, we have privacy preservation tech (next paragraph).  what has been happening in the past years?  first, iot markets, keep growing.
Big data needs to be accessible to institutions that can harness their potential and develop innovations, lest society fails to materialize their advantages.
Unfortunately, a significant share of the world's data is siloed and exploited solely by the institutions that host them~\cite{silos}, consequently locking the untapped potential of the data economy and hindering progress in science, business, and society. 
To surmount this obstacle, a paradigm shift towards openness emerged in the form of electronic data markets for the IoT, i.e., mediums for the trade of information across the Internet based on electronic infrastructure~\cite{data_market_definition}.
This paradigm brings potential benefits, such as increasing the efficiency of business processes, facilitating growth by unlocking new business models~\cite{data_exchange_pros}, and profiting from trading.
Decision-makers in governments and businesses have recognized the economic potential of data markets and hence recently supported significant projects that provide a shared digital infrastructure for data-sharing initiatives such as GAIA-X~\cite{noauthor_gaia_x_nodate} or the automotive-related Catena-X, which promote the collaboration of large enterprises in data markets.

%supply chain traceability, improved risk profiling of insurance companies in different domains, or allowing an application to optimize its service based on users' third-party-managed health data. 
% Open IoT data market ecosystems, coupled with data-sharing initiatives like GAIA-X~\cite{noauthor_gaia_x_nodate} or the automotive-related Catena-X in Europe, promote the collaboration of many large enterprises in data markets.
%Interestingly, these initiatives carry an undertone \jsnote{which is it?} similar to the long-established trade accords for goods other than data. 

Despite data markets' promise to benefit society by fostering innovation and collaboration across enterprises, these markets hold the risk of exposing individuals' and businesses' sensitive information~\cite{sweeney_identifying_2013, gao_elastic_2014}. Moreover, confidence in privacy protection is an essential driver of users' willingness to share their data~\citep{P8}. 
Similarly, businesses are unwilling to bear the risk of unintentionally leaking their customers' private or business-critical information. Consequently, the adoption of data markets is generally hampered. 
Additionally, while a corporation may have taken security measures to protect collected data from unauthorized access or unintended use, data buyers might not have the same standards. 
Hence, in this case, exchanging data entails an additional risk that the seller needs to mitigate \emph{before} the data are shared. Furthermore, blockchains are expected to play an essential role in the ability of institutions to trade data in tokenized form~\cite{sunyaev2021token}, but their inherent transparency further increases the need to make data exchanged in markets less sensitive~\citep{P25}.
Additional trends that aggravate the negative consequences of lacking data protection for institutions are recent privacy laws such as the GDPR in Europe or the CCPA in California with their increasingly expensive fines for data breaches~\cite{data_breach}. 
% Therefore, blackmailing with a data breach threat might become as common as ransom cyberattacks. \jsnote{Is there any reference where we could say "Some researchers even posit that..." }
%However, as this is an externality from the perspective of the business that discloses data, regulators should introduce countermeasures.

%P3 Most popular privacy-preserving tech. Keys to success. Make clear that the topic is relevant for practitioners as well as for academics.
%\alnote{the paragraphs below might be candidates for moving them into background to keep the introduction section short and to the point. Key challenges and contributions must be introduced, but not discussed at length}

As a first reaction to these risks, practitioners and corporations have increased their systems' security. 
However, if data are sold and, thus, replicated, confidentiality is not sufficient to protect privacy: Only managing or modifying the data in a way that enhances privacy while preserving as much utility as possible is effective~\cite{trask_structured_transparency_nodate}. 
Thus, institutions have started to allocate more resources to balance data utility and privacy, employing privacy-enhancing technologies (PETs)~\citep{P27}.
The term PET was coined in 1995 in a report by the Dutch Data Protection Authority and the Ontario Information Commissioner~\cite{hes_privacy_enhancing_1998} that explored a novel approach to privacy protection~\cite{PETs_coined}.
These technologies take the form of architectures built with privacy-by-design principles and policies~\citep{P39}\citep{P6}, or data modifications based on heuristics or mathematical privacy guarantees. 
Prominent examples of PETs are differential privacy~\cite{DP_original, dwork_algorithmic_2013}, syntactic anonymization definitions like $k$-anonymity~\cite{samarati_protecting_nodate}, homomorphic encryption~\cite{HomomorphicPaper, Theory1, P41R17}, trusted execution environments~\cite{OMTP_TEE}, secure multiparty computation~\cite{SMC_basics}, zero-knowledge proofs~\cite{first_introduced_ZKP}\cite{zkp_defs}, and a set of conventional de-identification approaches such as masking, rounding, or hashing~\cite{bondel_towards_nodate}.

The relevance of PETs in data markets also increases with the growing adoption of IoT devices, such as in vehicles, wearables, smartphones, and the applications that stream data daily from millions of individuals' private lives to data marketplaces~\cite{spiekermann_personal_2015}.
Despite their current relevance and growing attention~\citep{P27}, researchers and institutions still find PETs challenging to understand, integrate, and deploy in IoT data markets because most PETs are technically complex and have a wide range of variations and combinations with different tradeoffs~\cite{za_privacy_sensitive_2021}.
Regarding research addressing these challenges, primary studies are predominant, i.e., studies based on original designs developed or data collected by their authors, while secondary studies collecting and systematizing existing knowledge are less frequent.
The applications proposed by primary studies range from funneling data from markets into machine learning (ML) algorithms~\citep{P2}\citep{P25}\citep{P43}, crowdsourcing data into markets~\citep{P16}\citep{P22}\citep{P24}\citep{P47}, adopting data markets for smart mobility ecosystems~\citep{P3}, smart manufacturing~\citep{P19}, smart homes~\citep{P11}, and smart wearables in the health industry~\citep{P48}.
%supply chain traceability, improved risk profiling of insurance companies in different domains, or allowing an application to optimize its service based on users' third-party-managed health data. 
% Open IoT data market ecosystems, coupled with data-sharing initiatives like GAIA-X~\cite{noauthor_gaia_x_nodate} or Catena-X in Europe, promote the collaboration of many large enterprises in data markets.
%Interestingly, these initiatives carry an undertone \jsnote{which is it?} similar to the long-established trade accords for goods other than data. 

On the other hand, $9$ out of the $50$~studies that we identified in our systematic literature review (SLR) are secondary, and out of these, four studies~\citep{P19}\citep{P23}\citep{P35}\citep{P48} cover \emph{some} of the PETs available for data markets for the IoT, yet without giving a detailed comparison of their functionalities, benefits, and limitations.
The other five secondary studies~\citep{P4}\citep{P8}\citep{P14}\citep{P27}\citep{P38} perform high-level surveys revolving around challenges, non-technical privacy strategies, and user-centric perspectives on data markets for the IoT.
However, none of these secondary studies provided a rigorous, \emph{systematic} review that collected and mapped PETs and challenges comprehensively. 
Moreover, as we discuss in Section~\ref{sec:discussion}, we noted a low level of re-using existing components to build a more holistic architecture for data markets in related work, which may indicate the need for systematically analyzing the current seminal components, strengths, and weaknesses of solutions proposed for privacy-enhancing IoT data markets. 
% Consequently, a more comprehensive, detailed, and systematic paper on the technical characteristics and the different possibilities that PETs may bring to IoT data markets is necessary to fill these research gaps.
% we could identify nine secondary studies conducted in this area. 
% These publications are instances of secondary studies, and researchers have used them increasingly in software engineering over the years \cite{Systematic_mapping_SW_engineering} \cite{performing_SRs}. Our SLR takes the form of an overview, detailed summaries, and dissection of each filtered publication \cite{Systematic_mapping_SW_engineering}. % These studies synthesize primary studies in privacy and data markets for IoT devices.

% Why the need for an SR
Consequently, we tackle the research gaps mentioned above with a comprehensive and detailed SLR that aims to guide decision-makers, privacy officers, policymakers, and researchers in the challenge of employing PETs to build or participate in privacy-enhancing IoT data markets.
We guide these stakeholders by identifying, classifying, and describing how PETs are leveraged in the current body of scientific knowledge (see Sections~\ref{sec:PETS}~and~\ref{sec:AETS}) and presenting key findings from our SLR (see Section~\ref{sec:discussion}).
Moreover, for the benefit of the reader, we distill terminology from the extant literature to differentiate and navigate the concepts of PETs in the scope of this SLR (see Section~\ref{sec:terminology}). 
We also organize related work into a reference model for the use of PETs in IoT data markets in distinct categories (see Fig.~\ref{fig:reference_model_layers} and Fig.~\ref{fig:PETS_TREE_WHOLE}) and identify narrow and broad challenges that PETs can tackle or circumvent (see Fig.~\ref{fig:PETS_challenges}). 
Through mapping PETs to the distilled terminology and the identified narrow challenges, we want to support practitioners in making informed decisions about the appropriate PETs to employ in the context of IoT data markets (see Table~\ref{tab:Tech_Challenge_Master}). 

% Practitioners can use these definitions to differentiate PETs and their purpose clearly and discuss privacy-related concepts precisely. 

% Paper structure
The remainder of the paper is structured as follows.
Section~\ref{sec:background} introduces the main concepts of privacy, data markets, and the IoT. 
Section~\ref{sec:research_mehtodology} portrays how we conducted our SLR on publications dated before July 2020, followed by a discussion of related work in Section~\ref{sec:related_work} and a distillation of terminology in Section~\ref{sec:terminology}.
Sections~\ref{sec:PETS}~and~\ref{sec:AETS} present the main results from analyzing the content of the studies in our SLR, followed by a structured review of challenges in Section~\ref{sec:Challenges}. %, structuring challenges as well as the PETs to address them. 
Based on these results and the studies' metadata, we extract a set of key findings and artifacts in Section~\ref{sec:discussion}, where we also provide future work and discuss the limitations of our research.
Finally, Section~\ref{sec:advances_in_markets} updates our study by including new selected publications from May 2022 to July 2020, and Section~\ref{sec:conclusion} concludes with a summary of the results.

\begin{comment}
The popularity of SLRs goes hand in hand with the increasing popularity of the evidence-based research paradigm in software engineering, which promotes the use of empirical and systematic research methods and originally stems from medical research~\cite{Systematic_mapping_SW_engineering}. 
Systematic mapping studies are similar to SLRs because they take advantage of many elements of the SLR methodology. 
Systematic mapping studies differ from SLRs in their reduced complexity, as they examine broader research questions which do not require in-depth analysis but high-level analysis~\cite{Performing_SMs}. 
In contrast to SLRs, systematic mapping studies are very well suited to give an overview of a field of research~\cite{Systematic_mapping_SW_engineering}.
On the other hand, in contrast to systematic mappings studies, SLRs are very well suited to provide a deep assessment of a field of research~\cite{Systematic_mapping_SW_engineering}.
\end{comment}

\section{Background}
\label{sec:background}

\subsection{Privacy}
\label{sec:privacy}

\noindent Given the increased attention and relevance of privacy during the past decades, practitioners have provided many acknowledged definitions. 
For example, A.~F.~Westin~\cite{Westin2015} stated that ``[Privacy is] \textit{the claim of individuals} [...] \textit{to determine for themselves when, how and to what extent information about them is communicated} [...]''. Similar definitions have been given by other authors like G.~A.~Fink et al.~\cite{Fink_2018} or K.~Renaud and D.~Galvez-Cruz~\cite{Renaud}.
Despite these efforts, D.~J.~Solove argued that any attempt to distill a unique, timeless definition is infeasible due to privacy's multifaceted concept~\cite{Solove2015}.
However, in the field of computer science, a narrower definition may be possible by adopting an \textit{attack model} perspective, as the concept of privacy would likely not have emerged if transgressors would not exist: attackers of one's secrets tacitly give meaning to privacy.
% Furthermore, framing a practitioner's work with a user-centered definition, while helpful for writing laws, is not suitable for developing technical applications that aim to enhance user's privacy against malicious entities.
% % A user-centered definition obscures the reason privacy exists and leads to a starting point for developing technical solutions that do not address the underlying problem.
% Hence, we should first create a system that effectively deters attackers from invading the users' secrets.
% Consequently, the system reduces the user's responsibility, the risk, and the appeal of successful attacks.
Therefore, a helpful definition in the context of computer science may be F.~T.~Wu's~\cite{Wu2012}: \textit{``[Privacy] is defined not by what it is, but by what it is not -- it is the absence of a privacy breach that defines a state of privacy.''} 
F.~T.~Wu hence defined privacy as a product of a threat model, the one from M.~Deng et al.~\cite{deng_privacy_2011}, in which a practitioner needs to determine what information to hide, from whom, and what harms should be prevented before defining legal and technical tools.
%In summary of F.~T.~Wu's definition in the field of computer science, privacy is defined based on the failure or success of a specific attacker's goal.

%Finding the origins of privacy and a definition for computer scientists fitting the context of IoT data markets served as motivation and to discuss the core problems from the right angle during this SLR. 

% Throughout human history, the evolution of technology and the activities created around it affected how humans and institutions perceived and acted concerning privacy, shaping, in turn, human culture~\cite{dennedy_technology_2014}.
% However, it was not until the first ICTs such as the radio, television, first computers, or early mobile phones significantly impacted society that governments created the first data protection laws in the world. 
Once IT architectures and tools enhance privacy appropriately, other advantages emerge. 
For example, from an economic perspective, privacy enables data utilization across organizations and applications to create new fair products and services and prevent price discrimination~\cite{privacy_economic_good}.
Furthermore, employing PETs may increase the number of sources and data harvested by institutions because PETs help to overcome regulatory barriers~\cite{kaaniche2016abs}, in addition to mitigating the risk of fines and differentiating and appreciating a brand~\cite{data_exchange_pros}.
Moreover, political freedom and stability may only be achieved by unobtrusive forms of governments~\cite{panopticon}, privacy-enhancing journalism, and less pervasive forms of digital products such as social media that can enable malicious social engineering~\citep{P38}.
Moreover, research indicates that compromising privacy can result in negative long-term economic effects~\cite{mass_surveillance}.

Despite these benefits, and while consumers emphasize that privacy is important to them, they are typically not willing to make small additional efforts or pay for privacy~\cite{kokolakis_privacy_2017}, the so-called \textit{privacy paradox}.  
Thus, in the past decades, governments have enacted rules and laws to protect consumers against violations of their privacy, specifically for data captured through advanced ICTs such as personal computers, the world wide web, or smartphones. Examples include the European Data Protection Directive in 1995, the HIPAA Privacy and Security Rule of 1996, the APEC cross-border privacy rules of 2011, the GDPR of 2016, and the Consumer Privacy Act of 2020 in the USA, which comprises Acts such as the CCPA of 2018 in California.

\subsection{Data markets}
\label{sec:data_marketplaces}

\noindent According to F.~Stahl et al.~\cite{data_market_definition}, there is a misconception in everyday language between the terms ``market'' and ``marketplace'': A marketplace is the implementation of a market in terms of \textit{infrastructure}, time, and location (virtual or physical) where the participants transact. 
Markets are the \textit{environments} where buyers and sellers set the price and quantity of a particular good.
Marketplaces have evolved over millennia; however, the most drastic changes arguably have happened in the past few decades. 
ICT has driven the costs of instant and ubiquitous communication to an often negligible amount, which has led to the digitization of many existing transaction-based ecosystems, including marketplaces~\cite{data_market_definition}.
Moreover, ICT has enabled the creation of virtual marketplaces that did not exist before~\cite{noauthor_e-commerce_2000} -- the most prominent example being e-commerce. 
In this context, data have become goods themselves~\cite{data_market_definition}.
Data markets incentivize institutions to collect more data and to profit from trading, and, in turn, the resulting improvements and innovations benefit the public good~\cite{big_data_econmic_impact}.

Beyond the above formal definition, from the selected studies, we can carve out several characteristics of data marketplaces:
Y.~Li~et al.~\citep{P1} indicated that most of the data marketplaces in operation are centralized, where the platform is run by either a trusted third party (a broker) that coordinates buyers and sellers or by the data owner (e.g., a large institution) who is also selling the data.
Another $15$ selected studies also proposed decentralized architectures employing distributed ledger technology to counter the drawbacks of centralized systems (e.g., single point of failure or trust on a potentially malicious entity).
On the other hand, Z.~Guan~et~al.~\citep{P46} took another perspective, characterizing data trading platforms depending on the number and type of data domains: general platforms include data from any source type, while specialized platforms focus on one domain, e.g., financial, healthcare, or social media.
C.~Perera et al.~\citep{P27} identified two categories for data markets based on the type of participant: companies or private individual customers, e.g., owners of a smart home. 
Altogether, we distilled three dimensions for characterizing data markets: (i) the degree of centralization, (ii) the types and number of data domains, and (iii) the types of sellers and consumers. 
These dimensions permeate most of the identified solutions in this study, and all exhibit individual privacy trade-offs of which practitioners need to be aware (see Section~\ref{sec:discussion}).

% While we find these characterizations insightful, the focus of these selected studies was not to classify data marketplaces. 
% On the other hand, F.~Stahl et al.~\cite{data_market_definition} provides a classification as they compile similar studies to their own.
% F.~Stahl et al.~classify data marketplaces depending on their ownership and in turn on their business model: private (Buy or sell-side system, which is composed of one-to-many relationships or vice-versa), consortium (Buy or sell-side platform, or both, which are composed of few-to-many relationships or vice-versa), and independent (Two-sided marketplace, which is composed of many-to-many relationships). 
% Furthermore, the closer the marketplace is to functioning independently, the more the transactions resemble a market (market forces determine the price), and the closer it gets to private, the more the transactions resemble a hierarchical system (pre-determined prices).

%  V.~Koutsos et al.~\citep{P43} or S.~Kiyomoto et al.~\citep{P33}

\subsection{The Internet of Things}
\label{sec:iot}

%P1 Origin
% In 1999, K.~Ashton was the first practitioner to name a new paradigm emerging from the Internet. The name he proposed was the ``Internet of Things'' (IoT). He introduced this name to explain a new supply chain concept to his employer.
% His use case consisted of tagging parcels with RFID tags, so upon scanning with a device that is connected to the Internet, the parcel location would be shared for tracking. 
% The underlying concept of this simple use case (connecting things) could be applied to anything in the world. 

\noindent The IoT is considered a network of physical devices that leverages sensors to measure and collect information from the real world and support the access and exchange of data via the Internet instantly and ubiquitously~\cite{oberlander2018conceptualizing}.
IoT devices are considered essential for gathering big data~\cite{IoT_to_win}, which in turn brings new opportunities such as targeted advertisement, predictive maintenance, and quality improvements. Consequently, many companies have introduced the IoT in their strategy for participating in the data economy~\cite{oberlander2018conceptualizing} and make substantial investments in the technologies that make them possible: sensors, wireless networks, and cloud computing infrastructure~\cite{IoT_to_win}.
In this SLR, the definition of the IoT includes any device with a CPU connected to the Internet, including sensors in factories, supply chains, or vehicles, and devices such as smartphones, wearables, and computers that people use daily.
These devices act as data collectors and as the gateway to a plethora of applications that collect users' actions and behavior, such as browsers, social media, e-commerce, or media entertainment, as well as sensor data generated in business processes like manufacturing and predictive maintenance, and use them for analyses and predictions.

The design and implementation of data markets are dependent on the IoT.
The ubiquity of IoT devices generates many constellations for different degrees of decentralization, with a myriad of possible sources and prosumer types.
Furthermore, while such ubiquity will likely boost the data economy and its products and services, IoT devices also permeate many aspects of an individual's life, e.g., dealing with highly sensitive healthcare data or capturing sensitive information from a business perspective.
Hence, the sensitivity of the data gathered from IoT devices calls for the implementation of PETs.

\section{Research process}
\label{sec:research_mehtodology}

\subsection{Goal and research questions}
\label{sec:researchquestions}
\noindent We employed the Goal-Question-Metric paradigm~\cite{Basili1994} to formulate the focus of this study as follows: we systematically analyze peer-reviewed literature to provide an overview of the state-of-the-art concerning available research and trade-offs on privacy-enhancing data markets for the IoT as well as potential research gaps from the point of view of both scholars and practitioners. 
Based on this paradigm, the research questions (RQs) we pursued were: 

\paragraph{\textbf{RQ1}} \textit{What relevant PETs enable IoT data markets?} \newline
By answering this RQ, we aim to reveal, describe, and classify PETs in the context of data markets for the IoT based on their fundamentals and applications to give researchers and practitioners an overview of the PETs researched and employed so far.

\paragraph{\textbf{RQ2}} \textit{What challenges and trade-offs hinder privacy-enhancing IoT data markets?} \newline
Through answering this RQ, we account for explicit and implicit challenges depicted and tackled in existing work so that researchers may quickly identify pain points in the field and focus their research.

\begin{comment}
\paragraph{\textbf{RQ1}} \textit{How is research on privacy-enhancing data markets for the IoT characterized in the literature?} \newline
This research question aims to map the publication trends and characteristics of existing research on privacy-enhancing data markets for the IoT by employing the publications' metadata so that researchers may find the most acclaimed publications and authors to further their research.
\end{comment}

\subsection{SLR execution}
\label{sec:SLR_execution}

\begin{figure*}[t!]
\centering
\includegraphics[width=\textwidth]{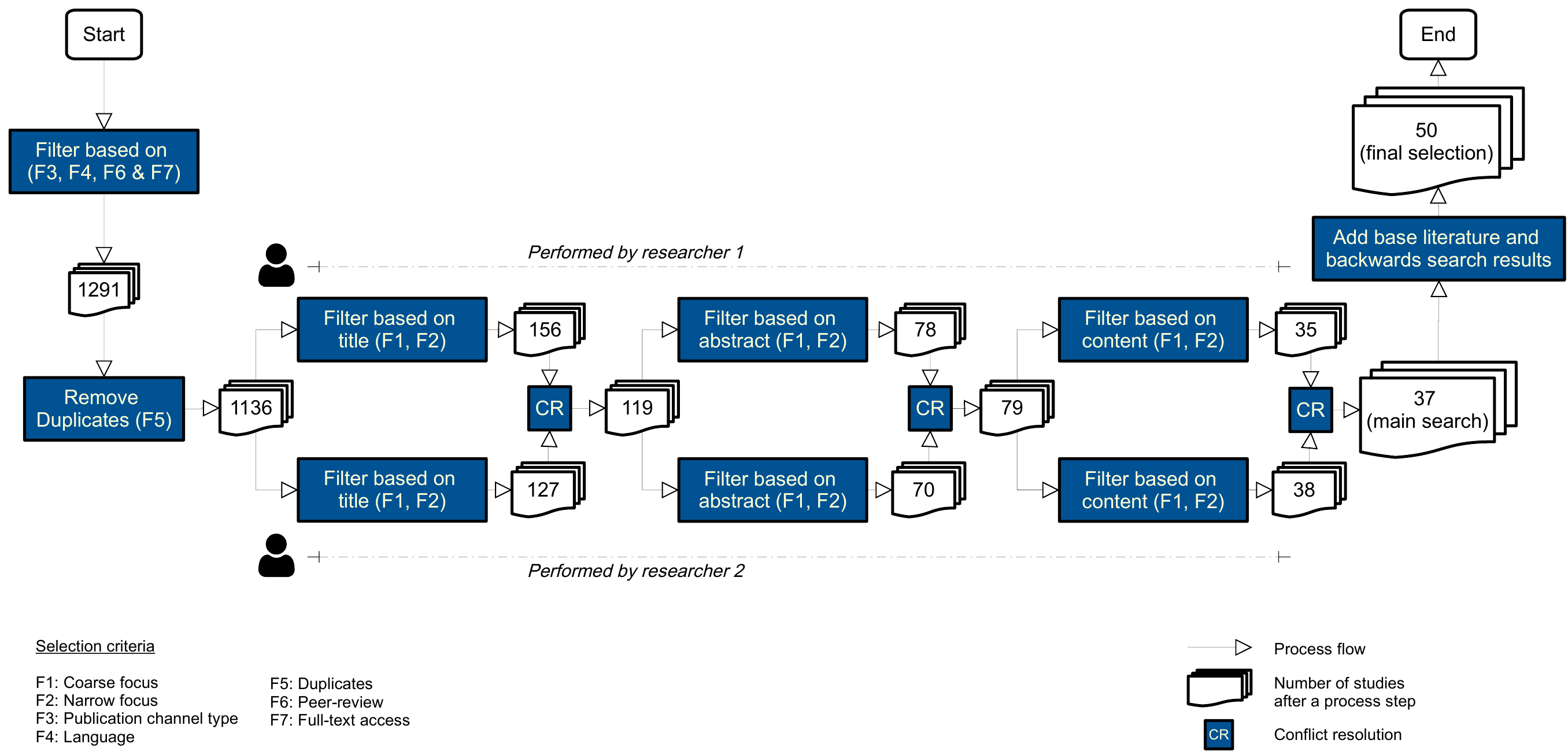}
\caption{Study selection process.}
\label{fig:study selection process}
\end{figure*}

\noindent We conducted a SLR based on the guidelines of B.~A.~Kitchenham and D.~Budgen~\cite{KitchenhamBook}. 
SLRs aim to collect, structure, and summarize the existing evidence and gaps in a particular research field to pave the way for future research. 
Furthermore, SLRs need to provide a rigorous and auditable methodology that can be reviewed and replicated~\cite{Kitchenham2004}. 
SLRs define research questions, and a set of predefined inclusion and exclusion criteria that assess potentially relevant primary studies to answer them~\cite{Kitchenham2007}\cite{Dyba2007}. 
Table~\ref{tab:selection criteria} of the Appendix contains the criteria for this SLR related to focus, quality, and accessibility. 

To conduct the \emph{study search}, we identified the most relevant publications in the field of privacy-enhancing data markets for the IoT to answer our research questions~\cite{Dieste2009}\cite{Zhang2011}.
To obtain a corpus of high-quality publications, we defined a search strategy based on the work of H.~Zhang~et~al.~\cite{Zhang2011}. 
Accordingly, our strategy consisted of three phases: 
%H.~Zhang et al.~proposed to answer four questions to steer the search in a SLR. 
% \textit{Which overall approach should be used in the search process?} 

(i) A preliminary search of the base literature. 
The base literature includes representative papers ($8$) in the field of privacy-enhancing data markets for the IoT known to the researchers before the SLR, and some other publications found manually in the digital library of the researchers' university, which also complied with the criteria described in Table~\ref{tab:selection criteria}.
We created preliminary search strings based on identified keywords and synonyms that we found in the base literature and research questions. 
Afterward, we parsed our base literature with a \href{https://www.sketchengine.eu/}{tool} to analyze frequent phrases and keywords. 
Using the results of this analysis, we refined our search terms~\cite{Kilgarriff2014}. 
Finally, we clustered the search terms into three strings based on the field of this SLR: Privacy, data markets, and IoT.
Altogether, we composed the following search strings: 

\indent\textbf{C1: }\textit{privacy OR private OR encryption OR encrypted OR encrypt OR data protection} \newline 
\indent \textbf{C2: }\textit{data market OR data marketplace OR data trading OR data broker OR data trader OR data auction} \newline 
\indent \textbf{C3: }\textit{Internet of things OR Internet of everything OR IoT OR sensor OR connected devices OR networked devices OR smart devices OR controller OR edge computing OR cloud infrastructure OR machine to machine OR M2M OR web-of-things OR WoT OR mobility OR automotive OR vehicle OR car OR automobile OR industry 4.0 OR smart grids OR V2V OR IIoT OR machine learning OR mobile OR cyber-physical OR microservice OR microcontroller OR micro-service OR micro-controller OR blockchain OR neural network OR smart learning OR automated driving OR autonomous driving OR smart city OR smart factory}. 

Consequently, we defined our final search string as \textbf{C1} AND \textbf{C2} AND \textbf{C3}.

(ii) The main search. Since no single source may contain all the high-quality, relevant publications~\cite{Brereton2007}\cite{Kitchenham2013}, we selected seven electronic databases (see Table~\ref{tab:databases} that focus on computer science or software engineering and, according to L.~Chen~et~al., cover the most relevant databases in these fields~\cite{Chen2010}. 
The time frame that we specified covers any publication included in the selected digital libraries before the 13\textsuperscript{th} of July 2020. 
With the defined search string, time interval, databases, and following the process Fig.~\ref{fig:study selection process} depicts, we collected~$1291$~studies ($1136$~after duplicates removal), which two researchers filtered independently and redundantly by title ($119$~selected out of~$1136$), abstract ($79$~selected out of~$119$), and body ($37$~selected out of~$79$) following the predefined inclusion and exclusion criteria of Table~\ref{tab:selection criteria} to reduce bias. 
After each of the three filtering phases, both researchers resolved conflicts in an informed discussion and attended to the criteria. 

(iii) A backward search of the references of the~$37$~studies resulting from the main search. 
After filtering by title, abstract, and body, considering our inclusion and exclusion criteria, we included another~$11$~studies in our corpus. The process resulted in a total of~$50$~studies from which we subsequently extracted and synthesized data. 
Hence, the SLR yielded a considerable but not excessive number of results. 
Furthermore, thanks to the multiple synonyms in the search string, the~$37$~studies only missed two studies from the base literature. Moreover, the backward search only added a modest number of new works~($11$). 
Thus, the process suggests that the choice of search terms was suitable. \newline

To answer the research questions, we performed a \emph{data extraction} of key information from the~$50$ publications in a structured manner~\cite{Wohlin2012}. 
To reduce the degree of bias, two scientists defined and independently followed an extraction card, which contained the following twenty fields: Authors, cite count, year, country, publication channel, publication type, publication source, research type, research approach, contribution type, tags, topic, subtopic, sub-subtopic, research goal, research questions, study findings, privacy-enhancing architecture or technologies, challenges, and future work.
After the two scientists completed the data extraction, they held an informed discussion to resolve any possible conflicts on the extracted information. 
For the \emph{data synthesis} necessary to answer RQ1 and RQ2, we adapted the ``narrative synthesis'' method described by B.~A.~Kitchenham and D.~Budgen~\cite{KitchenhamBook} and performed the following synthesis procedure: (i) we developed a preliminary synthesis of the findings, followed by (ii) exploring relationships in the data and (iii) refining the preliminary synthesis with the newly acquired knowledge. 
After the refinement, we returned to the second step until we deemed the RQs answered.

Finally, with the goal of including significant updates and reaffirm the findings extracted from our initial research process, we conducted the same systematic search process for publications dated between July 2020 and May 2022 ($24$ new publications) and included the findings in Section~\ref{sec:advances_in_markets}.

\section{Related work}
\label{sec:related_work}

\noindent In our SLR, we found nine secondary studies, i.e., studies that systematize and organize existing knowledge, conducted in the context of privacy enhancements for IoT data markets~\citep{P4}\citep{P8}\citep{P14}\citep{P19}\citep{P23}\citep{P27}\citep{P35}\citep{P38}\citep{P48}, which Table~\ref{tab:review} summarizes.
Some of these studies focused on privacy-related challenges in data markets for the IoT~\citep{P14}\citep{P19}\citep{P27}\citep{P38}, while others delved into PETs from a technical perspective and discussed their challenges and opportunities~\citep{P23}\citep{P48}\citep{P35}. 
Lastly,~\citep{P8} analyzed users' preferences in privacy-enhancing data markets, and~\citep{P4} listed technical design choices for data markets for the IoT. 

These secondary studies provided valuable contributions and built the foundation of our work; however, they focused on different aspects of IoT data markets and, therefore, lacked depth in the concepts we present in this study.
For example,~\citep{P23}\citep{P35}\cite{P48} briefly discussed, altogether, blockchain technology, secure and outsourced computation, $k$-anonymity, and differential privacy and sketched their implications without considering other PETs that we identified in our SLR. 
Furthermore, although~\citep{P19} provided an overview of the available technologies and their challenges, the authors did not discuss PETs in detail, e.g., the study mentioned anonymization but did not delve into $k$-anonymity or differential privacy.
Additionally, the authors only discussed three out of the six challenges we found: the recursive enforcement problem, the utility and privacy trade-off, and attacks on privacy. 
Moreover, J.~Pennekamp~et~al.~\citep{P19}~based their results on observations from exemplary use cases and, therefore, cannot provide the scientific rigor and comprehensiveness of a SLR.
The remaining secondary studies focused on privacy strategies instead of technology, discussed digital rights, described challenges at a high level, or provided a user-centric view of data markets.

\renewcommand{\arraystretch}{1.25}
\begin{table}[!b]
    \footnotesize
    \centering
    \begin{tabular}{p{1.7cm}p{6.25cm}}
        \hline
        \textbf{Layer} & \textbf{Description} \\ \hline
        \textit{Storage} & Persists data for future use. \\ %\hline
        \textit{Processing} & Uses data from storage and runs algorithms on it, typically to extract information. \\ %\hline
        \textit{Communication} & Exchange of data with other devices. \\ %\hline
        \textit{Verification} & Checking via processing whether the data received and the identities involved are authentic. 
        \\ %\hline
        \textit{Sovereignty} & Ability to govern which (sensitive) information the  communication with others exposes. \\ %\hline
        \textit{Consensus} & A special case of communication and verification in which data is compared and synchronized with other devices' data. \\ 
        \bottomrule
    \end{tabular}
    \caption{Short description of the involved layers.}
    \label{tab:short_description_layers}
\end{table}

Regarding PETs classification, some selected papers included in our SLR provided a framework.
Notably, S.~Sharma~et~al.~\citep{P48} considered two categories for classification (\textit{outsourced computations} and \textit{information sharing}), whereas~\citep{P19} provided a more involved classification than~\citep{P48} with five layers: \textit{data security}, \textit{data processing}, \textit{proving support}, \textit{platform capabilities}, and \textit{external measures}.
Furthermore, outside of our SLR, a notable framework developed by A.~Trask~et~al.~\cite{trask_structured_transparency_nodate}, which was heavily inspired by H.~Nissenbaum's work on contextual integrity~\cite{contextual_integrity}, dissected an information flow into input, computation, and output, and assessed privacy and verifiability in each step.
They also wrapped their framework with flow governance, i.e., the information flow rules upon which participants agree.
%To classify PETs in our study, we selected some of the components in the above studies.
%Specifically, we mapped parts of the more general framework of~\cite{trask_structured_transparency_nodate} to our \emph{systematic and structured} classification of PETs, and found overlaps with the categories \textit{platform capabilities} and \textit{external measures} of~\citep{P19}.
%Therefore, in our classification, we placed some of the technologies related to the latter categories under the umbrella of \textit{sovereignty}: policies and smart contracts.
%However, \citep{P19}~and~\cite{trask_structured_transparency_nodate} do not map PETs to the privacy aspects of IoT data markets. 
To provide an improved classification of PETs, we inspired some components of our classification from~\cite{trask_structured_transparency_nodate}\citep{P19} and distilled from the $50$ selected papers the set of layers necessary to build a privacy-enhancing IoT data market: \textit{verification}, \textit{storage}, \textit{communication}, \textit{processing}, and \textit{sovereignty}, which Table~\ref{tab:short_description_layers} describes succinctly. 
Moreover, we also considered important layers necessary for a functional data market that do not require PETs (see Fig.~\ref{fig:reference_model_layers}).

\begin{figure}[!t]
\centering
\includegraphics[scale=0.4]{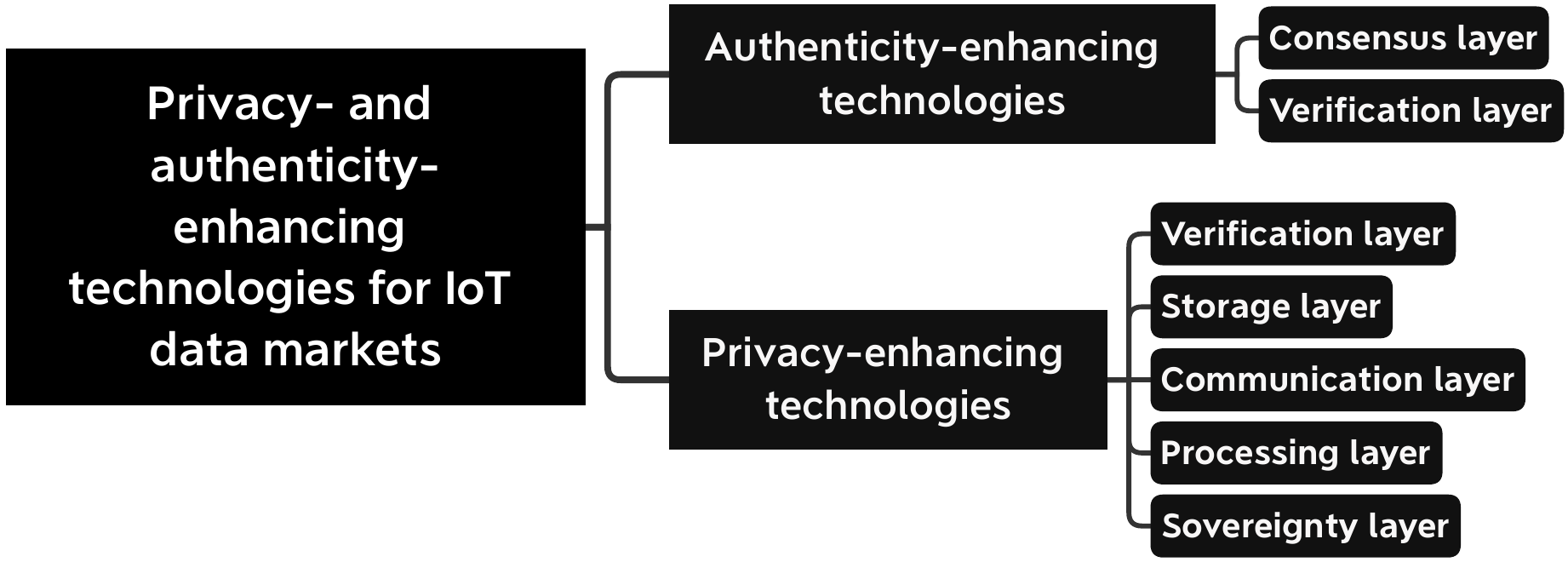}
\caption{Overview of the categories of our classification of the identified technologies among the selected studies. Note that some PETs also enhance authenticity.}
\label{fig:PETs_tree_high_level}
\end{figure}

Furthermore, not all of the technologies included in~\citep{P19} enhance privacy, e.g., version control and most distributed ledger technologies.
Therefore, unlike in~\citep{P19}, we have introduced another branch for technologies focused on authenticity, which we call authenticity-enhancing technologies (AET). 
Note that some PETs accomplish data authenticity or integrity while enhancing privacy or confidentiality, e.g., zero-knowledge proofs, homomorphic encryption, or some digital signatures (see terminology in Section~\ref{sec:terminology}).
Specifically, some PETs can also be AETs, but AETs are not always PETs.
Lastly, we classified the identified AETs into the \textit{verification} and \textit{consensus} layers, which are strongly associated with distributed ledger technology, as they coordinate entities and provide verification guarantees. 
We display our classification framework in Fig.~\ref{fig:PETs_tree_high_level}. 
Accordingly, we structure our key results into privacy-enhancing technologies (PETs, Section~\ref{sec:PETS}), and authenticity-enhancing technologies (AETs, Section~\ref{sec:AETS}). 
%In this paper, we delve into the different layers where we allotted the specific technologies within these two groups.
The authors of the selected $41$ primary studies jointly employed the technologies included in our classification to create holistic or parts of data market architectures for the IoT; Tables~\ref{tab:homomorphic_SMC_Table},~\ref{tab:AnonymousTable}, and~\ref{tab:DLTTable} describe the most salient architectures.

Overall, none of the related work conducted a SLR to create a holistic view of the body of scientific knowledge in privacy-enhancing IoT data markets, which, therefore, indicates the lack of an academically rigorous secondary study in this field~\cite{KitchenhamBook}.
Furthermore, despite the efforts in~\citep{P48}\citep{P19}\cite{trask_structured_transparency_nodate}, there is not yet a \emph{comprehensive classification and fine-grained analysis of technologies and challenges} that researchers have studied in the context of privacy-enhancing IoT data markets (see Sections~\ref{sec:PETS},~\ref{sec:AETS}, and~\ref{sec:Challenges}).
Lastly, unlike other secondary studies, we also provide a detailed mapping of technologies, IoT data market layers, and challenges in Table~\ref{tab:Tech_Challenge_Master}.

\begin{figure*}[htpb]
\centering
\includegraphics[scale=0.68]{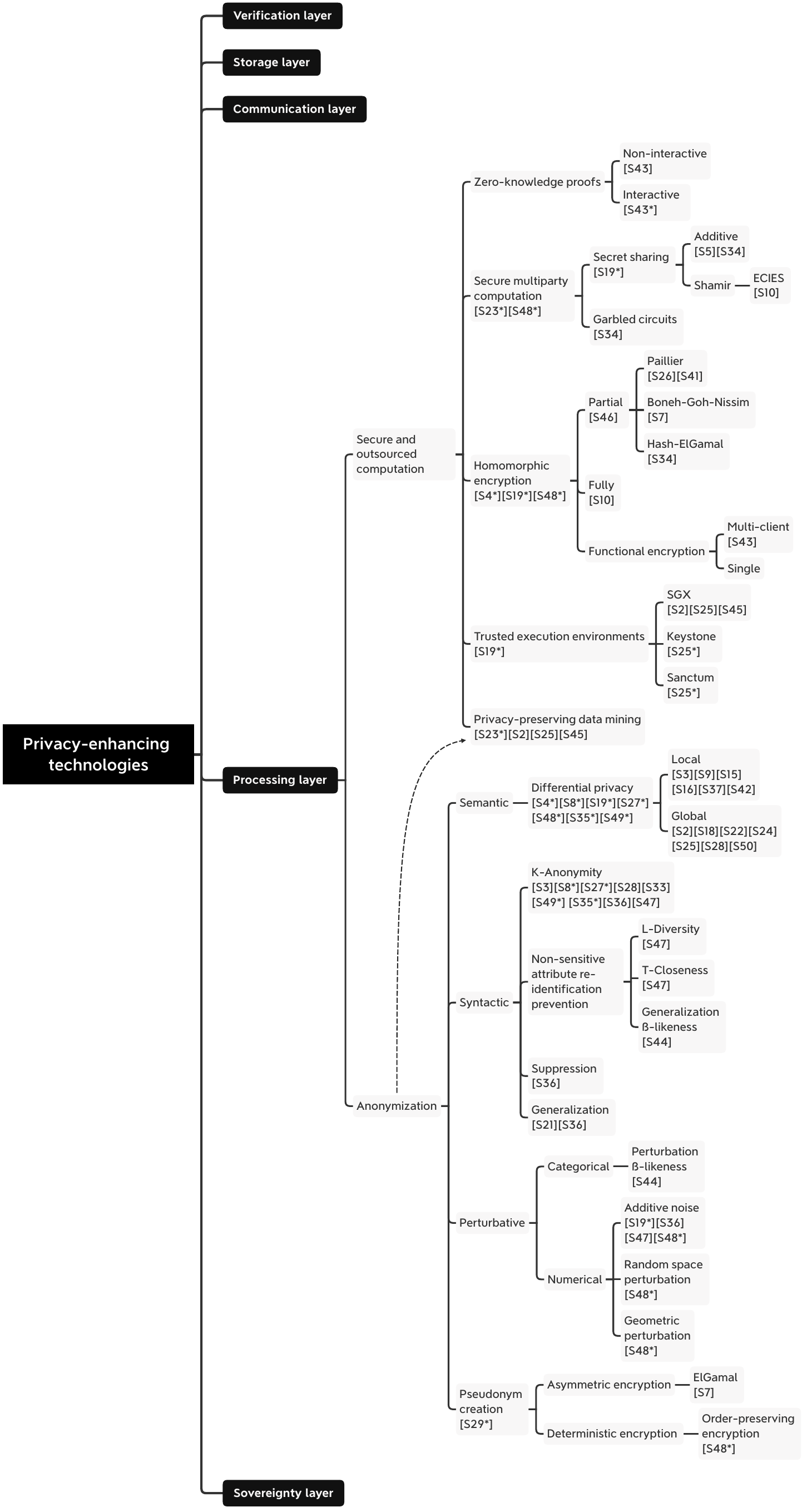}
\caption{Classification of PETs employed for data processing.
Any other privacy approach encountered in the SLR without explicit inclusion of the underlying technology was either not included in a leaf node but in a parent node or completely dismissed if too vague. 
*The publication reviews or briefly comments on the technology without delving in-depth or using it as a building block of the architecture concept, e.g., included for future work.}
\label{fig:PETs_tree_pro_processing}
\end{figure*}

% After our SLR, we also concluded that the terminology that describes PETs and challenges in data markets is quite heterogeneous in the literature, giving different interpretations and complicating the scientific discourse. 
% Consequently, we contribute with definitions distilled from structuring descriptions we found in our research (see Section~\ref{sec:terminology}). 
% Through clarifying a terminology, we aim to enable a more detailed discussion of privacy-related concepts and PETs, and, furthermore, the terminology helped us to build our reference model for IoT data markets (see Fig.~\ref{fig:data_market_framework}) and to map challenges to the corresponding PETs (see Table~\ref{tab:Tech_Challenge_Master}).

% As the privacy preservation domain of data marketplaces is gaining traction recently from scholars and practitioners, such as with initiatives like GAIA-X, a SLR to provide an overview into the current state of the art could advance their understanding of the state-of-the-art.

\section{Terminology}
\label{sec:terminology}

\noindent To help the reader follow our SLR, we first provide some terminology.
These definitions are the distillation of the concepts found in the $50$ selected studies and other seminal studies regarding utility and integrity~\cite{wang_beyond_1996}, and confidentiality and privacy~\cite{priv_secrecy_defs}. 
When we use the word \emph{assure}, a technology fully guarantees the quality of the data or computation. In contrast, when we use the term \emph{enhance}, a technology improves the quality of the data or computation to some extent.
These qualities concern with \emph{authenticity}, \emph{integrity}, \emph{confidentiality}, \emph{privacy}, and \emph{utility}. 
In line with the definition of privacy of Section~\ref{sec:privacy} in the context of computer science, we define these qualities by the absence of an attack against them, if applicable.

\emph{Data authenticity} is preserved when a malicious entity has not tampered with the truthfulness of the original data; truthfulness covers both \emph{provenance} and \emph{integrity}.
In the context of PETs, the degree of authenticity of data can be reduced to enhance privacy.
Correspondingly, \emph{identity authenticity} is preserved when a malicious entity has not impersonated another entity. 
If identity authenticity is assured, then the \emph{provenance} of the data is also assured.
In the context of PETs, the identity of an entity can be concealed to enhance privacy.
\emph{Data integrity} is preserved if data that have been copied and stored or are in motion are equal to the original~\cite{wang_beyond_1996}.
In practical scenarios where data are exchanged, if \emph{integrity} is not preserved, then \emph{data authenticity} is also inherently violated.
\emph{Computational integrity} is preserved when, even in the presence of malicious entities, the output of an algorithm that runs on data is computed correctly.
In the context of PETs, the computation can be concealed to enhance privacy.
Furthermore, some technologies \emph{enhance} \emph{confidentiality}, i.e., ensure that data or specific properties thereof are only shared only with the intended parties.
Furthermore, we refer to \emph{utility} as a measure of the usefulness of data for the successful completion of a task; it is high when the data is \emph{authentic}. 
While PETs reduce \emph{authenticity}, they are helpful in the balancing act between \emph{utility} and \textit{privacy}, as PETs may help to facilitate the sharing of data, which otherwise would not have been revealed (zero utility).

We briefly give some illustrative examples of the interplay of concepts:
A digital signature can \emph{assure} \emph{data integrity} provided the corresponding private key is not accessible to an adversary, and \emph{identity authenticity} if the signature includes a digital certificate that a trusted third party issued; otherwise, digital signatures \emph{cannot} assure \emph{identity authenticity}.
Distributed ledgers can \emph{assure} \emph{data and computational integrity} by replicated storage and computation~\cite{butijn2020blockchains}, but these ledgers cannot enhance \textit{data authenticity}; additionally, replication is often problematic regarding \emph{confidentiality} and, hence, \emph{privacy}~\cite{zhang2019security}.
Furthermore, zero-knowledge proofs can provide evidence for \emph{data and identity authenticity and computational integrity} without violating \emph{privacy}, and truth discovery can \emph{enhance} these qualities independent of \emph{privacy} considerations.
Moreover, privacy-preserving data mining can \emph{enhance} privacy; however, if the right PETs are not employed, qualities such as \emph{computational integrity} may not be \emph{enhanced} or \emph{assured}.
As last examples, technologies such as differential privacy or $k$-anonymization \emph{enhance} privacy by reducing \emph{data authenticity}, and onion routing or ring signatures \emph{enhance} privacy by forgoing or reducing \emph{identity authenticity}, respectively. 
These latter technologies consequently reduce \emph{data utility} in exchange for \emph{privacy}.

Lastly, we are mindful of the term \emph{tackling}, which refers to a technology that \emph{directly} and fully or partially solves a current challenge in the context of privacy enhancement, e.g., the copy problem or the recursive enforcement problem (REP) (see challenges in Section~\ref{sec:Challenges}). 
We use the term \emph{circumvent} when a technology bypasses a problem, i.e., the technology does not directly address the issue. However, the entities that leverage the circumventing technology are still not affected by the problem.
For example, obscuring the data and computation in a third-party server with homomorphic encryption (HE) does \emph{not tackle} the REP; instead, HE \emph{circumvents} such problem because the third-party server cannot see the contents.
On the other hand, distributed ledger technology \emph{tackles} the REP with a redundant and hence tamper-evident storage and execution.

% \input{04_4_Streams}
%\section{Identified technologies that enable IoT data markets (RQ1)}
%\label{sec:RQ2}

The following sections~\ref{sec:PETS}~and~\ref{sec:AETS} describe the technologies that we identified in our SLR.
We provide a new categorization of these technologies based on the characteristics emphasized in the corresponding selected publications and the technical properties described in this Section. 

\section{Privacy-enhancing technologies}
\label{sec:PETS}
\medskip
\subsection{Processing layer}
%\label{Data_processing_RQ2}
\smallskip
\noindent The PETs we included in data processing aim to enhance the privacy of either data inputs, outputs, the intermediate steps of a computation, or a combination thereof while maintaining a high degree of utility. This Section follows the structure of Fig.~\ref{fig:PETs_tree_pro_processing}.

\subsubsection{Secure and outsourced computation}
\label{secure_computation_RQ2}
\smallskip

\noindent Secure and outsourced computation comprises PETs that enhance privacy through confidentiality.
Furthermore, if the PET also employs digital signatures and their cryptography primitives, then the PET can also assure the integrity of the data and computation and identity authenticity in the presence of a digital certificate.

\paragraph{\textbf{Zero-knowledge proofs (ZKPs)}}
With ZKPs, a technology firstly conceived in the 1980s by S.~Goldwasser et al.~\cite{first_introduced_ZKP}, a \textit{verifier} can verify the authenticity of the data and the integrity of a computation conducted by a \textit{prover} without the need to access the data or replicate the computation itself~\cite{zkp_defs}. 
If the statement that is proven is about claims attested in a digital certificate signed by a trusted entity (e.g., age over $18$), ZKPs can verify identity authenticity while keeping the information leaked about the identity minimal.

Specifically, ZKPs exhibit (i) zero-knowledgeness, i.e., the \textit{verifier} learns nothing new from the \textit{prover} beyond the correctness of their statement, (ii) completeness, i.e., the prover can convince the verifier of a correct statement with high probability, and (iii) soundness, i.e., the \textit{prover} cannot convince the  \textit{verifier} of a wrong statement with high probability~\cite{simari_primer_nodate}\citep{P43}.
Furthermore, there are interactive and non-interactive ZKP protocols.
With the latter, there is no need to engage in sequential message exchange, and the prover can convince multiple parties of a claim with a single, potentially short, message~\cite{simari_primer_nodate}. 
These characteristics make non-interactive ZKPs highly attractive for use in blockchains~\cite{zhang2019security}.
ZKPs are also the building block of many anonymous credentials, which are also known as privacy-preserving attributed-based credentials~\cite{kaaniche2020privacy}. They allow the verification of information in a digital certificate without disclosing any unnecessary data, including the highly correlating value of the signature. 
Anonymous credentials were initially proposed in 1985 by D.~Chaum~\cite{chaum_security_1985}, and developed further with ZKPs and blind signatures~\cite{blind_signatures} chiefly by J.~Camenisch and A.~Lysyanskaya~\cite{goos_dynamic_2002} \cite{camenisch_efcient_nodate} and by S.~A.~Brands \cite{zkp_anonymous_credentials}. Lately, anonymous credentials have seen renewed interest also in the context of digital wallets for end users' identity management~\cite{sedlmeir2021digital,schlatt2021designing}.

Within our SLR in IoT data markets, V.~Koutsos~et~al.~\citep{P43} employ non-interactive ZKPs to verify the correct computation of outputs, which, in turn, unlocks the payment from a smart contract in the Agora blockchain, eliminating a third-party verifier.
While ZKPs have their limitations due to computational complexity, typically for the prover, and there is still a considerable gap between cryptographers and software engineers~\cite{zkp_not_efficient}, we expect to see more publications such as V.~Koutsos~et al.'s~\citep{P43}.
This projection is justified by the significant improvements in ZKPs' performance, and ease of use in recent years~\cite{hoffmann_efficient_2019}\cite{nakanishi_efficient_2020} and the availability of an increasing variety of domain-specific programming languages to implement ZKPs, such as \href{https://github.com/zkcrypto/bellman}{bellman} or \href{https://github.com/iden3/circom}{circom} in combination with \href{https://github.com/iden3/snarkjs}{snarkjs}. Recently, first research has emerged that uses ZKPs to prove that a machine learning model was trained correctly on specific data~\cite{zhang2020zkpml}, and there are many opportunities to leverage them in data markets, such as demonstrating that the input data of a computation was signed by a sensor that received a certificate from a trusted third party without revealing the sensor's identity or the data. 
In this case, the digital signature and certificate can be regarded as AETs, while their verification inside a ZKP enhances privacy and, hence, qualifies ZKPs as a PET.

\paragraph{\textbf{Secure multiparty computation (MPC)}}
In broad terms, MPC enables multiple parties to exchange information obliviously and jointly compute a function without revealing individual inputs to each other~\cite{SMC_basics}\cite{P19R43}.
The MPC implementations that we observed in our SLR employ either secret sharing~\citep{P5}\citep{P10}\citep{P34} or garbled circuits~\citep{P34}. 
In secret-sharing-based MPC, each party first obfuscates the input by splitting it into shares.
Secondly, this party distributes the shares among the other computing parties.
Afterward, each party executes arithmetic operations independently on these shares, and finally, all parties share the outputs to reconstruct the result~\cite{SMC_basics}. 

In Shamir's scheme~\cite{Shamir_secret_sharing}, one can specify a minimum of shares that the recipient needs to reconstruct the output, and any combination of fewer shares does not reveal anything about the secret to the receiving entity~\citep{P19} \cite{P19R47}\cite{P19R60}\citep{P48}. 
On the other hand, in additive secret sharing, all the shares are needed.
Outside MPC, Shamir's scheme has been commonly used for key management schemata for cryptographic systems so that if some shares that represent a private key are lost, one can still reconstruct the key with the remaining shares~\cite{Shamir_secret_sharing}.
On the other hand, MPC can also be implemented by garbled circuits~\cite{yao_garble_circuits}, for only two~\cite{garble_circuits_2_party} or multiple~\cite{garbled_circuits_SMC} parties. 
Garbled circuits are protocols that enable secure computation by using functions translated into Boolean circuits, i.e., a sequence of basic logic gates such as AND, XOR, and OR that may be combined to construct any function~\cite{yao_garble_circuits}\cite{GarbledCircuitsPaper}\citep{P48}. 
Garbled circuits make use of oblivious transfer~\cite{ObliviousTransferPaper}, which in turn utilizes asymmetric encryption, and symmetric encryption for encrypting and decrypting each gate's truth table.
Lastly, there are MPC hybrids that combine these approaches~\cite{secret_sharing_garbled_circuits}.

MPC allows computing functions without revealing the inputs to other participating parties. 
MPC protects inputs against brute force attacks and it is to date considered less computationally expensive than alternatives such as fully homomorphic encryption~\cite{yang_comprehensive_2019}.
Drawbacks of MPC include its high processing and communication costs~\cite{kaaniche2020privacy} and sensitivity to network latency, which can considerably decrease the performance~\citep{P5}\citep{P9}\citep{P48}.
Additionally, MPC protocols often need to be supplemented by mechanisms that prevent collusion~\citep{P5}. 
Moreover, since the individually provided inputs are only locally available, one cannot stop malicious entities from jeopardizing the authenticity of the input with false inputs. MPC can only prevent curious entities from learning information. 
A countermeasure for this reduction in accountability is zero-knowledge proofs to enforce the authenticity of participants' local computations while maintaining them confidential~\cite{boyle_practical_2019}. 

Three of the papers in our SLR implement MPC in their architecture: \citep{P5}~uses additive secret sharing, \citep{P10}~employs Shamir's secret sharing, and~\citep{P34} leverages a combination of garbled circuits an additive secret sharing.
Additionally, other publications acknowledge the importance of MPC schemata by including them in their review~\citep{P19}\citep{P23}\citep{P48}. We provide additional details in Table~\ref{tab:homomorphic_SMC_Table}. 
While several \href{https://github.com/rdragos/awesome-mpc#frameworks}{\textit{frameworks for MPC}} are available, the MPC solutions employed by these three publications were handcrafted. This may indicate that the integration of MPC into existing systems requires features that are not yet available with generic tools, such as performance aspects. 

\paragraph{\textbf{Homomorphic encryption (HE)}}
HE allows performing operations on encrypted data (ciphertext) as if they were not encrypted.
After the computation, the entities with the corresponding secret key can decrypt the output~\cite{P19R10}. 
There are variations of HE depending on the diversity of operations it can perform~\cite{HomomorphicPaper}\cite{Theory1}: Fully homomorphic encryption~(FHE) schemata support addition and multiplication, while partially homomorphic encryption~(PHE) schemata allow for only one of these alternatives -- typically in exchange for drastically improved performance. Any other schema in-between is called somewhat homomorphic encryption~\citep{P48}.

Five out of the six studies that use HE in our SLR use a PHE variation~\citep{P41}\citep{P26}\citep{P7}\citep{P34}\citep{P46}.
Each of the former four specifies the name of the employed schema, namely the Paillier cryptosystem in the first two~\cite{P41R17}, Boneh-Goh-Nissim~\cite{boneh_evaluating_2005}, and Hash-ElGamal~\cite{HashElGamal_HE}. 
The latter study only briefly mentions the additive homomorphic property of their handled data. On the other hand,~\citep{P10} uses FHE with a schema called fully homomorphic non-interactive verifiable secret sharing~\cite{FHE}.
Several other articles in our SLR underline the importance of HE~\citep{P4}\citep{P19}\citep{P48}.
On the other hand, V.~Koutsos et al.~\citep{P43} suggested the use of multi-client functional encryption~\cite{boneh_functional_2011}\cite{chotard_decentralized_2018} instead of HE so that the scheme combines data from some individuals with others, and, in turn, malicious entities cannot trace back the output of the computation to a single user, as it may happen in~HE. 
Furthermore, there is a related scheme called functional encryption~\citep{P43} that allows to retrieve a pre-specified function executed on a set of cyphertexts~\cite{boneh_functional_2011}, e.g., decrypt only the mean of a set of encrypted numbers by deriving a function-specific decryption key from the secret key that was used for encrypting the data. 
A summary of papers from our SLR that mention or use HE is given in Table~\ref{tab:homomorphic_SMC_Table}.

The major limitation of FHE is its high computational complexity and the comparatively large storage needs of its cyphertext, which poses a significant challenge for its use and is aggravated in the context of IoT devices' limitations~\citep{P41}\citep{P7}\citep{P48}\citep{P24}.
Therefore, the approach adopted by most authors is the use of PHE instead of FHE~\cite{FHE_limitation}, which, while still not as efficient as other PETs, consumes significantly more computing resources than PHE~\cite{wang_exploring_2015}.

As observed for the case of MPC, while there exist generic frameworks such as \href{https://github.com/microsoft/SEAL}{SEAL}, \href{https://github.com/homenc/HElib}{HElib}, or \href{https://github.com/tfhe/tfhe}{TFHE}, the authors of the publications in our SLR utilized handcrafted solutions, which may indicate the lack of framework versatility or performance.
Overall, practitioners and companies may use HE to perform lightweight functions on data privately on non-local resources, e.g., computing in the cloud, which otherwise would be too expensive to maintain in-house.
MPC would usually be preferred over HE when the inputs to the function belong to multiple parties. 
Nonetheless, some selected publications also employ HE in these cases, e.g., when data brokers determine the winner of an auction~\citep{P26}\citep{P34}\citep{P41}.

\paragraph{\textbf{Trusted execution environments (TEE)}}
TEEs were first defined in~2009 by the Open Mobile Terminal Platform as ``hardware and software components providing facilities necessary to support applications'' that are secure against attacks that aim to retrieve cryptographic key material or other sensitive information. These features include defense against more sophisticated hardware attacks such as probing external memory~\cite{OMTP_TEE} or measuring execution times and energy consumption. 
Moreover, TEEs defend against adversaries who are legitimate owners of the hardware or remote access to the operating system that can run the code themselves.
TEEs allow a user to define secure areas of memory (``enclaves'') that enhance confidentiality and assure data and computation integrity of the code and data loaded in the TEE~\citep{P2}, i.e., any other program outside the enclave cannot act on the data. 
Specifically, TEEs associate unique encryption keys to computer hardware, making software tampering at least as hard as hardware tampering and certifying the computation results within the TEE.
The reason is that the only way to hacking a TEE is physical access to the hardware and, consequently, performing manipulations so that the hardware provides false certifications to bypass remote attestation and sealed storage~\citep{P45}. 
Seal-stored data may not be accessed unless the user employs the correct hardware and software, and remote attestation is a process whereby a trusted third-party assures that the execution of a program in a specific piece of hardware is correct~\citep{P45}.

Four of the selected papers in our SLR leverage TEEs~\citep{P2}\citep{P12}\citep{P25}\citep{P45}, and a review mentions their importance~\citep{P19}.~\citep{P2},~\citep{P25}~and~\citep{P45} proposed TEEs to confidentially train and evaluate machine learning models on data available through a data market.
While the role of TEEs in data markets overlaps with the use of HE and MPC, authors have preferred the latter technologies to enhance confidentiality in auctions and data processing, which may be due to the limited memory TEEs offered at the time.
The reviewed four studies used Intel's Software Guard Extension (SGX)~\cite{sgx_infos}, where Intel is the trusted third party, and, therefore, the single point of failure.
However, practitioners should be mindful of the numerous vulnerabilities present in TEEs~\cite{TEE_vulnerabilities, TEE_attacks_1, TEE_attacks_2}, and Intel's SGX deprecation in 2022~\cite{sgx}, which affects many of the designs found in this SLR dated before July 2020.
Therefore, we suggest practitioners to explore Sanctum~\cite{sanctum}, Keystone~\cite{keystone} and AWS Nitro~\cite{nitro_AWS}.
Specifically, \citep{P2},~\citep{P25}, and~\citep{P45} used SGX for confidential computing, and L.~Ruinian et al.~\citep{P12}, employed SGX for their blockchain architecture to perform ``Proof of Useful Work''.
In this type of consensus mechanism, nodes perform useful computations instead of computing hashes like in Bitcoin or Ethereum mining. 
Moreover, N.~Hynes et al.~\citep{P2} decided to use TEEs to enhance data and computation confidentiality for machine learning algorithms because of the low performance of MPC and HE on machine learning~\cite{wei_framework_2020}. 

On the other hand, we noted that TEEs designed for resource-constrained devices -- potentially at the cost of offering less functionality -- were not prominently discussed in the selected papers. This includes, for instance, ARM~TrustZone~\cite{pinto2019trustzone}, which is relevant as many IoT devices run on ARM processors, and trusted execution modules~\cite{costan2008trusted}.

%Furthermore, while there are open-source proposals for TEEs, e.g., Sanctum~\cite{sanctum} and Keystone~\cite{keystone}, because of their lack of maturity~\cite{P25}, t

\paragraph{\textbf{Privacy-preserving data mining (PPDM)}}
J.~Du~et~al.~\cite{P23} describe PPDM as a means to enhance privacy while extracting useful information from data mining.
Data mining includes ML and conventional statistical analyses such as aggregations (e.g., mean or quantiles).
PPDM is achieved by performing the computation where the data reside, protecting the computation with cryptographic or data perturbation means, or a combination thereof. 
As a comprehensive example, suppose the clients' local data and computation are cryptographically protected and the clients have the capability to perturb data. In that case, the computation can run anywhere, which is accomplished by deploying a ML model and input data in a trusted execution environment (TEE) or implementing a ML model using MPC or HE. With input or computation perturbation, the clients also enhance the privacy of the outputs.

A popular tool for PPDM is federated learning (FL) \cite{konecny_federated_2015}\cite{FL_challenges}\cite{federated_learning}, as it avoids collecting users' data. 
Specifically, FL collaboratively trains a seed ML model across multiple clients' local data, after which a server aggregates the resulting weights to form a unique model (process repeated across rounds).
Researchers have increased the privacy of FL by sharing weights with secure aggregation protocols~\cite{sec_agg} (MPC, Shamir's secret sharing), and protected the privacy in client selection~\cite{FL_HE} and update parameter sharing~\cite{FL_HE_2} with additive HE (i.e., partial HE). 
Alternatively, split learning approaches~\cite{vepakomma2018split}\cite{gupta_distributed_2018} decompose neural networks' layers into elements and, thus, the input data and labels do not need to be within the same machine.
Split learning presents advantages over FL when the local hardware for computations belongs to different network speeds or hardware configurations~\cite{poirot_split_nodate}.
Additionally, gossip learning~\cite{9006216}\cite{ormandi_gossip_2013} proposes a framework whereby multiple models perform a random walk over clients, where they are trained and merged with other models they encounter.

Another way to achieve PPDM is by perturbing input data or weights of the ML model with anonymization techniques such as differential privacy (DP), resulting in privacy-enhancing optimization schemata like DP-stochastic-gradient-descent (DP-SGD)~\cite{DP_SGD}.
DP-SGD perturbs the weights' updates with noise and, therefore, one may not reconstruct the inputs based on the outputs, which may happen in ML or stand-alone FL~\cite{wei_framework_2020}.
Practitioners can plug in weight DP perturbation in central ML, FL, gossip learning, or split learning,  in combination with MPC as well. 
We depict such leverage of anonymization technologies for PPDM with the dashed line connecting both elements in Fig.~\ref{fig:PETs_tree_pro_processing}.

With PPDM, individuals may enjoy a higher degree of privacy than outsourcing the computation transparently to a trusted third party.
Data markets can offer an infrastructure leveraged by PPDM, where data prosumers and consumers only need to provide the input data and ML models, much like the studies in our SLR propose~\citep{P2}\citep{P25}\citep{P45} using TEEs to train models with DP. 
Like data, trained ML models could also be exchanged in markets.

% gossip learning \cite{ormandi_gossip_2013}
% " The basic idea of gossip learning is that many models perform a random walk over the network, while being updated at every node they visit, and while being combined (merged) with other models they encounter." 
% "This paper identifies the conditions in which gossip learning can and cannot be applied, and introduces extensions that mitigate some of its limitations." \cite{9006216}

\subsubsection{Anonymization}
\label{Anonymization_RQ2}
\smallskip

\noindent While the previously presented PETs hide sensitive data from unsolicited parties and, thus, provide confidentiality while enhancing or assuring data and computation integrity, the authorized receiver of the plaintext may reverse engineer the output and correlate data records with individuals (re-identification attack). 
Consequently, employing only secure and outsourced computation PETs is insufficient to provide the required degree of privacy in cases where the recipient may not be fully trusted.
Anonymization technologies can help in these situations by protecting non-explicit identifiers and sensitive attributes~\citep{P47}\citep{P44}. The cost of this protection is forgoing data authenticity and thus decreasing utility.
Given the frequency of re-identification attacks, anonymization should be a critical element of any survey or modern online application, and, in particular, IoT data markets~\citep{P8}.
One may observe that anonymization technologies rely on statistics, probability theory, and heuristics, while secure and outsourced computation usually employs cryptography and trusted hardware.

This sub-section describes our findings for the most employed anonymization technologies identified in our SLR.
We categorize most of them into two groups~\cite{P28R9}. 
\emph{Syntactic} technologies provide a numerical value to the degree of individuals' protection in a dataset, resulting in a perceptible perturbation of data, e.g., generalizing the values $42$, $44$, and $45$ to the interval $[40, 45]$ such that it is harder for an attacker to distinguish between the three individuals.
\emph{Semantic} technologies enforce a privacy definition to a learning mechanism executed over a dataset, namely differential privacy, whereby the output distribution of the mechanism should be insensitive to the removal or addition of an individual in the dataset.
Typically, the property is fulfilled by adding calibrated noise to the output of a mechanism, yielding a result that is not syntactically different from the original value, e.g., $42$ could become $45$ after noise addition.

Semantic technologies have an advantage over syntactic technologies, as they provide a mathematical guarantee of privacy agnostic to background information, i.e., an attacker cannot use related information to re-identify an individual in the dataset.
Additionally, we discuss other anonymization technologies not covered in these two groups, namely noise perturbation and pseudonym creation. Perturbation, in this context, is not classified as semantic because its process does not necessarily provide a formal semantic privacy guarantee (such as in differential privacy) and, simultaneously, the outputs are not syntactically modified.
In essence, anonymization techniques obfuscate information by perturbing the data during measurement or processing~\cite{kaaniche2020privacy}. 
From this perspective, anonymization can be understood as a kind of statistical disclosure control~\cite{DP_critic}, and, thus, also encompasses semantic techniques such as differential privacy. 

\paragraph{\textbf{Syntactic technologies}}
We identify the implementation of the privacy definitions of $k$-anonymity and its variations $l$-diversity and $t$-closeness, a newly proposed model called $\beta$-likeness, and also their building-blocks: generalization, and suppression.
The most frequently utilized model for syntactic anonymization in our SLR is $k$-anonymity~\citep{P3}\citep{P28}\citep{P33}\citep{P36}\citep{P47}, which was also reviewed or highlighted by~\citep{P8}\citep{P27}\citep{P49}\citep{P35}.
$K$-anonymity is a privacy model that guarantees any individual in a dataset to be indistinguishable from at least $k-1$ others.
$K$-anonymization, i.e., altering a dataset to fulfill $k$-anonymity, clusters a set of sensitive attribute values into equivalence classes of size $k$.
However, finding an optimal value of $k$ for minimum information loss is NP-hard. 
Thus, researchers have proposed alternative heuristics~\cite{P44R23}.
Nonetheless, some of the selected studies used the building blocks of $k$-anonymization (transformations): generalization~\citep{P21}\citep{P36} and suppression~\citep{P36}.
Suppression deletes selected data points, while generalization substitutes data points for others that belong to a higher level in a manually pre-defined hierarchy, e.g., substituting a city by a country to make the location less detailed.

The selected studies~\citep{P28}\citep{P33} applied $k$-anonymization to aggregate data from a set of entities.
M.~A.~Alsheikh~et~al.~\citep{P47} innovated upon~\citep{P28}\citep{P33} by also employing the $l$-diversity model to ensure at least $l$ different values in sensitive attributes, and $t$-closeness so that the distribution of the sensitive attributes within each equivalence class was at most at a distance $t$ from the overall dataset distribution of that attribute.
These two models have their own limitations, they prevent homogeneity and external knowledge attacks ($l$-diversity) and skewness and similarity attacks ($t$-closeness) \cite{P28R9}, to which $k$-anonymity is vulnerable.
Furthermore, D.~Sánchez~et~al.~\citep{P36} tailored the use of $k$-anonymity based on record history, privacy policies, and disclosure context.
Their new approach prevented a significant decrease in the data utility compared to homogeneously applying $k$-anonymity to all individuals' records equally.

Nonetheless, there are detractors of syntactic technologies in data markets because of the need for a centralized intermediary that sees and aggregates the data in a, e.g., $k$-anonymous fashion~\citep{P18}. 
%However, practitioners could (theoretically) circumvent the need for a trusted third party by applying HE, MPC, or TEEs with a semantic definition of privacy (differential privacy) before the aggregated data is released.
Moreover, J.~Cao~et~al.~\citep{P44} stated that these conventional syntactic approaches are not sufficient because they lack an attacker perspective in the model. 
For this reason, they designed a novel model called $\beta$-likeness that explicitly bounds the additional knowledge that an adversary gains from seeing the released data.

In the context of this SLR, $k$-anonymization is mainly employed before sharing data in an IoT data market.
However, researchers also employ $k$-anonymity for privacy-enhancing location-based services that exchange location data in IoT data markets, whereby similar fake locations hide the real ones.
This type of approach fits well with IoT devices embedded in phones, vehicles, and laptops, among other mobile \emph{things}. 
Some of the approaches named by~\citep{P3} were \textit{cloaking}, which consists of sending a more extensive region that encompasses the real one, and \textit{geomasking}, whereby the real location is randomly displaced outside of an inner circle but within an outer one.
D.~Lopez~et~al.~\citep{P3} adopted \textit{geomasking} for situations where low accuracy is sufficient, and a high degree of privacy is required.
A modern alternative to releasing anonymized data is synthetic data generation, which creates data by randomly sampling from a distribution representative of the real data. 
Practitioners can employ generative adversarial networks (GANs)~\cite{dikici2020constrained}, or GANs with differential privacy for a higher protection~\cite{torfi2020differentially} to synthesize data.
Synthetic data could be helpful in some contexts as they ``look'' similar to the real data (unlike fulfilling $k$-anonymity), e.g., developing applications before testing them with the real data. \smallskip

Altogether, anonymization technologies and PPDM compose the building blocks of \emph{statistical disclosure control}~\cite{SDC}, which organizations may leverage \emph{internally} in their privacy-preserving data management and analysis solutions, or \emph{externally}, by using privacy-enhancing publishing solutions~\cite{R5_2} in, e.g., data markets.
Among past surveys focused on the latter solutions (namely $k$-anonymity and related models, in addition to a few cryptographic primitives), the reader may refer to~\cite{R5_2}\cite{R5_3}\cite{R5_4}\cite{R5_5}\cite{R5_6} for further specialized reading. 
Most notably, C.~C.~Aggarwal~and~P.~S.~Yu~\cite{R5_4} provide a comprehensive survey of syntactic models and differential privacy and their \emph{attacks}, and M.~Cunha~et~al.~\cite{R5_6} compile a helpful \emph{mapping} of data types to the appropriate syntactic and semantic techniques for anonymization.

\paragraph{\textbf{Semantic technologies}}
Introduced by C.~Dwork~et~al.~\cite{DP_original} in 2006, differential privacy (DP) proposes a formal guarantee of privacy that has become the golden standard for researchers~\citep{P37}.
DP appears in its pure form or one of its flavors in $13$ of the $35$ studies that propose a solution in our SLR.
Furthermore, another seven studies refer to DP to underline its importance or drawbacks. 
The potential reasons behind the high number of references and use of DP are multifaceted. 
While HE or MPC may protect the inputs' and computations' confidentiality, they do not prevent reverse-engineering the outputs (re-identification attacks). 
Moreover, syntactic technologies or other conventional anonymization technologies, e.g., additive noise, lack a mathematical guarantee of privacy and are subject to background knowledge attacks.
DP, however, tackles these issues.
%, DP has become the de facto standard to address privacy for many researchers~\citep{P37}.
% The local version of DP, e.g., randomized response or using the Laplacian or Exponential mechanism~\cite{dwork_algorithmic_2013} on a single individual, can be deployed without the need for a trusted third party to aggregate the data, unlike with $k$-anonymity and its variations~\citep{P18}. 

In broad terms, DP guarantees that the output distribution of an analysis (a statistical query or a ML model) over a dataset is ``essentially'' identical, irrespective of the presence or absence of an individual in the dataset.
Additionally, DP is agnostic to auxiliary information available in the present or the future.
DP is \emph{typically} achieved by adding random noise sampled from a probability distribution such as the Laplacian or the Gaussian.
Specifically, the noise limits the output distribution \emph{difference} of an analysis executed over two datasets (one \emph{with} and one \emph{without} an individual) to be no greater than an upper bound, making the outputs ``differentially'' indistinguishable. 
Overall, DP bounds the amount of new information gained by an attacker after observing the output of an analysis.

The set of selected studies of our SLR that used DP in their proposed solutions are~\citep{P2}\citep{P3}\citep{P9}\citep{P15} \citep{P16}\citep{P18}\citep{P22}\citep{P24}\citep{P25}\citep{P28} \citep{P37}\citep{P42}\citep{P50}. Tables~\ref{tab:AnonymousTable} and~\ref{tab:DLTTable} summarize their proposed architectures.
Six of these studies employ DP locally~\citep{P3}\citep{P9}\citep{P15}\citep{P16}\citep{P37}\citep{P42}, i.e., the noise is added to the data of an individual on the client-side. 
In contrast, the rest of the studies apply DP centrally, i.e., on aggregated data on the server-side.
Furthermore, we can cluster the studies into those that focus on a data trading design for data markets~\citep{P22}\citep{P37}\citep{P15}\citep{P50}, crowdsensing data markets~\citep{P42}\citep{P9}\citep{P16}\citep{P24}, and architectures that host a data market in an attempt to achieve end-to-end privacy~\citep{P2}\citep{P3}\citep{P18}\citep{P25}\citep{P28}.

While DP offers a mathematical privacy guarantee, DP is not a panacea. 
DP still holds flaws in its real-world implementations~\cite{dwork_exposed_2017} that the research community and practitioners should address. 
Moreover, DP's combination with ML needs further improvements regarding balancing privacy and accuracy~\cite{DP_critic}.
In our SLR, S.~Sharma~et~al.~\citep{P48} and D.~Sánchez and A.~Viejo~\citep{P36} identify two specific problems with DP: firstly, DP cannot be used when a high level of accuracy is required~\citep{P48}, e.g., analyzing data from the brakes of vehicles to improve safety.
Secondly, releasing an entire dataset with current DP approaches is troublesome.
Despite these challenges, the authors of~\citep{P2}\citep{P50} argue that the benefits of DP predominate, as DP can adapt to many use-cases and allows a practitioner to fine-tune the added noise to enhance privacy.

As we already noted with ZKP, MPC and HE, the authors of the selected publications that used DP did not employ open-source DP libraries such as \href{https://github.com/opendp}{OpenDP},
\href{https://github.com/google/differential-privacy}{Google-DP},
\href{https://github.com/IBM/differential-privacy-library}{diffprivlib}, \href{https://github.com/tensorflow/privacy}{TensorFlow privacy}, or \href{https://github.com/uvm-plaid/chorus}{Chorus}. Instead, they used handcrafted implementations of DP. 
Aside from syntactic and semantic technologies, other anonymization technologies are simpler to implement, e.g., sampled data release, character masking, truncation, rounding, top and bottom coding, data swapping, randomization, creating pseudonyms, character scrambling, microaggregation, or noise perturbation~\cite{bondel_towards_nodate}\cite{ISO20889}. 
Moreover, N.~Li~et~al.~\cite{DP_k_anonymity} designed an algorithm that combines the syntactic definition of $k$-anonymity with DP.
The two other anonymization technologies employed by another three studies were noise perturbation~\citep{P36}\citep{P47} and pseudonym creation~\citep{P7}. \newline

\paragraph{\textbf{Perturbative anonymization}}
Perturbation relies on the use of noise to obfuscate sensitive information.
One of the simplest forms of perturbation is additive noise, employed in~\citep{P36}~and~\citep{P47}.
Additive noise consists of adding to a deterministic value a random value sampled from a uniform distribution whose bounds are set by a specific percentage of the deterministic value.
%Additive noise consists on adding to a true value $V$ a random value $\mathcal{U}$ sampled from a uniform distribution whose bounds are determined by a specific percentage of $V$, i.e., $\mathcal{U} \sim \mathrm{unif}[- \alpha V, \alpha V]$, where $0 \leq \alpha \leq 1$, and the output $O = V + \mathcal{U}$.
Furthermore,~\citep{P48} reviews two novel perturbative technologies: 
First, random space perturbation~\cite{rasp}, which strives to protect the privacy of cloud-stored data by utilizing a confluence of order-preserving encryption, dimensionality expansion, random noise injection, and projection. 
Second, geometric perturbation~\cite{chen_geometric_2011}, which is motivated by the idea of protecting the geometric transformations that a machine learning model may perform on a dataset rather than the data itself.
While perturbative technologies aim to tackle the same problems, unlike DP, they do not provide mathematical guarantees of privacy, even though some are also based on noise addition.

\paragraph{\textbf{Pseudonym creation}}
Pseudonym creation is applied to direct identifiers, e.g., names or social security numbers, to enhance privacy while uniquely identifying each record.
Practitioners create pseudonyms by hashing or deterministically encrypting an identifier, e.g., using order-preserving encryption~\citep{P48}, or by applying asymmetric key encryption like ElGamal~\citep{P7}.
However, researchers have demonstrated that pseudo-anonymization falters against some attacks like profiling, task tracing, or re-identification~\citep{P35}.

\begin{figure}[t!]
\centering
\includegraphics[scale=0.4]{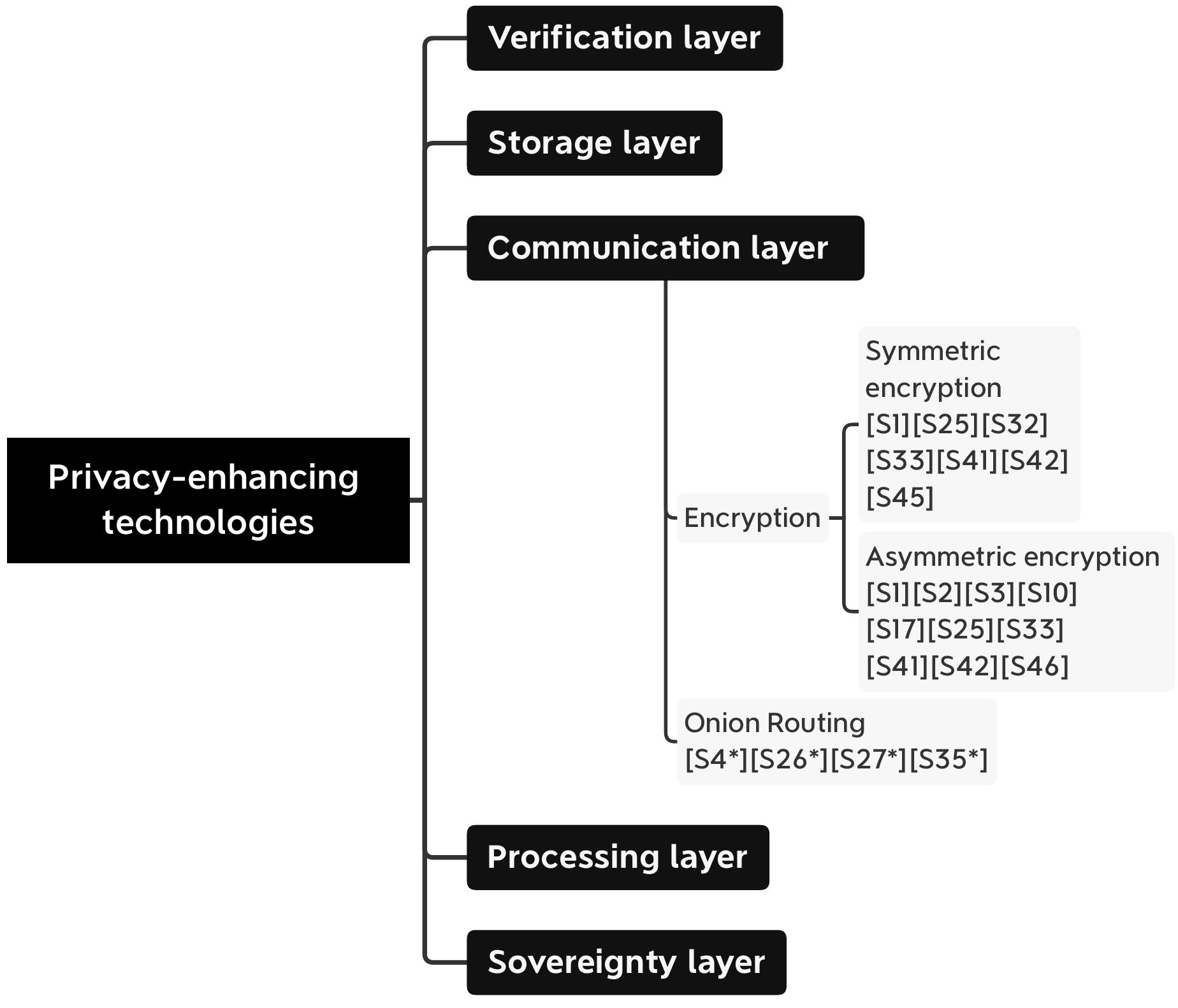}
\caption{Classification of PETs employed for communication.}
\label{fig:PETs_tree_comm}
\end{figure}

\subsection{Communication layer}
%\label{Data_processing_RQ2}
\smallskip

\noindent The PETs included in this Section enhance the confidentiality of data in transit or of the sender's identity (see Fig.~\ref{fig:PETs_tree_comm}).
These PETs rely on cryptography.

\paragraph{\textbf{Encryption}}
Encryption is one of the most fundamental technologies to enhance confidentiality~\citep{P19} because after encrypting a piece of data (cipher), only the anointed holders of a decryption key can decipher such data.
We underline that encryption cannot guarantee privacy because nothing stops an intended receiver from publicly sharing the decrypted message; this also emphasizes employing anonymization PETs.
Encryption may be symmetric (one key to both encrypt and decrypt data) or asymmetric, known as public-key cryptography (two keys, a public key to encrypt and a private key to decrypt, or vice versa).
Encryption is the building block of virtually every secure communication established through a network and takes a key role in digital signatures.

\begin{figure}[!tb]
\centering
\includegraphics[scale=0.4]{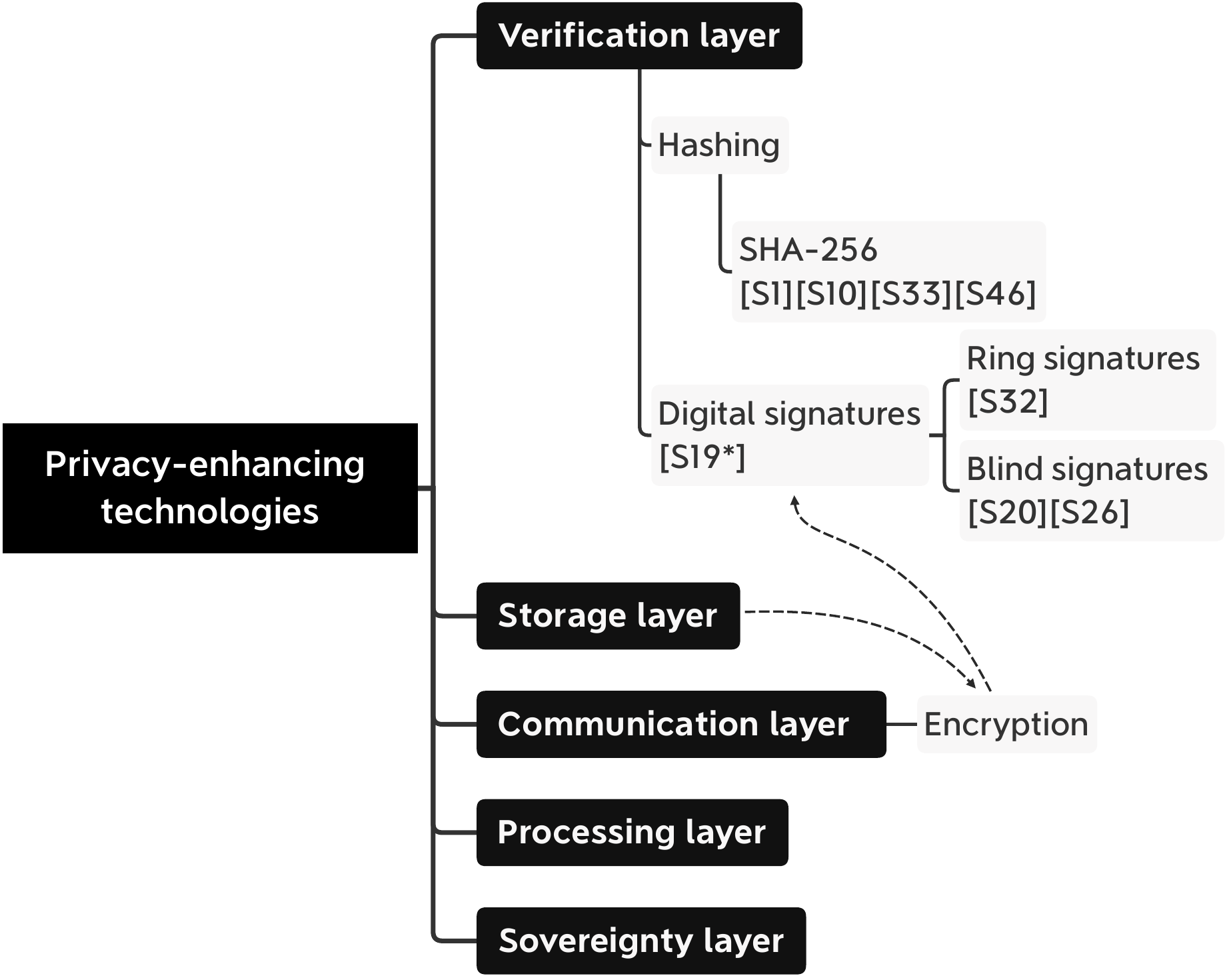}
\caption{Classification of PETs employed for verification.} 
%Refer to the caption of Fig.~\ref{fig:PETs_tree_pro_processing} for details on the inclusion criteria.}
\label{fig:PETs_tree_ver_processing}
\end{figure}

While some publications from our SLR employed asymmetric encryption for the confidential communication of data \citep{P1}\citep{P2}\citep{P3}\citep{P10}\citep{P17}\citep{P25}\citep{P33}\citep{P41}\citep{P42}\citep{P46} (most of these publications also employed asymmetric encryption for digital signatures, hence the high frequency of digital signatures in Fig.~\ref{fig:PETs_distributions}), other publications such as \citep{P1}\citep{P25}\citep{P32}\citep{P33}\citep{P41}\citep{P41}\citep{P45} employed symmetric encryption to also confidentially store data.
Naturally, encryption is also a building block for digital signatures (verification layer) and for the storage layer to maintain data at rest confidential. We depict this relationship with the dashed lines connecting these elements in Fig.~\ref{fig:PETs_tree_ver_processing}. 

\paragraph{\textbf{Onion routing}}
Onion routing, the backbone of the P2P network resulting from the Tor project \cite{TorPaper}, consists of a series of re-transmission steps through the network's nodes.
A sender's message is encrypted once for each step. The intermediaries decrypt only their appointed encryption layer. Thus, the node only knows the immediate sender and receiver but not the origin of the chain of messages.
As the messages are encrypted, the nodes cannot see the contents either.
Overall, onion routing renders one's messages unreadable and untraceable. 
The paper that suggests employing onion routing in a data market context is~\citep{P26}, which some of the identified reviews equally appreciate~\citep{P4}\citep{P27}\citep{P35}.

However, some drawbacks exist. 
Implementations backed by Tor have high-latency and redundant communication that challenges bandwidth, which can be hard to align with high transactional environments, such as IoT data markets.
Moreover, if an architecture decides to use Tor, the network is often blocked by IT departments within organizations or even subject to state-level censorship by some governments~\cite{TorPaper}. 
Therefore,  practitioners can use alternative technologies such as a VPN to enhance entities' privacy in a network in these contexts. However, these typically centralized alternatives generally offer lower anonymity guarantees, e.g., a VPN provider can identify a user~\cite{kaaniche2020privacy}. 

Because there is no central authority to set privacy policies unilaterally, one must remember that onion routing enhances privacy by preventing malicious entities from collecting IP addresses to identify users. Onion routing will not help if the data that users submit to the network is intrinsically sensitive or correlating.  
To tackle these limitations, practitioners may use onion routing as a building block of a more extensive privacy-enhancing system that leverages other PETs~\citep{P26}. 

Notably, the publications surveyed in our SLR did not include many other untraceability protocols, which would include mixnet-based alternatives to onion routing (e.g., anonymous remailers, Chaum’s mixes~\cite{chaum1981mixes}), DC-nets~\cite{chaum1988DCnets}, or peer-to-peer anonymous communication systems. An extensive overview of these systems is given by~Ren~and~Wu~\cite{ren2010survey}.

\subsection{Storage layer}
%\label{Data_processing_RQ2}
\smallskip

\noindent The authors of the selected papers that propose confidential storage functionality in their architecture leverage symmetric encryption, mostly AES \citep{P1}\citep{P25}\citep{P32}\citep{P33}\citep{P41}\citep{P41}\citep{P45} (encryption is described in the \textit{communication layer}).
Furthermore, researchers could leverage InterPlanetary File Systems (IPFS)~\cite{IPFS} to compensate for the lack of storage capacity in blockchains to some extent.
Specifically, IPFS is a peer-to-peer protocol for data storage and access in a distributed file system.
Among the PETs in the processing layer, practitioners could employ homomorphic encryption~\cite{9582232} to encrypt and store certain types of data, so that data are readily available to compute confidential operations.
Furthermore, unless strictly necessary, practitioners should store encrypted data that are, in turn, \emph{anonymized} with syntactic or semantic technologies.
In case of a breach that leaks the decryption key, anonymized data would reduce the likelihood of attackers re-identifying individuals. 

% IPFS~\cite{9311120}

\subsection{Verification layer}
\label{Data_verification}
\smallskip

\noindent Some of the PETs that support data processing cannot verify the authenticity of data, identities, or the integrity of data~\citep{P19} by themselves.
The PETs we include in this Section accomplish these verifications with different levels of privacy enhancement.
The data processing PETs that can assure identity authenticity and data integrity use the digital signatures of the \textit{verification layer} and the encryption technologies of the \textit{communication layer} as building blocks.
Furthermore, the credibility associated with verifying the information exchanged, analysis outputs, and identities can increase the willingness of users to share data~\citep{P8}. 
To navigate this Section, we refer to Fig.~\ref{fig:PETs_tree_ver_processing}.

\paragraph{\textbf{Privacy-enhancing digital signatures (DSs)}} 
DS schemata assure data integrity and identity authenticity if accompanied by a digital certificate. As a consequence, DSs also provide non-repudiation~\citep{P1}, i.e., actions that an entity cannot deny later. 
The steps that usually constitute a DS scheme are private and public key generation, encrypting a digest of data with a private key, and a signature verifier that employs the public key to check whether the sender signed the data with the private key. 

DSs and the encryption primitives of the communication layer are so fundamental that one of the selected studies solely relies on HTTPS for their data market architecture~\citep{P17}. 
However, this architecture does not consider privacy beyond data in transit. Hence, most selected studies rely on multiple PETs.
Moreover, although not all of the selected studies explicitly mention DSs, we can safely assume that since DSs are already a living part of virtually any enterprise IT system, most selected studies employ them in their architectures (hence the high frequency of DS utilization in Fig.~\ref{fig:PETs_distributions}). 
Nonetheless, while DSs allow verifying the integrity of data or the authentic identity of the sender, users still need to trust the sender with the authenticity of the data.

So far, we have only described DS as an authenticity-enhancing technology. However, some of the studies selected in this SLR employed two DS schemata based on asymmetric encryption primitives that make DSs privacy-enhancing:
\begin{itemize}
    \item \textbf{Ring signatures}~\citep{P32}, whereby any party within a pre-defined set of parties could have been the signer of a message. Thus, the identity of the authentic signer is kept hidden~\cite{P32R26}\cite{GroupSignaturesPaper}. 
    %These signatures provide some degree of privacy that conventional DSs cannot provide, yet they might not suit contexts where the verification of the authentic identity is necessary.
    \item \textbf{Blind signatures}~\citep{P20}\citep{P26}, whereby the signer does not have access to the content being signed~\cite{blind_signatures}. 
    It is possible to use blind signatures in combination with zero-knowledge proofs to convince the signer that the content to be signed has the expected properties.
    Also, one can make an entity sign multiple contents and allow for spot checks to detect fraud.
    The latter procedure has been employed in the first approaches toward privacy-enhancing payments~\cite{chaum_security_1985}.
\end{itemize}

\paragraph{\textbf{Hashing}}
Hashing is a tool to deterministically map data of an arbitrary length to a fixed output length.
In the context of privacy and verification, and aligned explicitly with some of the selected studies~\cite{P1}\cite{P10}\cite{P33}\cite{P46}, hashing is used to verify the integrity of transferred data by hashing the data and making the hash public before transferring the data.
In this manner, the recipient can verify the integrity of the confidentially transferred data by comparing the hash of the received data with the previously published hash. Provided the entropy of the data is sufficiently high, nobody except the intended recipient can determine the data from the published hashed value. Hence, hashing can be considered a form of version control with a privacy component.

The hash function employed by the publications mentioned above was SHA-256. Their authors commonly use the published hashed data on distributed ledger technologies to ensure immutability and availability.
In this setting, hashing enhances the confidentiality of the sender's data while the parties (network nodes) ensuring the ledger's integrity (and inherently the persisted hash) cannot unveil the original data.
The original data is only viewed by the intended receiver, which validates the integrity of the data received through another channel with the hash persisted in the ledger.

\begin{figure}
\centering
\includegraphics[scale=0.4]{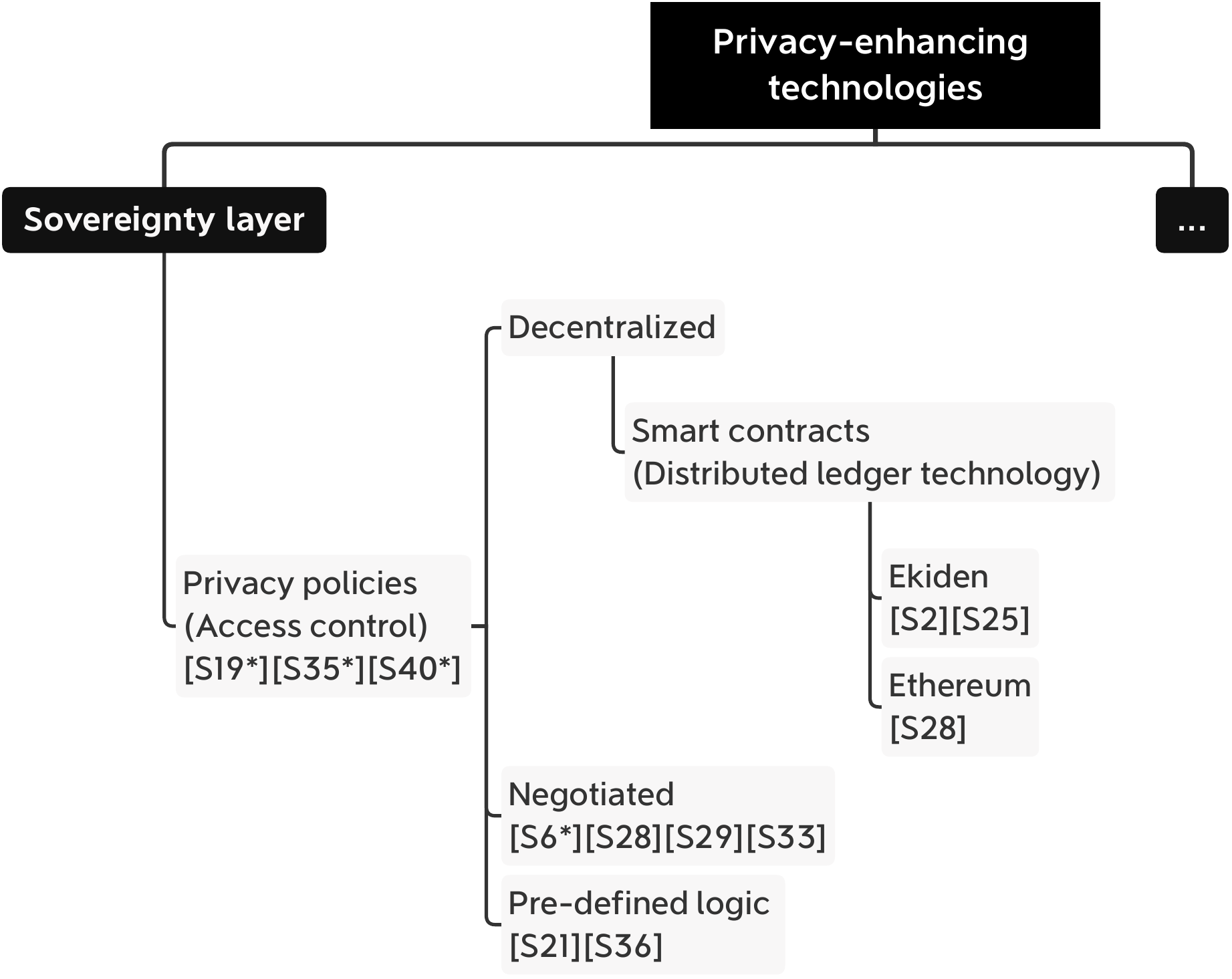}
\caption{Classification of PETs used for sovereignty purposes.}  
%Refer to the caption of Fig.~\ref{fig:PETs_tree_pro_processing} for details on the inclusion criteria.}
\label{fig:sovereignty_figure}
\end{figure}

\subsection{Sovereignty layer}
\label{Data_psovereignty_RQ2}
\smallskip

\noindent The \textit{sovereignty layer} deals with the concept of information control, the perceived ability to govern what is exposed from one's data~\cite{priv_secrecy_defs}.
Specifically, based on an entity's requirements, this layer defines the entity's rules and guidelines regarding ownership and management that can indirectly govern data processing, verification, and other IoT data market layers.
Furthermore, to prevent entities' violation of privacy, practitioners should map these rules to the PETs capable of fulfilling them.
For example, GAIA-X's high-level architecture contemplates privacy policies in their data \textit{sovereignty layer}~\cite{noauthor_gaia_x_nodate}. 
Privacy policies (closely related to access controls), have been predominant across publications, and, thus, dominate the sovereignty layer depicted in Fig.~\ref{fig:sovereignty_figure}, which illustrates the three identified types.

\paragraph{\textbf{Privacy policies and privacy by design}}
Privacy policies embody the requirements and guidelines of a data governance model and are meant to be part of any privacy-enhancing application.
To define them, given the regulatory and human aspects of privacy policies, it is also helpful to adopt perspectives from definitions beyond computer science, such as the one underlined in Section~\ref{sec:background}.
A.~F.~Westin~\cite{westin_privacy_nodate} indicates that the privacy requirements depend on the recipient of the information, e.g., an individual can have different reservations when disclosing information to a family member than to the government.
% \textit{``Privacy is the claim of individuals, groups, or institutions to determine for themselves when, how, and to what extent the information about them is communicated to others"}.
Privacy policies should reflect this definition, which means that individuals should express the privacy policies they expect. Several studies in our SLR explicitly proposed policies as part of their solution~\citep{P6}\citep{P21}\citep{P25}\citep{P28}\citep{P29}\citep{P33}\citep{P36}, while many others reviewed privacy policies~\citep{P19}\citep{P35}\citep{P40} or mentioned similar ideas.
For example, E.~M.~Schomakers et al.~\citep{P8} did not provide a concrete implementation or explicitly named privacy policies. However, they mentioned that, in data sharing scenarios, the data owners should be able to control some fundamental aspects: data types to share, with whom to share, the required degree of trust in another party, the purpose of sharing, and for which benefit. 

Among the publications discussing privacy policies, there is a discernible classification.
Four of these publications~\citep{P6}\citep{P28}\citep{P29}\citep{P33} considered privacy policies as a negotiation between the user and a third party, such as the data consumer or data broker.
S.~Spiekermann~et~al.~\citep{P6} provided a set of legal requirements and high-level technical solutions that facilitated the introduction of policies in international data markets, e.g., writing policies in a standard language. 
On the other hand, S.~Duri~et~al.~\citep{P29} presented an approach where the data owner could choose among a set of four privacy policies, which included how data was aggregated, and S.~Kiyomoto~et~al.~\citep{P33} relied on a \textit{privacy policy manager} that acted as a gatekeeper and managed the privacy settings from a set of users.
Furthermore, another set employs a pre-defined logic to execute PETs based on the desires and track record of the data shared by the individual~\citep{P21}\citep{P36}, while the last set relied on smart contracts for decentralized pre-defined~\citep{P2}\citep{P25} or negotiated~\citep{P28} policies.

Nonetheless, the implementation of policies faces challenges. 
Firstly, there may be multiple colliding policies, i.e., applications must prioritize policies depending on the context~\citep{P21}. 
Secondly, there is no uniformly accepted global standard for electronic privacy policies~\citep{P6}. 
C.~Perera~et~al.~\citep{P40} investigate how practitioners model privacy policies in different domains, focusing on IoT applications. They point to the lack of a uniform standard and propose to utilize ontology-based privacy-knowledge modeling.
Thirdly, policy enforcement also causes overhead and an increase in latency due to the need for compliance checks and a lack of automation~\citep{P21}. 
Furthermore, conventional users should include their privacy preferences with minimal manual effort, as they could be overwhelmed otherwise.
C.~Perera~et~al.~\citep{P40} suggested using recommender systems based on similar users' data to address this issue. 
However, this solution may incur a biased recommendation. 
Moreover, data acquisition expenditure for privacy policies should not incur costly computational resources as they scale to a growing number of transactions~\citep{P40}. 

Privacy policies are crucial to protect users' privacy; however, they are not enough.
Organizations must consider privacy issues at each stage of the data pipeline (i.e., processing data end-to-end with the extract-transform-load framework~\cite{ETL_pipeline}), contemplating aspects that escape user-defined or mutually-agreed policies, and taking into account that typically, neither users nor data brokers will be privacy experts.
If a user does not know the potential harms of sharing sensitive information such as DNA data, a data consumer may take advantage of the user.
Therefore, while privacy policies are a stepping stone toward end-to-end privacy, practitioners must develop systems with a privacy-by-design philosophy~\citep{P36}\citep{P4}. 

Privacy by design is a term coined in the '90s by the former information and privacy commissioner for the Canadian province of Ontario, A.~Cavoukian, who created seven principles \cite{priv_by_design_7_foundations}.
Privacy by design claims that privacy goes beyond current regulations and must be an ever-present concern in the minds of organizations~\cite{priv_by_design_7_foundations}.
Following privacy-by-design principles entails, for example, preventing sensitive information extraction by default~\citep{P36}, minimizing the amount of shared data at each exchange (proportionality)~\cite{P4R106}, and increasing the price of large data packages~\cite{P4R106}, among others. 
However, adopting these design principles comes with effort, forcing developers to adapt their design patterns.
For example, current homomorphic encryption techniques force data scientists to express their analysis in terms of additions and multiplications, and differential privacy requires new software engineering design patterns that track the privacy budget of individuals or data scientists. 

\paragraph{\textbf{Smart contracts (SCs)}}
A SC alone is mainly equivalent to conventional scripts. Nevertheless, because SCs are executed in distributed-ledger-technology-based architectures (DLT) (see Section~\ref{subsec:consensus}, SCs inherit from DLT their enhanced availability and integrity guarantees~\cite{butijn2020blockchains}.
DLTs execute SCs synchronously on every node of a P2P network if an arbitrary transaction demands a function's execution. Once deployed, no one can change the script, not even the creators (unless there is an intended call of the script that enables modification), and the script will remain in the network as long as the network exists unless specified differently (e.g., through a self-destruct call).
This inherited integrity property of SCs makes them a unique tool to specify and enforce policies between parties or any other process where no trusted third party is available.

Within this review, all the studies that used the Ethereum, Quorum, Hyperledger Iroha, or Ekiden blockchains relied on SCs to declare privacy policies \cite{P2}\cite{P25} (Ekiden) \cite{P28} (Ethereum), fair auctions \cite{P3} (Hyperledger Iroha), or payments or incentives~\cite{P32} (Ethereum) \cite{P43} (Agora)~\cite{P3} (Hyperledger Iroha).
However, while SCs ease verification and enable democratic proposals of privacy policies, SCs also inherit the privacy flaws of DLT, i.e., SCs by default imply the disclosure of data and computations to all DLT network nodes~\cite{zhang2019security}\cite{sedlmeir2022transparency}.
For example, the architecture from R.~Cheng~et~al.~\citep{P25} employed SCs to set user-defined policies, yet it relied on trusted execution environments to enforce them.
\emph{SCs alone cannot enforce privacy policies without relying on other PETs.} The only privacy-related feature that a SC can offer to an IoT data market is declaring privacy policies.

\paragraph{\textbf{Data access control}}
Data access control refers to allowing an organization or an individual to choose \emph{who} has access to \emph{which} data. Access control represents a subset of privacy policies in data markets and may utilize different PETs to enforce access rights.
While access control is a long-established approach, R. Cheng~et~al.~\citep{P25} propose a novel method, using a key-rotation system~\cite{P25R33} in combination with a key manager. Thus, the potential impact of a leaked key is only temporal, with the downside of shorter access permissions.

\section{Authenticity-enhancing technologies}
\label{sec:AETS}
\medskip

\noindent The included authenticity-enhancing technologies (AET) focus on enhancing the authenticity of data and identities and also cover data integrity as described in Section~\ref{sec:terminology}. Some of the AETs that we describe incorporate privacy-enhancing features, while others do not address or even aggravate privacy protection issues and, thus, need to be combined with PETs.

\subsection{Consensus layer}
\label{subsec:consensus}
\smallskip

\paragraph{\textbf{Distributed ledger technology (DLT)}}
While DLT may take different forms, most architectures follow the blockchain design pattern, except for IOTA, which uses the so-called Tangle~\cite{P27R24}.
A blockchain is a tamper-proof distributed database whose state is stored, synchronized, and replicated by nodes in a P2P network following a consensus algorithm~\cite{butijn2020blockchains}.
By its distributed nature, the shared ledger becomes a medium to verify claims, data, payments, or contracts, as once an entity writes something on the ledger, it is practically impossible to modify or erase this record in the future. This property makes blockchain a decentralized and highly reliable alternative to conventional auditing methods like version control~\citep{P19}.
%This consensus algorithm solves the Byzantine generals problem, which describes the challenge of finding agreement in a network under the presence of malicious entities. 

Benefits of DLT in IoT data markets are the ability to represent the governance, distribution, and roles of authorities on a technical basis~\citep{P3}, and the enforcement or transparent storage of pre-defined rules by the architects of the respective platform~\citep{P25}. Other benefits include eliminating the need for a trusted third party, which removes a single point of failure, improves censorship resistance, and provides more robust data and computational integrity guarantees. DLTs also enable payments through their often built-in cryptocurrencies or other payment systems implemented via smart contracts~\citep{P32}\citep{P46}. 

However, some of the studies in our SLR also point at the challenges of current DLT designs:
IOTA fails to deliver regarding throughput~\citep{P28}, is still centralized~\citep{P20}, and provably has security flaws~\cite{heilman_cryptanalysis_2020}. 
Furthermore, blockchains exhibit low transaction throughput~\citep{P25}, high latency~\citep{P11}, limited storage~\citep{P1} and scalability~\citep{P11}\citep{P25}, computational overhead~\citep{P25}, high energy consumption~\citep{P12}, and, most importantly, excessive information exposure that can entail a privacy violation~\citep{priv_in_blockchains}. 
However, some of these aspects can be mitigated. 
For example, the energy consumption issue only concerns proof-of-work blockchains~\cite{sedlmeir2020energy}, and performance can be improved to some extent by private permissioned blockchains that restrict participation in consensus and read access to a small number of nodes in a consortium~\cite{kannengiesser2020trade}.

Despite the possible operational improvements, employing a DLT for a privacy-enhancing IoT data market needs in-depth consideration.
Firstly, through highly replicated storage, a DLT is not suitable for storing large amounts of data produced by IoT devices, not even in a privacy-compliant manner. 
Consequently, most architectures of the selected studies transfer data through interplanetary file systems~\citep{P28}, employ a hashing verification approach as described in the \textit{communication layer}~\citep{P1}\citep{P10}\citep{P33} or use Merkle trees~\citep{P46}.
Secondly, while DLT allows for disintermediation and verification in a trust-less manner, it exposes to the network whatever information someone writes on the ledger for as long as the network exists, which may, among others, violate GDPR's Article~17 "\textit{Right to be forgotten}" for personally identifiable information~\cite{GDPR_EU2016}. 
Lastly, even if an organization uses a DLT only for the matching and clearing steps of an auction, potentially sensitive business information such as turnover can become available to other network participants, which can conflict with antitrust regulation.

Despite the privacy and performance issues of DLT, $31$\% of the included papers implemented a DLT as the backbone of IoT data market architectures, employing the Ethereum blockchain \citep{P1}\citep{P13}\citep{P20}\citep{P28}\citep{P32}\citep{P45}\citep{P46}, Quorum \citep{P18}, the Agora blockchain \citep{P43}, Hyperledger Iroha \citep{P3}, Hyperledger Fabric \citep{P10}\citep{P33}, IOTA \citep{P20}\citep{P28}\citep{P30}, Intel's TEE-based consensus Rem~\citep{P12}, and Ekiden \citep{P2}\citep{P25}. Other publications only considered them agnostically \citep{P11}\citep{P49} or in a review \citep{P19}\citep{P23}. 
The most salient architectures are described in Table~\ref{tab:DLTTable}.

Some of the selected studies included privacy-enhancing features in their stack.
For example, \href{https://github.com/ConsenSys/quorum}{Quorum} supports private transactions and private contracts through a public-private state separation and P2P encrypted message exchange for the direct transfer of private data~\citep{P18}. However, the interaction between the private and public ledgers is thus naturally limited and cannot be directly applied, for example, to an on-chain payment system.
Another example is Ekiden, which offers a horizontally scalable blockchain potentially capable of hosting end-to-end privacy-enhancing applications through key management protocols and Intel's TEEs~\citep{P25} (Note that~\citep{P12} uses these TEEs only for consensus, not privacy). 
Like the solution that W.~Dai~et~al.~\citep{P45} presented, Ekiden allows for smart contracts to execute data analysis in TEEs.
However, it is essential to note that these DLTs accomplish the described privacy and integrity functionalities not because of the DLT characteristics but by leveraging the PETs described throughout Section~\ref{sec:PETS}.

\subsection{Verification layer}
\smallskip

\begin{figure*}[t!]
\centering
\includegraphics[scale=0.7]{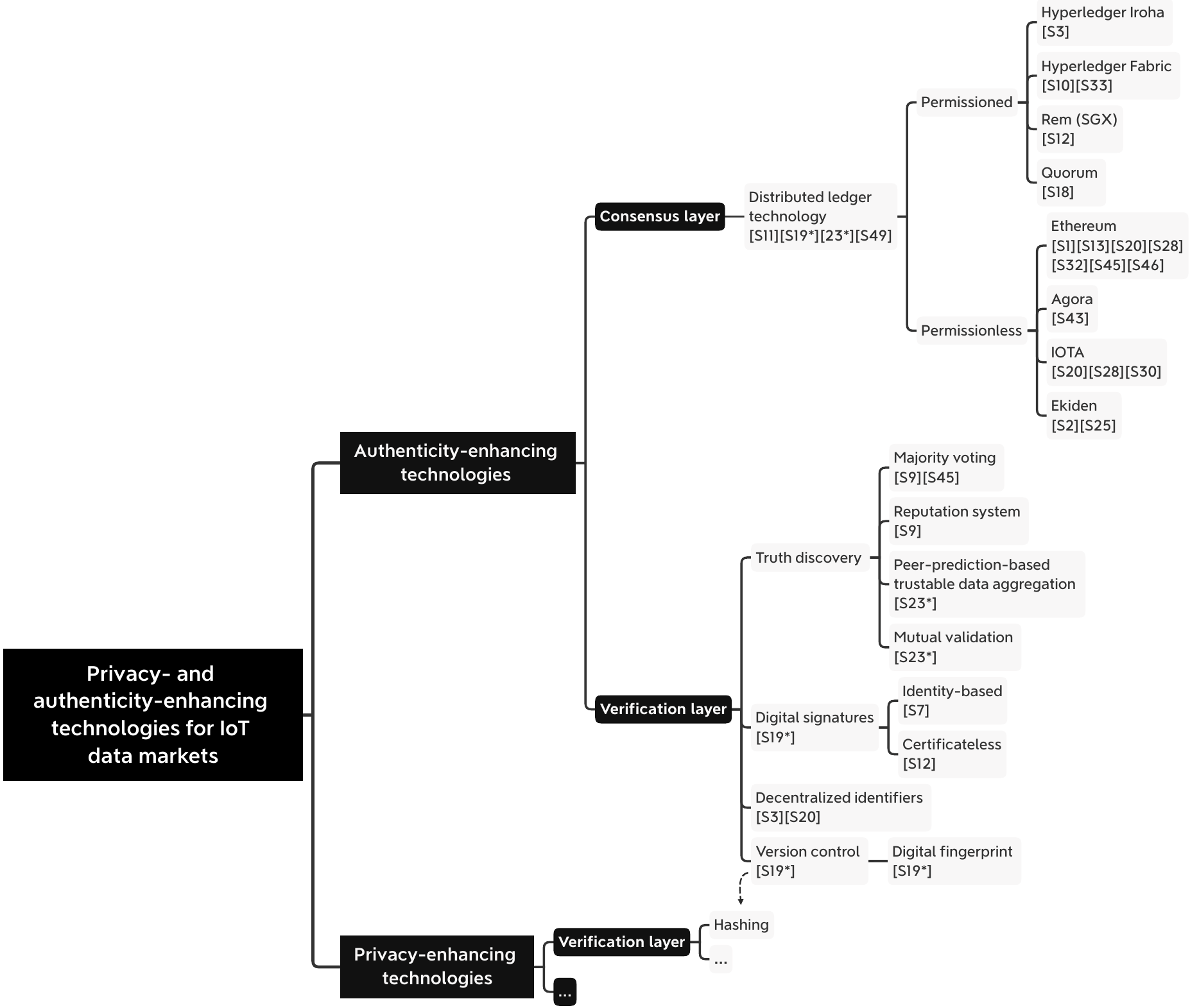}
\caption{Classification of the identified authenticity-enhancing technologies in the selected studies of this SLR.}
%Refer to the caption of Fig.~\ref{fig:PETs_tree_pro_processing} for details on the inclusion criteria.}
\label{fig:verification_figure}
\end{figure*}

\noindent This verification layer corresponds to AETs that can be employed for the verification of data and identities.
We structure this Section according to Fig.~\ref{fig:verification_figure}.

\paragraph{\textbf{Truth discovery (TD)}}
TD encompasses algorithms aiming to find the authentic value when different data sources provide conflicting information. As a consequence, TD enhances data and computation integrity and can also enhance identity authenticity, e.g., through reputation systems~\citep{P9}. 
In our SLR, we found that TD takes different forms. 
For example, the survey by J.~Du~et~al.~\citep{P23} mentioned a mechanism called peer-prediction-based trustable data aggregation~\cite{P23R12}, in which the system administrator rewards participants for predicting outcomes of arbitrary events based on other participants' data. This design created incentives for honest reports and therefore enhanced data correctness, resulting in almost all participants choosing to report their bids truthfully~\citep{P23}.
Moreover, J.~Du~et~al.~\citep{P23} also proposed mutual validation in which an IoT device compares its data with that of other nearby IoT devices. 
However, this only applies to specific measurements that are positively correlated for neighboring devices, e.g., temperature, speed of a vehicle, or location in particular settings. It also seems challenging to establish generic handling of differences.
Other TD approaches are majority voting, implemented by Y.~Li~et~al.~\citep{P9} in their crowdsourcing architecture and by W.~Dai~et al.~\citep{P45} in their data processing-as-a-service model. 
Specifically, given the use of differential privacy in the former approach, they systematically discovered high-quality data with an \emph{estimated} measure of utility that compares individual data points with an aggregate (the ``majority''). 
The latter publication created a reputation system based on the quality of previously sold data. 

While most TD approaches leverage transparency to enhance data and identity authenticity and data and computation integrity, TDs are flexible to include PETs such as ZKPs, MPC, HE, TEEs, and DP such as in~\citep{P9}.
Furthermore, TD can also tackle the oracle problem of DLT, i.e., nodes within the network cannot assure the authenticity of data from outside the network, e.g., the price of a physical asset or the result of an election. For example, ChainLink~\cite{chainlink} is an initiative that utilizes incentives to create a trusted oracle network and incorporates many of the principles of TD.

\paragraph{\textbf{Digital signatures (DS)}} While DS\footnote{We introduced the fundamentals of DSs in the \textit{verification layer} within the PETs branch.} schemata are commonplace for authentication purposes in today's IT architectures, we have found in selected studies the use of two notable public key cryptography (PKC) schemata that offer some convenience-related advantages over conventional PKC systems:

\begin{itemize}

    \item \textbf{Identity-based}~\citep{P7}, where a key generation center (KGC) creates a secret key in a way that the entity's public key can be a publicly available unique string, e.g., the entity's email address. 
    The KGC must be trusted because it holds the master secret key from which all parties' secret keys can be derived. 
    This digital signature assures identity authenticity as the signature is digitally certified by the KGC.
    
    \item \textbf{Certificateless}~\citep{P12} DS schemata are a special form of identity-based PKC whereby an entity's private key is generated by both the entity and a KGC so that the KGC is not aware of the private key of the entity.
    However, the entity can prove that the KGC was involved in the key generation~\cite{certificateless_crypto}. 
    This approach assures the authenticity of an entity while tackling the single point of failure of the KGC.
\end{itemize}

\paragraph{\textbf{Decentralized identifiers (DIDs)}}
Identifiers can link an entity electronically across multiple IT systems, such as mobile phone numbers, ID cards, user names, or emails.
These links are sometimes but not always unique and are facilitated by identity providers that centrally host registries of these identifiers'~\cite{DIDs}. 
In contrast, DIDs are globally unique (with certainty through publishing them on a DLT or probabilistically through randomized generation) identifiers decoupled from centralized registries. DIDs essentially correspond to URLs linked to a file containing one or several public keys and associated metadata that specifies the policies of controlling or interacting with the associated identity.  
There are two studies in our SLR that employed DIDs in combination with DLT in their conceptual frameworks~\citep{P3}\citep{P20}, described in Table~\ref{tab:DLTTable}. 

\paragraph{\textbf{Digital fingerprints (DF)}}
DFs are unique physical identifiers that can be attached to or are inherent of items, and thus, one can be sure to interact with, e.g., the right IoT device~\citep{P19}.
DFs can be seen as a form of version control at a high level.
However, attaching an identifier securely to a physical object is difficult unless it has a unique property, e.g., unique metal patterns in the soldering of a chip.
However, even if the attachment is relatively tamper-proof, e.g., with a crypto-chip, the same problem also pervades the items that interact with the digital fingerprinted item, e.g., tracking scanners.
Therefore, despite the authenticity assurance of DFs, their authentication can only be as truthful as the honesty of the devices that scan the DF.
\section{Privacy challenges in IoT data markets}
\label{sec:Challenges}

% RQ3: What are the challenges in the implementation of privacy-enhancing data markets for the IoT?

\begin{figure*}[t!]
\centering
\includegraphics[scale=0.5]{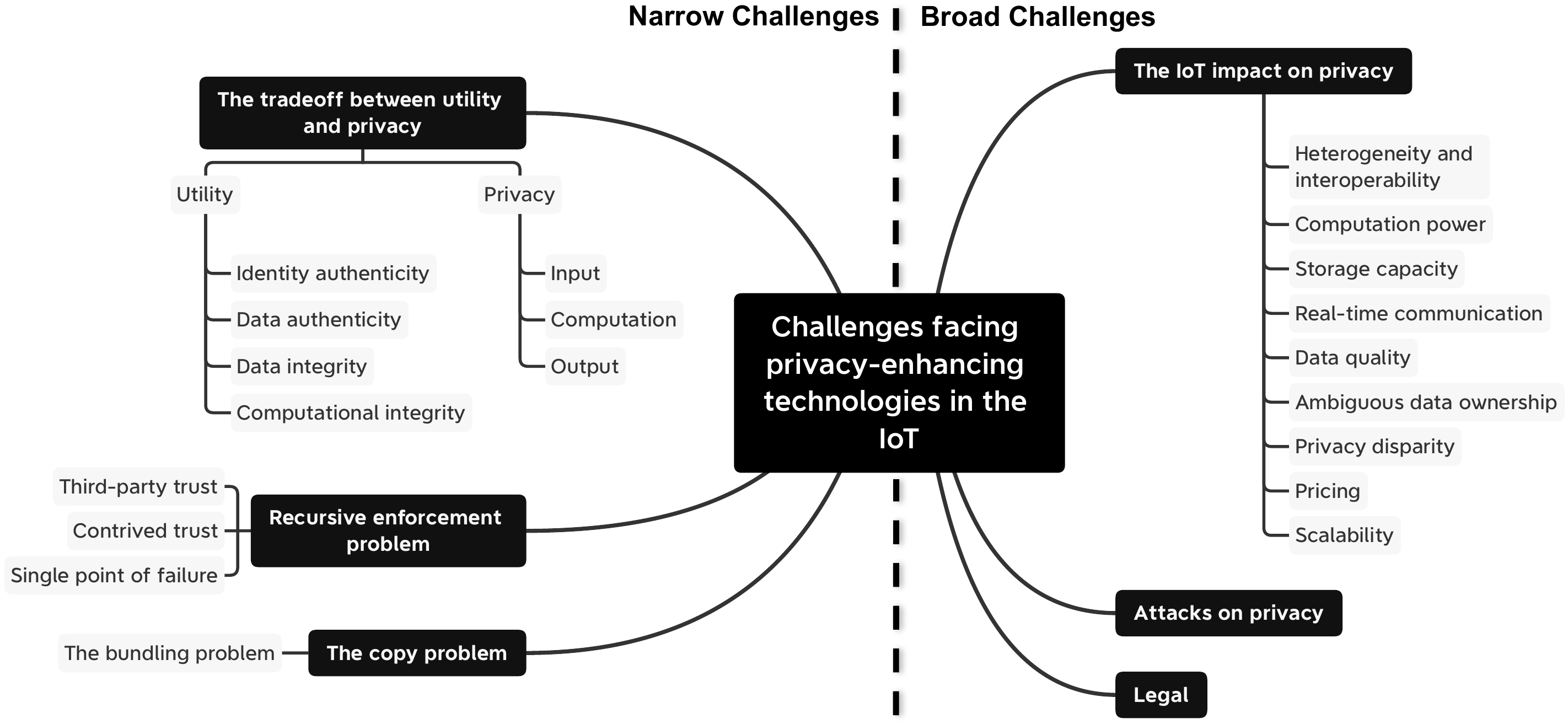}
\caption{Overview of the narrow and broad challenges facing privacy-enhancing IoT data markets.}
\label{fig:PETS_challenges}
\end{figure*}

\noindent This Section aims to answer RQ2 by distilling the implicit and explicit challenges unveiled in our SLR and other seminal studies~\cite{trask_structured_transparency_nodate}\cite{za_privacy_sensitive_2021} concerning privacy in the context of IoT data markets. 
We further classify them into narrow and broad challenges depending on the scope of their definition.
Fig.~\ref{fig:PETS_challenges} summarizes and outlines the structure of this Section.

\subsection{Narrow challenges}
\smallskip

\noindent Aside from the inherent complexity and low maturity of some PETs and the compatibility issues with legacy systems~\cite{za_privacy_sensitive_2021}, we identified another specific set of challenges tackled or circumvented by the selected studies. 

\subsubsection{The trade-off between utility and privacy}
Practitioners working with personal data face the challenge of balancing the enhancement of individuals' privacy with the preservation of data's utility~\citep{P14}. This challenge is explicitly mentioned by some of the selected studies~\citep{P6}\citep{P9}\citep{P13} and implicitly tackled by others~\citep{P2}\citep{P24}\citep{P25}\citep{P35}\citep{P47}.
This dichotomy is the underlying reason behind the tension between data owners and consumers: the former aim to maximize privacy while the latter intends to maximize utility, which, in turn, is frequently determined by data authenticity (see terminology in Section~\ref{sec:terminology}). 
Furthermore, privacy officers should consider balancing this trade-off at each stage of an information flow~\cite{trask_structured_transparency_nodate}: input, computation, output, in transit, and at rest, which increases the complexity of the task.
On the other hand, decision makers' or data scientists' quality of judgment depends on computational integrity and the authenticity of data and identities, which is affected by the privacy-utility trade-off.

While PETs from the secure and outsourced computation category seem to circumvent the utility-privacy trade-off by concealing inputs, computation, and outputs, the anointed recipients of these outputs can still perform a re-identification attack.
Thus, anonymization PETs, such as differential privacy, should also be included in the stack as they lower the probability of successful re-identification attacks~\cite{trask_structured_transparency_nodate}. 
%Overall, there will always be a trade-off between privacy and utility. However, practitioners may move along the Pareto frontier by employing synergistic combinations of PETs.

Data and identity authenticity and accountability bring another problem in the utility-privacy trade-off.
Some PETs, namely anonymization technologies, increase plausible deniability at the expense of reducing authenticity and, therefore, accountability~\citep{P14}\citep{P43}.
If the data is fuzzy, the data owner may claim that such a result is not resembling the truth, which is favorable for individual users. However, such protection is not beneficial for society in some contexts, e.g., in criminal contexts.
Regarding authentic data from fuzzy identities, if an authority cannot trace data back to the origin, an individual could try to claim plausible deniability, which would hinder processes such as tracking COVID-19 patients to improve pandemic countermeasures.
For practitioners to find a balance in these contexts, the requirements of any application or platform should first define the accountability of the involved entities to strike an optimal balance between utility and privacy.

% potentially delete paragraph
Regarding accountability in data markets specifically, mechanisms to punish misbehavior, such as banning an entity for re-selling or not selling authentic data~\citep{P7}, can be beneficial to enhance the utility of the market.
While AETs such as truth discovery, e.g., majority voting~\citep{P9}~\citep{P45} or reputation systems~\citep{P9}, incentivize market participants to report honestly about the exchanged data, their identity, and computation integrity, among others, the privacy of the entities is not necessarily enhanced. 
Additional related issues that may arise are verifying the purchased data's authenticity without violating the individuals' privacy~\citep{P7}. 
For example, if a data broker sells analysis outputs (insights) and not the (privacy-enhanced) original data, the original data owner's digital signature is not valid to authenticate the insights~\citep{P7}. 
Nevertheless, systems could use zero-knowledge-proof-based authentication of data and computation, coupled with value deposits locked in smart contracts to hold participants accountable through enforcing reimbursement across intermediaries. 
Moreover, other nascent solutions exploit the ubiquity and proximity of IoT devices in specific contexts because the data gathered is likely to be correlated, which allows for mutual data authenticity verification among IoT devices~\citep{P23}.  
However, exploiting correlation can only be used for specific measurements, e.g., weather conditions, vehicle speed, and location, among others.

\subsubsection{The recursive enforcement problem}
%A.~Colman~et~al.~\citep{P49} define trust as a ``\textit{[...] directional relationship between two entities – a trustor and a trustee – where a trustor trusts a trustee to perform a specific action within a given scope.}''
\emph{Trust} is an essential component in distributed systems that involve different stakeholders and, thus, trust is specifically relevant in the context of IoT data markets. Definitions of trust generally refer to a ``\textit{[...] directional relationship between two entities}''~\cite{P49} where one entity (the trustor) has subjective expectations on the behaviour of the other entity (the trustee) based on previously observed behaviour (reputation-based trust)~\cite{mui2002computationaltrust} or the belief in competencies and corresponding actions -- often within a specific context~\cite{grandison2000surveytrust,artz2007trustsurvey} and incentivized by joint interests~\cite{cook2005cooperation}. Following this definition, we can consider users as trustors of application owners protecting their data when they engage in digital activity.
However, the number of data breaches~\cite{data_breach} and privacy scandals such as Cambridge Analytica indicate that this trust is not always deserved. 
The recursive enforcement problem (REP) encompasses the underlying problem of third-party trust with more nuance: Given a third-party authority~($A$), there ought to be another authority~($B$) to supervise~$A$, so that~$A$ can be trusted. In turn, there should be yet another authority~$C$ to supervise~$B$~\cite{trask_structured_transparency_nodate}, and so forth.

The REP is a significant challenge that has been covered and tackled implicitly by some of the studies in our review~\citep{P1}\citep{P3}\citep{P4}\citep{P7}\citep{P10}\citep{P16}\citep{P25}\citep{P41}.
Additionally, others tackle a sub-set of the REP, which is the single point of failure of trusting a unique third party~\citep{P6}\citep{P12}. 
According to E.~Schomakers et al.~\citep{P8}, the hesitation in trusting third parties is one of the main reasons for the slow adoption of IoT data markets. 
Indeed, it is hard to technically ensure and prove that the third party will not use one's data for purposes other than those agreed~\citep{P8}.
Additionally, adoption is further slowed down because the use of third parties to supervise other parties incurs costs~\citep{P46}.
Furthermore, users' daily interactions with ``trusted'' third parties can be regarded as a product of \textit{contrived trust}, another form of the REP. For instance, applications from large service providers with negligible competitors push users to accept the sometimes poor privacy conditions, e.g., GPS apps.
Note that \textit{contrived trust} is different from the trust users have on cryptography, open-source code, or consensus mechanisms that the broad scientific community has audited over the years.

Tackling the REP requires reducing the power and the responsibility of the third party in a particular aspect of a specific service by, e.g., distributing such responsibility among other parties or distributing the power among multiple parties that enforce rules on each other.
These measures can ease the hesitation to trust a single third party, tackle contrived trust, and reduce the single point of failure because the third party would be supervised and held accountable by other third parties in a flat hierarchy.
Fortunately, the PETs included in this study can also circumvent---only onion routing can tackle---the REP, which, in turn, reduces the need for third-party trust and, therefore, reduce contrived trust and a single point of failure. 
Additionally, $5$ of the $7$ AETs included in this study can tackle the REP, primarily distributed ledger technology, whose architecture was purposefully built to tackle the byzantine generals' problem~\cite{ismail_towards_2019}\cite{byzantine_generals}, a manifestation of the REP.

\subsubsection{The copy problem}
Once an entity releases data freely or for profit-seeking, the data is no longer under the original owner's control.
Consequently, the recipients of such data can \emph{copy} and, e.g., re-sell or use the entity's data for a non-agreed purpose without informing or acknowledging the original owner~\cite{trask_structured_transparency_nodate}. 
Beyond the privacy threats the copy problem (CP) entails for users of service providers under poor privacy conditions, the CP is a major obstacle for organizations to engage in data markets, which some of the selected studies implicitly tackle~\citep{P2}\citep{P25}\citep{P45}.
The CP leads companies to either hoard or sell data as fast as organizations obtain the data, lest its value drops~\cite{trask_structured_transparency_nodate}.
Nonetheless, secure and outsourced computation PETs such as trusted execution environments or homomorphic encryption can tackle the CP by allowing other entities to extract value without losing control over the input data beyond the specific information sold, such as an algorithm's evaluated output on this data. 
This paradigm is profound because tackling the CP makes data scarce (to some extent), as the original data is not shared, and the data owner would not allow a non-agreed computation.
Thus, selling the \emph{access} to data can be more attractive to companies, as data would preserve their value longer than releasing the data.

A subset of the CP is the \textit{bundling problem} (BP)~\cite{trask_structured_transparency_nodate}, which is an attack vector different from re-identification that occurs when an entity requests actively or passively more data than strictly needed to (i) prove a claim or (ii) perform an analysis. 
Harvesting more information than needed worsens data breaches' consequences for individuals and companies and indicates questionable business ethics.
For instance, (i) to prove one's age with an ID card, the prover usually shares all the information in the ID instead of only the age and proof of the card's authenticity.
The BP is a subset of the CP because if one tackles the CP, neither necessary nor additional information is released beyond the required computation.
For example, tackling the CP by restricting verification and processing to a trusted execution environment also tackles the BP. In this setting, the data consumers cannot copy the necessary data or metadata for other unsolicited analyses, despite being able to process metadata to verify the authenticity of the data and the integrity of the computation and obtaining the desired outputs of the analysis. 
Additionally, (ii) anonymization-based PETs such as differential privacy or $k$-anonymity reduce data authenticity to tackle the BP.
For instance, in a demographic analysis that only requires the first digits of the ZIP code to perform clustering, data curators can generalize the ZIP codes with $k$-anonymity, so only the strictly necessary information is revealed to data scientists.
Nonetheless, anonymization can suffer from background-knowledge-based attacks~\citep{P48}\cite{narayanan_robust_2008} and does not solve the CP because the data consumers can replicate the privacy-enhanced data.
\bigskip

%If data is hoarded, the companies that siloed the data may miss the opportunity to economically profit from or exchange data with other organizations for synergistic projects or innovation.
%On the other hand, if the data is released once, recipients can copy the data and the value can drop dramatically, reducing the incentives of engaging in open-data ecosystems initiatives such as GAIA-x.

\subsection{Broad challenges}
\smallskip

\subsubsection{The IoT impact on privacy}
The paradigm brought by the IoT brings significant amounts of data to markets. However, this paradigm also bears some of the shortcomings of IoT devices~\cite{P39R32}.
Table~\ref{tab:Challenges_IoT_Table} contains an overview of these challenges and briefly discusses their impact on privacy.
In summary, privacy is always affected by the context and employed technologies, which underlines the importance of adhering to privacy-by-design principles \cite{priv_by_design_7_foundations} and the need for practitioners in other fields such as software engineering, economics, law, and politics to tackle together the diverse issues that IoT entails for privacy.

\begin{table*}
\centering
\footnotesize
\begin{tabular}{%
>{\RaggedRight}p{2cm}%
>{\RaggedRight}p{1cm}%
>{}p{6.5cm}%
>{}p{7cm}}
\toprule
\textbf{Challenge}  & \textbf{Studies} & \textbf{Description} & \textbf{Impact on privacy} \\
\midrule
\addlinespace
\textit{Heterogeneity and \break Interoperability} & \citep{P36} \citep{P39} & 
The IoT consists of billions of IoT devices from different manufacturers, running different software on different local networks and geographic regions, with different computation power and storage capacity~\cite{P39R30}.
Furthermore, different communications standards, connectivity and availability aggravate the interplay of IoT devices. &
An IoT data market should be agnostic to these differences and minimize any additional requirements; however, it is unclear how global data markets should harmonize data coming from different jurisdictions with different privacy regulations and how an IoT device can interact with another whose, e.g., verification schemata are considered inadequate. 
In addition to these obstacles, a lack of interoperability may restrain PETs that involve the communication between many devices, e.g., MPC.  
\\ \addlinespace

\textit{Computation power } & \citep{P5} \citep{P11} \citep{P48} & 
Manufacturers produce many IoT devices designed to consume low energy and require minimal volume, limiting these IoT devices to the core functionalities of monitoring and communication~\citep{P11}. &
Any additional computation requires a higher investment in resources and manufacturing, and running some PETs becomes infeasible without this extra investment. 
Consequently,  a set of PETs is excluded without more computation power, e.g., cryptography-based PETs such as HE, MPC, ZKP, some digital signatures, or consensus algorithms.
This limitation, however, may only apply to contexts where it might not be possible to connect IoT devices acting as clients with proprietary or trusted third-party nodes where these PETs are executed.
\\ \addlinespace
\textit{Storage capacity and real-time communication} & \citep{P1} \citep{P5} \citep{P11} \citep{P17} \citep{P29} \citep{P48}  & 
Minimizing the physical volume of an IoT device reduces their price but limits their storage capacity, forcing IoT devices to transmit the data to a data warehouse or a data market as quickly as possible. 
This tendency intensifies in some IoT applications where the time delay tolerance is low to enhance the utility of real-time information~\citep{P17}. &
Processing time constrains the number of usable PETs, excluding those that require long execution times, such as fully HE or creating a ZKP. 
\\ \addlinespace

\textit{Data quality} & \citep{P14} & 
An unreliable IoT design may afflict thousands of IoT devices mass-produced by a manufacturer, which at deployment may lead to millions of unreliable data points. 
Furthermore, networks may also be unreliable, further worsening the quality~\citep{P14}. &
The impact may seem beneficial in terms of privacy; however, unreliable data leads to verification and secure computation schemata to fail and anonymization technologies to over-perturb the data as the underlying data is not entirely truthful. 
\\ \addlinespace

\textit{Ambiguous data ownership} & \citep{P4} \citep{P10} \citep{P14} & 
When purchasing a device or a cluster of IoT devices, e.g., a phone, consumers also expect to own the data they are generating. However, the phone manufacturer and service providers expect to receive parts of this data nowadays with meager consent. 
In addition to this clash of interests, there are scholars that ponder whether data belongs to anyone in the first place, like L. Determann~\cite{no_one_owns_data}.  &
Having unclear data ownership leads to a misguided deployment of PETs, which may cause detrimental consequences if the privacy measures fall short. 
On the other hand, if the practitioner knows who has the right to the data and what the owner is reticent to share with a third party, then selected PETs and their privacy tuning can be optimized accordingly.
\\ \addlinespace

\textit{Privacy disparity} & \citep{P4} \citep{P40} & 
Depending on the IoT devices' deployment location, the degree of privacy measures should be higher or lower, e.g., sensors in vehicles, smart homes, phones, and wearables. 
Furthermore, IoT deployments should adapt the monitoring time to an adequate amount depending on the context~\citep{P40}. &
Some PETs, such as semantic and syntactic technologies, allow adjusting the degree of privacy; however, others are more rigid. 
Selecting and adapting a PET to the IoT devices' deployment context requires expertise.
\\ \addlinespace

\textit{Pricing} & \citep{P4} \citep{P14} \citep{P17} \citep{P50} & 
There are multiple variables imposing the price of data aside from supply and demand: the truthfulness, the source, either purchasing the data or the access~\citep{P17}, and the privacy level. 
These factors add additional complexity to pricing, e.g., the sources have become disparate with the IoT, which drives pricing to a more granular task than before, when aggregated data could be sold as a unit~\citep{P17}. &
Aside from payment enforcement mechanisms, pricing involves negotiations, which frequently must ensure privacy. This adds an extra layer of complexity to the deployment of PETs.
Furthermore, as data markets trade with more granular data points, PETs that need aggregation might be excluded in some contexts, e.g., syntactic technologies such as k-anonymity.
\\ \addlinespace

\textit{Scalability} & \citep{P25} \citep{P36} & 
The number of IoT devices and streamed data grow exponentially across industries~\cite{iot_exp_growth}\cite{P39R30}, which extends data collection and improves analytics across different domains, e.g., health, insurance, or finance.
To gain these benefits, there is a need to increase networks' communication and overall storage capacity as well as interoperability and security efforts.
% However, scale comes with challenges in, e.g., maintaining the quality of message delivery and manufactoring moe, which $5$G~\cite{5G_ioT} technology and tackle, respectively.
&
As the IoT scales, analysts will access more datasets from different domains to create new products and services, e.g., linking driving behavior with insurance in pay-how-you-drive schemata~\cite{insurance_driving}.
Such innovations stem from the ``\textit{mosaic effect}''~\cite{pozen_mosaic_2005}, where disparate datasets with limited information value can obtain significance when combined with other datasets.
However, malicious entities can leverage such an effect to extract sensitive personal information not explicitly contained in a dataset ~\cite{archie_anonymization_nodate}.
\\ \addlinespace

\bottomrule
\end{tabular}
\caption{Overview of challenges brought by the IoT paradigm into data markets explicitly covered by some of the studies included in this SLR.}
    
\label{tab:Challenges_IoT_Table}
\end{table*}

\subsubsection{Attacks on privacy}
Adversaries can be malicious, actively trying to breach users' privacy through hacking, or honest but curious, passively gathering data from users to reveal hidden insights~\citep{P35}.
Both of these entities can carry re-identification attacks with the collected information.
Within the context of the IoT, the list of security and privacy attacks is extensive (sniffing, cache poisoning, DoS/DDoS, sinkhole attacks, replay attacks, among others)~\cite{IOT_Attacks}. Furthermore, within our SLR, W.~Dai et al.~\citep{P45} discuss some of the additional attack vectors these malicious or curious entities may execute in the context of IoT data markets to learn sensitive information from users.
Notable ones include: \textit{Data forwarding}, which is one way the \textit{copy problem} materializes; \textit{roles collision}, where data brokers and buyers may be the same or collaborating entities, and, therefore, the broker could rig the auction for its benefit and access the sold data; and
\textit{side channel attacks}, where attackers exploit the physical properties of the hardware or its power consumption to extract knowledge from the hidden computations (trusted execution environments suffer mainly from this attack). 

Such attacks make the possession of data intrinsically risky because if attackers are successful, data re-identification is possible~\citep{P38}, even if data have undergone some form of privacy enhancement~\cite{kondor_towards_2020}. 
There are common attacks used to re-identify data, e.g., reconstruction, tracing, or linkage attacks~\cite{dwork_exposed_2017}\cite{wood_differential_2018}.
Some of the most famous re-identification \textit{white-hat} attacks involve
A.~Narayanan and V.~Shmatikov~\cite{narayanan_robust_2008} who deanonymized the Netflix Prize dataset with IMDB's public dataset in 2008, M.~Archie et al.~\cite{archie_-anonymization_nodate} who performed the same feat with Amazon's public review data, and L.~Sweeney et al.~\cite{sweeney_identifying_2013} (the inventor of $k$-anonymity) who re-identified participants within a genome sequence dataset in 2013. 
Furthermore, in 2014, X.~Gao et al.~\cite{gao_elastic_2014} tracked drivers with home address and vehicle speed as inputs, and in 2020, D.~Kondor et al.~\cite{kondor_towards_2020} matched users with large-scale mobility datasets from a mobile network operator and transportation smart card usage. 

Overall, IoT data markets will facilitate access to large quantities of data from different domains, including biometrics, which will increase the impact of these attacks and the potential harms to individuals, e.g., insurance, employment, or price discrimination.
Therefore, IoT data markets require a more robust adoption of PETs and security standards.

\subsubsection{Legal challenges}
Progressively along the past decades, governmental institutions have released laws to protect the privacy of their citizens (see Section~\ref{sec:privacy}).
These laws also refer to an individual's and businesses' right to exploit their data commercially, which provides leeway for data markets~\citep{P28} and aims to uncover the untapped potential of data for innovations.

Nonetheless, research points out the sometimes unrealistic expectation to monitor the entirety of the Internet for privacy violations~\citep{P45}, and the dexterity of hackers to find novel deception methods~\citep{P3}, and that laws are more reactive than preventative.
Well-known networks of illegal proprietary digital asset exchanges, e.g., scientific works and how users of digital services give away data, tacitly provide testimony of the failure of data-related legal measures today, and the problems will likely increase with the accruing number of IoT devices~\citep{P38}.
Moreover, privacy regulations can strangle free markets and innovations if they are too stringent~\citep{P45}.

Aligned with these deficiencies, J.~Henrik~et~al.~\citep{P38} introduced privacy regulation pitfalls that the IoT unfolds in data markets in~2013.
They note that (i) definitions of personally identifiable information will be deprecated as unprecedented amounts of data can be aggregated, easing re-identification, (ii) the development and audit of PETs is costly, which may limit business models and potentially make disregarding privacy regulation profitable~\cite{P38R25}, (iii) privacy violations result on small fines or remain unpunished, (iv) technology tends to outpace regulation, and (v) the ubiquity of IoT devices will yield more illegal secondary personal data markets. 
After almost a decade of further research, (i) seems valid, at least in some scenarios. The ambiguity of privacy regulation is a barrier in some cases, as practitioners may default to weaker forms of privacy if their architecture appears to comply. This leads to re-identification -- an attack that is also more practical with the increasing number of IoT devices~\cite{narayanan_robust_2008} -- being more likely to succeed.
However, in defense of these practitioners, while PETs have improved since 2013, some PETs that offer better privacy enhancements are still complex and not yet performant in 2021. 

Based on the prior arguments, pitfall (ii) seems to hold; however, (iii) is no longer a strong pitfall.
Since the enforcement of GDPR~\cite{GDPR_EU2016} in 2018, GDPR has punished multiple corporations with considerable fines ranging between \euro$20$ million and up to $4$\,\% of a corporation's annual worldwide turnover of the preceding financial year. 
As of the writing of this publication, GDPR has harvested considerable fines assigned to Google in France on two occasions~\cite{Porter2019}, Amazon~\cite{Lomas2020}, H\&M~\cite{BBC2020} or the telecommunications operator TIM~\cite{italian}. These fines alone accumulate to \euro$282$ million. 
These statistics are a sign that PETs are not appropriately introduced in production applications even by \textit{big} technology companies and that not complying with privacy regulations in an IoT data market has dire economic consequences.
While these fines could indicate how profitable it still is to violate privacy regulation (iii), one can no longer vigorously defend (iii).
Pitfall (iv) seems to materialize as long as the nature of law-making does not change.
Lastly, pitfall (v) is concerning, given the existence of legal personal data markets that store up to $750$~million user profiles and trade $75$~million online auctions daily like BlueKai~\cite{spiekermann_personal_2015}, whose data could leak to the increasing number of illegal \textit{shadow markets}~\citep{P6}.

\begin{table*}[ht!]
\caption{A mapping of pivacy- and authenticity-enhancing technologies (PET and AET) to the narrow challenges using the terminology defined for this review. The extent of enhancement of privacy, utility, and characteristics of the different PETs and AETs varies from significantly increasing~++, over~+,~+-,~- to significantly decreasing~-~-. na denotes \textit{not applicable}. w/ denotes \textit{with}. \textsuperscript{*}Considering a digital 
certificate when using digital signatures, if applicable. The privacy column assumes data and identity are authentic.}
    \centering
    \vspace{0.25cm}
    \begin{tabular}{c}
        \includegraphics[width=\linewidth, height=\linewidth, keepaspectratio]{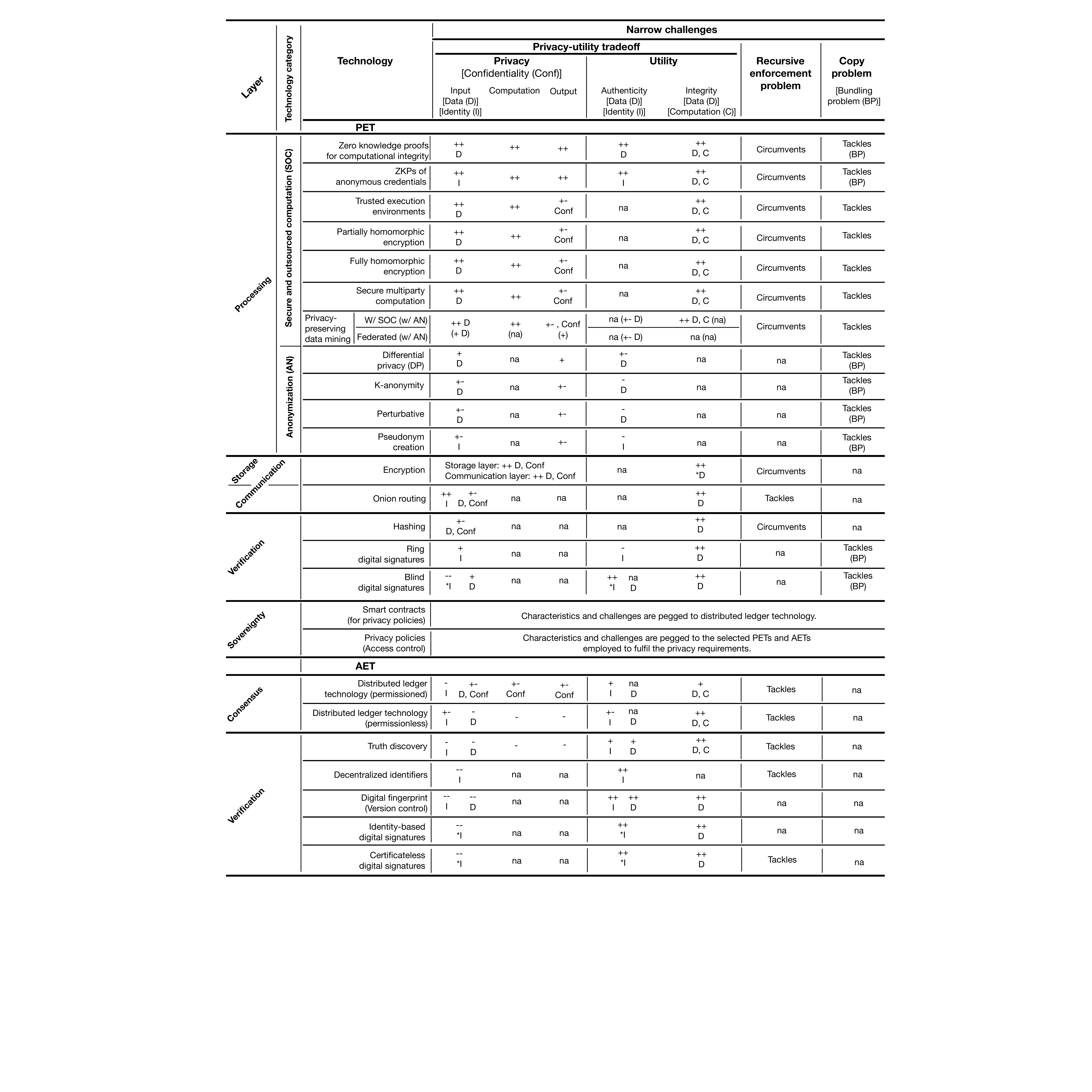}
    \end{tabular}
\label{tab:Tech_Challenge_Master}
\end{table*}

\section{Discussion}
\label{sec:discussion}

\noindent This Section presents a set of key findings (KF) distilled from the two research questions answered in Sections~\ref{sec:PETS}~and~\ref{sec:AETS} as well as~\ref{sec:Challenges}, the content and metadata of the $50$ publications included in our SLR, and other seminal studies that we encountered throughout our SLR but which do not necessarily address IoT data markets directly.
Lastly, we cover the limitations of this study and future work.

\subsection{Key findings}
\label{sec:key_findings}

\begin{figure*}[!htb]
\centering
\includegraphics[scale=0.35]{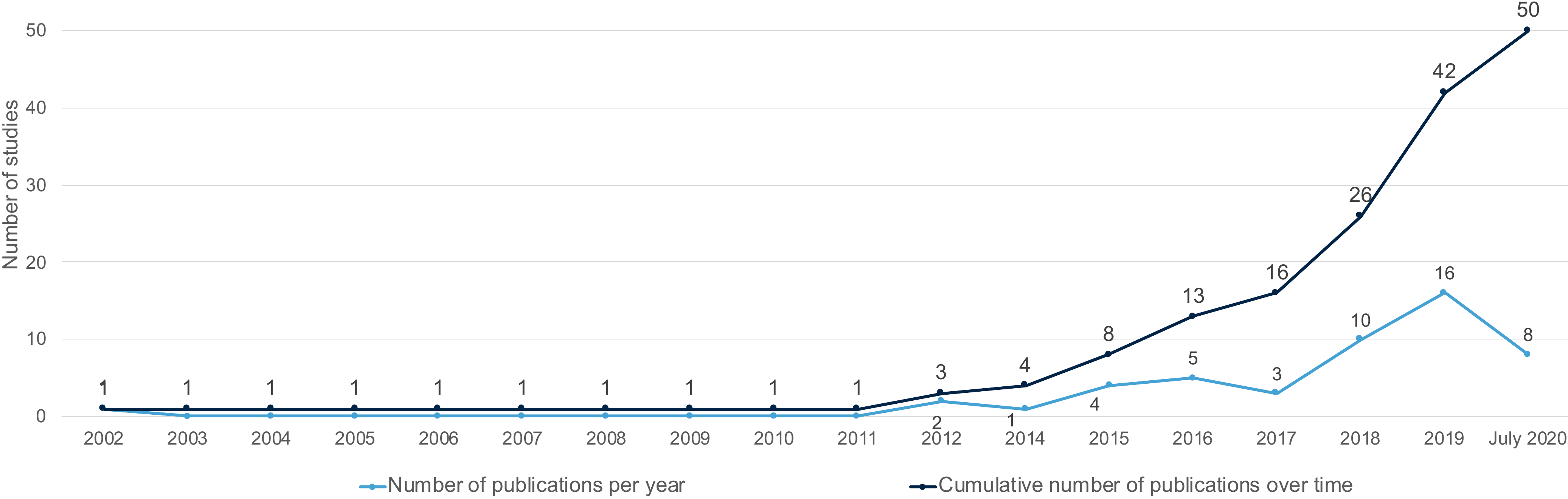}
\caption{Publications in the field of privacy-enhancing data markets for the IoT from January 2002 to July 2020.}
\label{fig:studiesoveryears}
\end{figure*}

% \cite{Kitchenham2010b}
% growing interest
\paragraph{$\mathrm{(KF1)}$ The attention of scientists towards privacy-enhancing technologies in the field of data markets for IoT devices has increased notably in recent years}
The selected publications are modern, as $49$ of the $50$ studies were published between 2012 and 2020, and $34$ of them ($68$\,\%) were published either in 2018, 2019, or during the first half of 2020. 
While the absolute number of publications in 2020 is lower than in 2019 because we captured only the first seven months of 2020, Fig.~\ref{fig:studiesoveryears} illustrates the arguably accelerating trend of the cumulative curve of publications in the field of privacy-enhancing IoT data markets.
%Moreover, the surge of publications employing blockchain technology ($18$ of the $50$ studies, see Fig.~\ref{fig:PETs_distributions}) may be due to the \textit{blockchain-hype} also experienced in this time frame. 

% \paragraph{China and the USA are the countries with the most contributions within the selected studies}
% To determine the geographical distribution of studies, we assigned each study to the location of the first author's research institutions. 
% Figure~\ref{fig:world_map} displays that the countries with the largest number of publications were China ($14$) and the USA ($13$), followed by Germany ($4$) and the UK ($3$). The rest of the countries held either one ($8$ countries) or two ($4$ countries) publications.
% Overall, APAC holds most publications ($22$), followed by Europe ($14$) and North America ($14$). 
% Overall, one may observe that $54$\,\% of the studies were written by (first) research institutions located in the USA ($26$\,\%) or China ($28$\,\%). Coincidentally, both countries are known for their mass surveillance programs~\cite{mass_surveillance_1}\cite{mass_surveillance_2}.
% Furthermore, Europe as a whole accrues $28$\,\% of the total number of publications. 

% stack in design pahse
\paragraph{$\mathrm{(KF2)}$ The most frequent research type (\textit{design and creation}) and least common research contribution (\textit{lessons learned}) suggest that privacy-oriented IoT data markets are still maturing and have not faced many production-grade implementations yet} 
According to Fig.~\ref{fig:contribution_type_research_type_research_approach}, around $76$\,\% of the publications use a \textit{design and creation} research approach, while only $4$\,\% perform a case study.
A further indication of field novelty is that only one out of the $50$ publications had the contribution type \textit{lessons learned}~\cite{P40}. 
Furthermore, while $35$ studies ($70$\,\%) were of research type \textit{solution proposal}, to the best of the author's knowledge, only one solution appears to have an implemented \href{https://oasisprotocol.org/}{system that is applied in production}~\citep{P25}.

% not reusing building bcloks
\paragraph{$\mathrm{(KF3)}$ The selected studies rarely leverage existing libraries that provide PETs and often only build upon architectures developed in previous work to a small degree. Therefore, to gain more practical relevance, it may be beneficial for researchers to improve and extend existing work instead of reinventing the wheel}
The research community and industry have developed many open-source libraries to employ zero-knowledge proofs, homomorphic encryption, secure multiparty computation, or differential privacy (see Section~\ref{sec:PETS}). However, none of the studies have indicated their use.
Furthermore, studies often do not build upon each other, leading to overlapping further. 
For example, W.~Gao~et~al.~\citep{P41} and Z.~Chen~et~al.~\citep{P34} both showcase an auction that obscures the bids by employing partially homomorphic encryption. However, W.~Gao~et~al.~only refers to the work from Z.~Chen~et~al.~in one line, noting that ``[...] \textit{there is only few literatures on designing privacy-preserving schemes in data market auctions.}'' Moreover,~\citep{P42} builds upon~\citep{P27}, and~\citep{P2} upon~\citep{P25}, but each of these two sets belongs to the same group of researchers. In conclusion, it may be beneficial for researchers to incorporate building blocks from previous data market architectures to advance privacy-oriented research.
%For example, \citep{P22} and \citep{P24} propose auctions in which sensitive data is protected with global differential privacy, but their mechanisms differ. \jsnote{Aren't there any improvements?}

Moreover, many studies included in Table~\ref{tab:DLTTable} aim to create an IoT data marketplace employing distributed ledger technology (DLT).
However, there seems not to be a consensus about which DLT to use for IoT data markets, as the authors build upon Ethereum, IOTA, Hyperledger Iroha, Fabric, Agora, or Quorum, among others. Specifically, as an example, \citep{P32} uses Ethereum smart contracts for payments while~\citep{P28} only uses these contracts for safelisting and employs IOTA for payments instead.

%$25$ studies were published in \textit{journals}, $20$ in \textit{conferences} and $5$ in \textit{workshops} (see also Figure~\ref{fig:publicationdomain}).

\paragraph{$\mathrm{(KF4)}$ The content of the selected studies can be categorized into two main orthogonal research streams within the context of privacy-enhancing IoT data markets: architectures and data trading schemata}
The first research stream is dedicated to the design of privacy-enhancing architectures for the exchange of data in IoT data markets ($25$~studies,~$50$\,\%), and the second one focuses on the design of privacy-enhancing data trading such as auctions ($12$~studies,~$24$\,\%). The remaining studies can be associated with domains like legal~\citep{P6}, user preferences~\citep{P8}, or IoT data market challenges~\citep{P14}\citep{P19}.
% Consequently, privacy-oriented data market architectures and trading designs are the predominant deliverables of the studies in this SLR. 
The selected studies, and also international initiatives such as the European GAIA-X~\cite{noauthor_gaia_x_nodate}, hence envision data markets beyond matchmaking and auction capabilities. 
Specifically, the studies that we analyzed structure the software, hardware, abstract entities, and their coordination, data processing, storage, communication, and the offered services to build a holistic or part of a privacy-enhancing IoT data market that includes PETs to tackle some of the challenges described in Section~\ref{sec:Challenges}. 

% architectures 1 2 3 5 7 9 10 11 12 16 17 20 23 25 27 28 30 32 33 39 43 45 46 47 49
% trading 4 (survey) 13 15 18 22 24 26 34 37 41 42 50
% other 6 8 14 19 21 29 31 35 36 38 40 44 48

\paragraph{$\mathrm{(KF5)}$ Despite the acknowledged need for combining anonymization and secure and outsourced computation techniques, none of the researchers behind the $12$ studies proposing data trading schemata, and only two publications out of the $25$ designing data market architectures employ both PET categories in combination}
Although PETs such as homomorphic encryption (HE) or secure multiparty computation conceal inputs and computation, the outputs can leak information about the underlying data and hence may be exposed to re-identification attacks~\cite{trask_structured_transparency_nodate}. Combining secure and outsourced computation techniques with anonymization-based PETs like differential privacy (DP) can help to make the outputs less sensitive. Moreover, leveraging only anonymization PETs does not sufficiently address the copy problem.

Within the $12$ selected studies focused on data trading, DP is the most frequently used PET in auctions to enhance the privacy of the exchanged data. 
\citep{P16}\citep{P22}\citep{P24}\citep{P37} employ DP in various forms to set the privacy levels and, subsequently, the price of the traded IoT data.
Researchers might choose DP over other anonymization technologies because DP is the only PET with a mathematical guarantee of privacy~\cite{dwork_algorithmic_2013}.
At the same time, partially HE (PHE) is the PET of choice to enhance the privacy of the bidding process.
A group of authors~\citep{P26}\citep{P34}\citep{P41} chose PHE primarily for hiding the bids, confidentially computing the winner, and only revealing the output to the auction's winner. 
Researchers might decide to use PHE over other forms of HE, secure multiparty computation, or trusted execution environments (TEEs) despite PHE's significantly less general scope because PHE has relatively high performance and is conceptually simple. 

Together, DP and PHE can holistically enhance the privacy of auctions, which is a contribution we have not found in this review. S.~Sharma et al.~\citep{P48} emphasize that some HE schemata, such as Paillier's, must complement other methods to guarantee more protection. Moreover, while HE protects the input and the computation itself, if the intended recipients of the decrypted output are malicious, they may reverse engineer the output to learn properties about the input. An additional modification employing, e.g., differential privacy, of inputs or decrypted outputs before sharing may help prevent this attack in exchange for accuracy and thus utility. The same argument applies to other secure and outsourced computation methods when used in isolation. We consequently point to a lack of combination in the research stream of data market architectures, except for two publications from the same group of researchers~\citep{P2}\citep{P25}, which use TEEs to train machine learning models with DP. 

% ML also brings privacy issues, which two selected studies tackled directly with TEEs and DP~\citep{P2}\citep{P25}.

\paragraph{$\mathrm{(KF6)}$ The selected studies employ three dimensions to characterize data markets that entail privacy concerns: the degree of decentralization, the types and number of data domains, and the types of sellers and consumers}
Each of these dimensions, for example, characterized by~\citep{P1}\citep{P27}\citep{P46} respectively, brings privacy concerns.
Data may be stored by the seller, the platform provider, or a decentralized platform using, e.g., a combination of commercial cloud storage, interplanetary file systems, or blockchains.
Depending on the degree of decentralization and replication, practitioners need to consider different leakage risks. In particular, if the architecture relies on a blockchain, PETs are particularly important~\citep{priv_in_blockchains}.

An increase in the number and types of data domains opens additional attack vectors and more possibilities for malicious entities to link an individual's data across databases. 
This hyper-connectivity between datasets can render the definitions of de-identified data, such as \href{https://www.dhcs.ca.gov/formsandpubs/laws/hipaa/Pages/1.00WhatisHIPAA.aspx}{HIPAA's}, obsolete and suggests that privacy enhancements in the data economy should be defined globally and not locally. 

The degree of privacy enhancement should depend on the type of seller and consumer, e.g., consumers may expect higher privacy guarantees when a health insurance company gathers their data than when the collector is a renowned health research institution. 

\begin{figure}[htpb]
\centering
\includegraphics[scale=0.45]{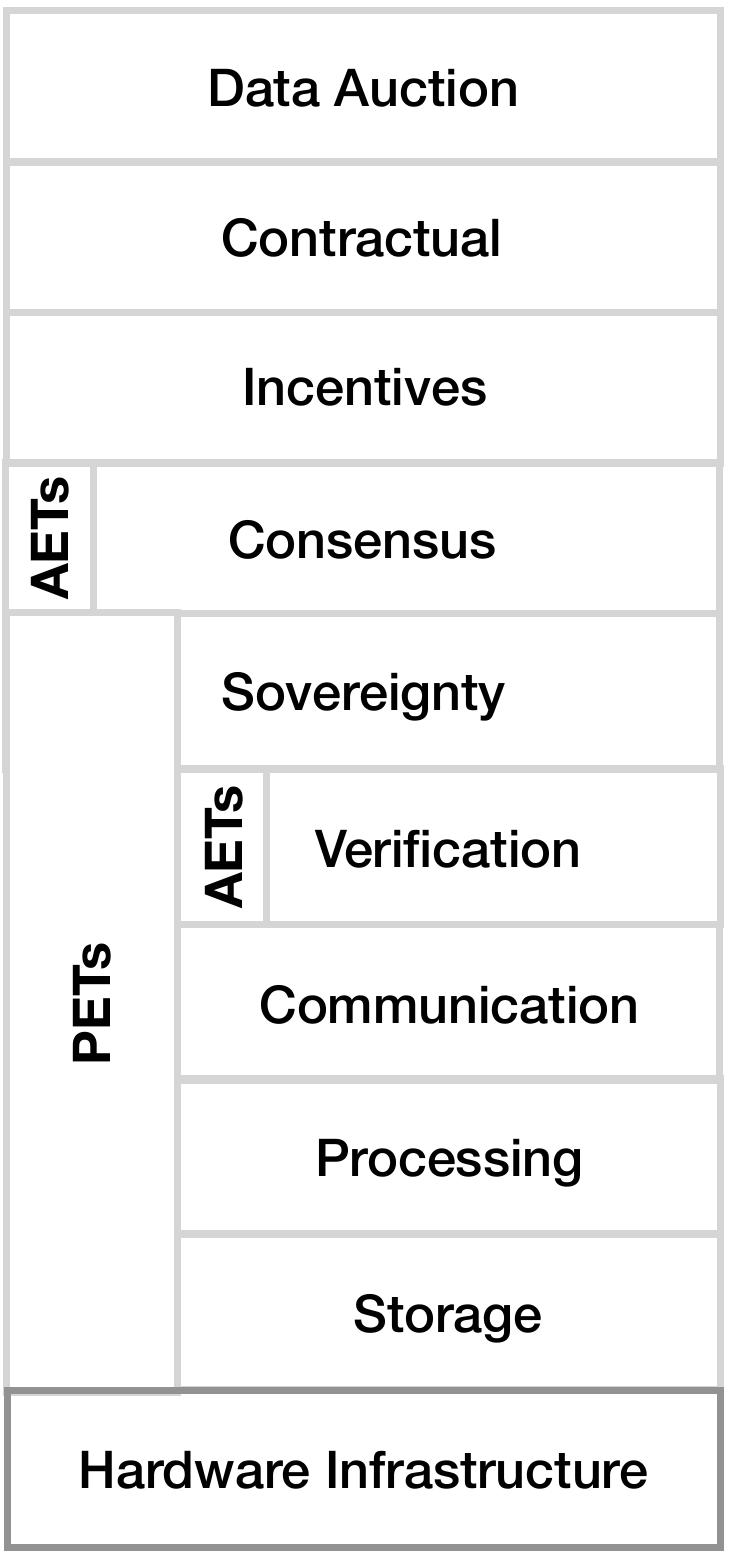}
\caption{Reference model for the layers of a privacy-enhancing IoT data market.}
\label{fig:reference_model_layers}
\end{figure}

\paragraph{$\mathrm{(KF7)}$ Based on our classifications in Sections~\ref{sec:PETS}~and~\ref{sec:AETS} and inspired by a set of seminal selected studies, we have created a reference model for the design of IoT data markets in Fig.~\ref{fig:reference_model_layers}, and detailed in Table~\ref{tab:Tech_Challenge_Master}}
Most of the studies included in this SLR proposed solutions without following a reference model, except for D.~López and B.~Farooq~\citep{P3} and C.~Niu et al.~\citep{P7}, who developed their own without a systematic research. 
C.~Niu et al.~\citep{P7} condense their architecture into two layers: data acquisition and trading.
On the other hand, D.~López and B.~Farooq~\citep{P3} present a more holistic view of privacy-enhancing IoT data markets with six layers (\textit{identification, privacy, contractual, communication, consensus, and incentive}) inspired by the Open System Interconnection model and heavily conditioned by the use of blockchain technology.
This model, however, lacks essential steps of an IoT data market that several publications in our SLR focused on, namely storage~\citep{P1}\citep{P12}\citep{P25}\citep{P28} and processing~\citep{P7}\citep{P19}\citep{P20}\citep{P45}.
Furthermore, the \textit{identification layer}~\citep{P3} can be regarded as a subset of verification, which also includes data verification. 
Other studies base their market design on the type of participants~\citep{P15}\citep{P24}\citep{P27}\citep{P33}\citep{P43}, e.g., sellers, aggregators, brokers, among others, and the type of data domain~\citep{P46}, e.g., health, financial, or a combination.
However, these categories cannot be transferred to other contexts as easily as a reference model agnostic to entity and data domain types.

Our reference model hence combines and generalizes some of the layers from~\citep{P3} and \citep{P7} and complements them with additional layers such as the data auction, storage, verification, processing, and sovereignty layer (see Fig.~\ref{fig:reference_model_layers}).
Most of these layers need multiple PETs, as there is no ``one-size-fits-all'' technology to enhance privacy. 
To navigate these layers in detail, refer to Table~\ref{tab:Tech_Challenge_Master} and Fig.~\ref{fig:PETS_TREE_WHOLE}.
Furthermore, we distinguish between a contractual and sovereignty-related design to separate formal agreements from privacy and ownership policies.
Furthermore, given the distinct purpose and implementation that auction schemata play in a data market, they should be respected by a unique IoT data market layer (auction dedicated studies: ~\citep{P16}\citep{P22}\citep{P24}\citep{P37}, among others).
Lastly, incentives are necessary to encourage behavior that preserves the pre-defined qualities of the IoT data market, e.g., optimized prices~\citep{P3}\citep{P16}, data authenticity~\citep{P16}\citep{P23}, or maintaining the infrastructure like a permissionless DLT. 

% different types of participants \citep{P15}\citep{P18}\citep{P19}\citep{P24}\citep{P33}\citep{P43}
% auction dedicated studies \citep{P22}\citep{P24}\citep{P26}\citep{P34}\citep{P37}\citep{P41}
% Moreover, in our reference model, the characteristic regarding the degree of centralization proposed in~\citep{P1}\citep{P18} will depend on the practitioner's selection of PETs and AETs from our proposed market layers, e.g., 

\paragraph{$\mathrm{(KF8)}$ Aside from the ubiquity of digital signatures in IT systems, in this SLR, distributed ledger technology (DLT) is most frequently employed as the backbone of IoT data market design (see Fig.~\ref{fig:PETs_distributions}), despite the lack of consensus on its use and DLT-based applications in production}
Although centralized systems seem more efficient and easier to deploy, and despite the seemingly few industrial applications running on blockchain today, many researchers in this SLR still advocate for distributed systems using DLT. 
Within the $35$ solution proposals, around $31$\,\% chose permissionless DLT, $14$\,\% consortium DLTs, and the authors of the remaining $55$\,\% either reviewed DLT, implemented a centralized solution, or focused on designing narrow features.
However, we noted that within the $45$\,\% of DLT-based designs, many authors still relied on single entities for data processing or storage.
Specifically, only one of the $50$ studies~\citep{P25} has a public blockchain-based ecosystem in production, yet without a real-world use case running. These statistics indicate a lack of adoption despite substantial research efforts.

Furthermore, while blockchains enhance authenticity, assure integrity, and enable payments without the need for a trusted third party, blockchains are limited in storage capacity~\citep{P1}, computation power~\citep{P25}, and can exacerbate privacy issues because of their tamper proof-quality and inherent data and computation replication~\cite{zhang2019security}\cite{sedlmeir2022transparency}\cite{kannengiesser2020trade}\cite{feng2019survey}.
Consequently, almost all studies that include blockchain technology to support an IoT data market require PETs to protect users' data and identities. These studies go as far as creating innovative privacy-enhancing blockchain architectures with other PETs as building blocks, e.g., trusted execution environments~\citep{P25}\citep{P45}, or adding a privacy layer to their market design based on differential privacy~\citep{P3}.
However, within the literature, there are also questionable statements such as ``[...] \textit{researchers and technologists have found that blockchain can be a potential solution to the privacy problem by decentralizing information} [...] \textit{Blockchain can be used to securely share private information} [...]''~\citep{P3}, ``\textit{Blockchain-based approaches provide decentralized security and privacy} [...]''~\citep{P11}, or ``\textit{Blockchain has been proven to possess security, immutability, and privacy properties, which has caused a lot of researchers to introduce it into the privacy and security concerned IoT}''~\citep{P23}. 
These statements, coupled with the current excitement around blockchain, can lead practitioners in the industry to wrongfully push blockchain for ``privacy''. Therefore, the community would benefit from clear explanations of why authors employ blockchain and clearly state the need for other technologies to enhance privacy.

\subsection{Limitations}
\label{sec:Limitations}
Even though we have adopted a rigorous research design and paid particular attention to the selection and analysis of published studies, SLRs have limitations that may have undermined our effectiveness. These threats include (i) incompleteness of study search, (ii) bias in study selection and (iii) inaccuracy of data extraction.

(i)~Some relevant publications might be absent. To mitigate this limitation, we searched in several highly reputed digital libraries, performed a preliminary search to determine suitable search strings, conducted a backward search to identify additional related work, and included studies in advance that met the standards and filters of this SLR. These measures reduce the probability of missing relevant publications. 
(ii)~The experience and knowledge of the researchers may drive the study selection with an inherent bias. 
Nonetheless, following Kitchenham~\cite{Kitchenham2007}, we aimed to create a set of explicit inclusion and exclusion criteria to maximize the degree of objectivity. To mitigate different appreciations of these criteria, we conducted a preliminary search to ensure researchers have a consistent understanding of the requirements. Furthermore, two researchers conducted the selection process independently and resolved the conflicts between their decisions interactively. (iii)~There might be a bias in selecting the extracted data, which may affect the classification results of the selected studies. To mitigate this potential limitation, the two researchers specified a set of data extraction cards (see Section~\ref{sec:SLR_execution}) to eliminate any misalignment in the data extraction process results. 

\subsection{Future work}
\label{sec:future_work}

The opportunities and need for future work in the context of privacy and data markets for the IoT highlighted by the selected studies resonate with the challenges covered in Section~\ref{sec:Challenges}.
Most notably, there is a need to solve the copy problem~\cite{trask_structured_transparency_nodate}\citep{P4}\citep{P17} and to lessen IoT devices' limitations regarding computation~\citep{P5}\citep{P11}, storage and capacity~\citep{P29}\citep{P48} to tackle or circumvent the constraints PETs may induce.
Moreover, to decrease the probability of re-identification attacks, further work is needed to advance the maturity of PETs and combine them, e.g., bringing together differential privacy and secure and outsourced computation efficiently.
Additional research is also necessary to create standards for data markets, such as a language to describe privacy requirements, universal APIs to interact between different IoT devices with various degrees and techniques for privacy protection, and machine-readable definitions of privacy, e.g., using ontologies~\citep{P40}. In this context, a more detailed description and classification of the layers that we found relevant for classifying privacy and authenticity enhancing technologies (see Figure~\ref{fig:reference_model_layers}) constitutes a promising and relevant avenue for future research. Nonetheless, we want to emphasize that privacy is not the only challenge that needs to be addressed, as future research must also consider, for instance, scalability. 

If society considers privacy a necessity, it should be enhanced by default and optimally in any system without attaching price tags to one's privacy, as some of the selected studies pursued suggest~\citep{P16}\citep{P15}\citep{P22}. We find this posture a worthy research endeavor and encourage researchers to ponder whether monetizing privacy in a competitive market ultimately benefits society. Furthermore, legal practitioners have ample ground to develop legislation specifically around privacy in IoT data markets and for economists to delve into data pricing and decentralized market interactions.
Legal researchers could investigate how stringent privacy regulations should be, as heavy regulation may strangle free markets and innovations~\citep{P45}. Additionally, the legal, pricing and privacy aspects hinge around data sovereignty. As long as ownership is ambiguous, researchers' efforts will struggle to maximize impact. Furthermore, the relevance of our results may reach beyond IoT data markets, as the analysis of PETs and derived insights, e.g., how IoT impacts privacy, can permeate other research areas such as privacy-by-design software engineering, policy-making, and data governance, politics, and economics. Moreover, most PETs have specific performance-, complexity- or utility-related shortcomings (which we describe in Section~\ref{sec:PETS}) that researchers can address.

Lastly, we recommend that researchers derive decision trees based on Table~\ref{tab:Tech_Challenge_Master} to enhance the decision-making of privacy officers beyond our work. Moreover, we could not find any formulation of an information-theoretic quantification of the data leaked from a data market.
We also encourage social scientists to focus on questions related to data sovereignty. 
To realize a vision of data markets that benefit society, we suggest researchers concentrate on roadblocks such as the copy problem. Finally,  institutions should consider updating their privacy-enhancing processes to effectively participate in IoT data markets.

\section{Reassessment of the results}
\label{sec:advances_in_markets}

This section provides and discusses new key publications since the research process ended.
Accordingly, we conducted a research process as per Section~\ref{sec:research_mehtodology} for studies dated between July~2020 and May~2022 and, among them, picked for discussion the ones providing the most significant updates to our systematic literature review or, on the contrary, underlining our previous findings.
Note that the references included in this section correspond \emph{only} to the newly found publications.

In our new search, we again selected primary and secondary studies. Overall, our new search resulted in $24$~publications: $3$~more from~2020, $14$~from~2021, and---as of May---$7$ from 2022. These statistics indicate that the trend depicted in Fig.~\ref{fig:studiesoveryears} (consolidating KF1) has not reversed. 
Notably, we could still not find publications discussing production-ready deployments of privacy-enhancing architectures or auction schemata and no reference to open-source tooling despite PETs being more mature since July~2020 (underlining KF2 and KF3). 

Among the \emph{secondary studies}, Wang~et~al.~\cite{wang2021managing} explored the concept of privacy in the digital economy more broadly and pointed out the need for interdisciplinary research to supplement the purely technical PET constructions with the economic (tradeoff between accuracy and privacy) and governance perspectives (privacy policies) that we elaborate on in our paper.
M.~Akil~et~al.~\cite{9194705} systematically investigated privacy-oriented identity management in the context of the IoT, such as anonymous credentials and other techniques that our review covers. Moreover, T.~Gebremichael~et~al.~\cite{9169653} conducted a less systematic survey of standards and future challenges, including discussions regarding authentication and access control, and highlighted a subset of the privacy challenges of IoT that we present in Table~\ref{tab:Challenges_IoT_Table}.
Additionally, S.~W.~Driessen~et~al.~\cite{9739681} presented a recent systematic literature review on designing data markets. However, their work did not focus on privacy.

More secondary studies, such as from Deepa~et~al.~\cite{deepa2022survey} considered blockchain and smart contracts beneficial for privacy.
Furthermore, Perez~et~al.~\cite{perez2022secure} reviewed PETs in the context of crowdsensing and emphasized the privacy issues with smart contracts, along with practical challenges in security and feed-in of reliable data. They suggested a subset of the anonymization techniques that we present in Section~\ref{sec:PETS}, such as privacy-oriented digital signatures, anonymous networking, 
$k$-anonymity, $l$-diversity, $t$-closeness, and differential privacy (DP), and some more specific ones in the context of location. While they also mentioned ZKPs, there is no detailed discussion of the secure computing techniques we survey. Gon{\c{c}}alves~et~al.~\cite{goncalves2021critical} considered privacy mechanisms in data sharing for collaborative forecasting and discussed the tradeoff between privacy and accuracy. They distinguished between perturbative techniques (``data transformation''), MPC-based protocols, and distributed or federated approaches (``decomposition'') combined with DP.
Lastly, Y.~Wu~et~al.~\cite{wu_deep_2021} provided a survey that examined the privacy risks that machine learning poses on IoT data markets supported by blockchains. 
Hence, our SLR, with its comprehensive focus on privacy, still fills the gaps that we discussed in Section~\ref{sec:related_work}. 

The \emph{primary studies }followed a similar pattern to the previously collected studies.
Above all, many still employed blockchains and often did not provide clear explanations the corresponding benefits and acknowledgements of the corresponding challenges, specifically regarding privacy (reaffirming KF8). We again encountered questionable claims such as ``[...] a decentralized approach based on distributed ledger technologies (DLT) enables data trading while ensuring trust, security, and privacy''~\cite{IoT_trade_benchmark}, without discussing why DLT enhances privacy in the rest of the publication about benchmarking IoT data trading protocols in blockchains.
Others followed suit on the use of blockchain to support electric vehicle trading marketplaces with IPFS and a scheme to hide payment sources~\cite{9311120} and cloaking location with $k$-anonymity~ \cite{9220870} or proposing a new blockchain architecture with permissioned domains to enhance privacy for data market places~\cite{Xu2021FedDDMAF}. Another presented several building blocks (blockchain, trusted execution environments, gossip learning) without an evaluation of the proposal~\cite{9438799}. 

A notable exception is the comprehensive details provided by Manzoor~et~al.~\cite{manzoor2021proxy} in their blockchain architecture. Their architecture stores encrypted sensor data in cloud storage, and smart contracts support sensor registration, data auctioning, and payments. 
While the smart contract emits notifications and displays the endpoint for retrieving proxy re-encrypted data, the data are exchanged off-chain confidentially via proxy re-encryption. This construction addresses transparency and scalability issues regarding sensor data. Nevertheless, bidding and payment processes may still reveal sensitive information and require future research by combining this approach with some of the PETs we surveyed. Rückel~et~al.~\cite{rueckel2022fairness} also acknowledged the aggravation of privacy issues on blockchains and combined federated learning with DP to obfuscate clients' weights and use ZKPs to prove the integrity of the training and evaluation process, which they required for providing fair incentives managed by a smart contract. 
Another related publication by Gupta~et~al.~\cite{gupta2022trailchain} presented a blockchain-based solution for tracking IoT sensor data across marketplaces and, thus, only detecting but not preventing illegitimate replication and resale.

These publications fall into the category of \emph{architectures} identified in KF4. We also found papers in the \textit{data trading schemata} category (or related): two new data auction schemata enhanced with DP~\cite{9381206}\cite{101109}, a task assignment scheme in crowdsensing that hides the tasks' content with homomorphic encryption (HE) for crowdsensing~\cite{9408590}, and another where they employ DP on billing data~\cite{9756499}.
The latter publication, however, does not discuss \emph{fairness}, which is critical in monetary use cases as a noisy bill can make data prosumers profit less from their data on some occasions. 
Furthermore, Shen~et~al.~\cite{shen2022personal} and Hu~et~al.~\cite{hu2021trading} focused on determining fair prices for end users' data sets that are anonymized with DP according to their accuracy and, correspondingly, risks of revealing sensitive information.

One interesting development that explored the paradigm in ML markets comes from Q.~Song~et~al.~\cite{3485832}. They developed a privacy-enhanced framework to evaluate the quality of ML models and data for sale with functional encryption, achieving improvements over similar schemata implemented with HE.
Another novel concept by M.~N.~Alraja~et~al.~\cite{alraja_integrated_2021} strives to empower users with tools to help them determine the risk of sharing their information and, accordingly, make an informed decision about their data framework. 
It is thus closely related to enforcing privacy policies in the sovereignty layer.
Except for Rückel~et~al.~\cite{rueckel2022fairness}, who 
focused narrowly on federated learning, the rest of the new (notable) primary studies did not leverage the combination of anonymization and outsourced computation technologies, underlining KF5.
Notably, the new publications have not altered the data market characterization of KF6 or the reference model of KF7.

\section{Conclusion}
\label{sec:conclusion}

With this review, we reveal the landscape of PETs in data markets for the IoT. 
We have conducted a systematic literature review (SLR) to identify and filter the studies aiming to solve this landscape's challenges.
Consecutively, we formulated terminology to dissect the selected studies' architectures and findings and identified the PETs that related work employed and which specific challenges they addressed.

%The benefits of data trading have become evident to institutions and businesses~\cite{kaaniche2016abs}, but the increasing number of data breaches also demonstrates that the privacy threat is real~\cite{data_breach}. 
%Society's progress thus partly depends on the former, but it is hindered by the latter. This conundrum has sparked the interest of researchers around the globe.
The authors of the selected studies in this SLR have devised proposals for privacy-enhancing IoT data marketplaces to comply with privacy requirements while maintaining utility, profitability, and fair and seamless data exchange.
Since this is a relatively new, multidisciplinary research field, the optimal combination of technologies and theoretical foundations employed in these proposals is still in the development phase.
Therefore, no proposal has established itself as canonical yet.
Moreover, we observed that the research community needs to further explore the balancing act of utility and privacy before data markets flourish.
We conclude that the practicality of PETs needs to advance further to positively impact data markets for the IoT.
Additionally, we suggest researchers solve the copy problem and improve privacy-enhancing verification as their absence discourages data markets from forming. 
We also discovered that research on privacy-oriented data markets could benefit from increased reuse of components from previous articles and existing open-source libraries and a more explicit description of critical objectives. 
For example, the benefits of utilizing distributed ledger technology (DLT) in data markets for IoT architectures often remain unclear, and authors do not sufficiently consider DLT's lack of maturity and inherent privacy challenges.

The IoT's particular characteristics bring new challenges for privacy enhancement, most notably, the consequences of a lack of interoperability, computation and storage constraints, and the privacy disparity across jurisdictions.
We have also observed the importance of \emph{first} determining the sovereignty layer in data market design, as the participants' ownership and management rules impact the PETs in the rest of the layers. 
We also must underline that there is no ``one-size-fits-all'' PET. Only a combination may tackle the various privacy challenges facing data markets for the IoT.
Lastly, we recommend that institutions invest resources in the research and adoption of PETs to remain competitive in the advent of a more privacy-enhancing IoT.

\section*{Acknowledgements}

\noindent We would like to thank the Bayerisches Forschungsinstitut für Digitale Transformation for supporting our research on differential privacy, and the Bavarian Ministry of Economic Affairs, Regional Development and Energy for their funding of the project ``Fraunhofer Blockchain Center (20-3066-2-6-14)'' that made this paper possible.

%elsarticle-num

\ifappendix

\begin{btSect}[elsarticle-num]{selected_papers.bib}
\renewcommand{\bibprefix}{}
%\section*{Selected studies}
\label{appendix:selectedstudies}
\btPrintAll
\end{btSect}

%\clearpage

%%% AITSAM commented this part %%%
%\begin{btSect}{references.bib}
%\section*{References}
%\label{sec:references}
%\btPrintCited
%\end{btSect}

%\bibliographystyle{unsrt}
% IEEEtran

%TC:ignore
%plainnat for author and year info
\bibliographystyle{elsarticle-num}
\bibliography{9_references}

\newpage

\appendix

\clearpage
\onecolumn
\section{Acronyms}

\footnotesize
\begin{tabular}{ l l }

AET & Authenticity-enhancing technologies \\
BP & Bundling problem \\
CP & Copy problem \\
DF & Digital fingerprint \\
DID & Decentralized identifier \\
DLT & Distributed ledger technology \\
DP & Differential privacy \\
DS & Digital signature \\
FHE & Fully homomorphic encryption \\
FL & Federated learning \\
GAN & Generative adversarial networks \\
GDPR & General Data Protection Regulation \\
HE & Homomorphic encryption \\
ICT & Information and communication technology \\
IoT & Internet of things \\
KGC & Key generation center \\
ML & Machine learning \\
PET & Privacy-enhancing technology \\
PHE & Partially homomorphic encryption \\
PKI & Public key infrastructure \\
PPDM & Privacy-preserving data mining \\
REP & Recursive enforcement problem \\
RQ & Research question \\
SC & Smart contract \\
SLR & Systematic literature review \\
MPC & Secure multiparty computation \\
TD & Truth discovery \\
TEE & Trusted execution environment \\
ZKP & Zero-knowledge proof \\

\end{tabular}

\clearpage

\section{Summaries of the selected secondary studies}
%\label{sec:summaries_architecture_papers}
%TC:ignore
% Table: Listing of experience reports, and mapping to case organizations
\begin{table}[htpb]
%\begin{landscape}
\begin{centering}
\footnotesize
\begin{tabularx}{\linewidth}{%
>{\RaggedRight}p{0.5cm}%
>{\RaggedRight}p{0.5cm}%
>{\RaggedRight}p{3.55cm}%
>{}p{10.95cm}}

\caption{Secondary studies on privacy in data markets.} \label{tab:review} \\
\toprule
\textbf{Year}  & \textbf{Study} & \textbf{Topic} & \textbf{Description}  \\
\midrule
\endfirsthead
\caption{Secondary studies \emph{(contd.)}} \\
\toprule
\textbf{Year}  & \textbf{Study} & \textbf{Topic} & \textbf{Description} \\
\midrule
\endhead
\midrule
\endfoot
\bottomrule
\endlastfoot

2020 & \citep{P8} & Privacy-enhancing design of data markets &
Analyzes internet users’ preferences for privacy in data sharing to uncover mental models of these preferences and their motives, barriers, and conditions for a privacy-enhancing data market. 
It provides a set of key findings, the two most notable ones being that the primary barrier to creating data markets is privacy and moral concerns and that the level of anonymization has the largest effect on the willingness to share. \\ \addlinespace

2019 & \citep{P19} & Privacy and security data flow challenges in an internet of production &
Introduces the internet of production and illustrates its inter-organizational data flows.
It also identifies security and privacy demands and challenges within these data flows: authenticity, data access scope, and anonymity. 
Furthermore, it provides a survey of PETs to tackle these challenges: provide confidentiality, hide information during computation (data processing), verify the authenticity of information (providing support), deploy mechanisms that enforce rules (platform capabilities), and support approaches that focus on the security of data flows (external measures). \\ \addlinespace 

2019 & \citep{P14} & Challenges and research opportunities in data markets for the IoT  &
A short study that identifies three research opportunities in IoT data markets: Procurement, pricing, and privacy. 
Significant identified challenges are: ambiguity in data ownership that hinders trading, the difficulty to detect data piracy, and that privacy must be considered before trading.  \\ \addlinespace

2018 & \citep{P23} & Privacy enhancing in IoT applications  &
Introduces and surveys privacy-enhancing technologies in the processes of data aggregation, trading, and analysis; in particular, it discusses outsourced computation, data validation, and blockchain technology.
Additionally, it describes types of privacy breaches and their countermeasures. 
Furthermore, it reviews relevant aspects of pricing procedures as well as game-theoretical approaches and auction schemes.\\ \addlinespace

2018 & \citep{P4} & Pricing, trading, and protection in data markets for the IoT  &
Surveys the three fields of pricing models and strategies, design of platforms and data trading, and digital copyright mechanisms with a focus on privacy enhancement. \\ \addlinespace

2018 & \citep{P48} & Privacy-enhancing analytics for IoT and cloud-based systems &
Summarizes privacy-enhancing technologies in the specific use case of a health data collecting app in the health industry.
More specifically, it separates privacy-enhancing technologies into two scenarios: Outsourced computation and information sharing. \\ \addlinespace

2016 & \citep{P35} & Privacy enhancing in crowd-sourcing task management & 
Surveys privacy-enhancing technologies and the challenges of crowdsourcing task management.  
The proposed technologies are anonymization, such as k-anonymity, spatio-temporal privacy approaches, such as spatial cloaking or aggregated location via differential privacy, and policy-based privacy preferences.
The challenges that they present revolve around trust and credibility, reward-based tasking, utility, efficiency, enforcing privacy-enhancing technologies, and raise privacy awareness.\\ \addlinespace

2015 & \citep{P27} & Privacy enhancing and challenges in data markets for the IoT  &
The study introduces privacy enhancing for data markets for IoT devices, focusing on sensing-as-a-service (data analysis of user-aggregated data). 
It identifies three challenges: Developing IoT middleware for data analysis and autonomous privacy enhancing, autonomous end-user consent acquisition and negotiation, and the autonomous modeling and negotiation of privacy risk and economic reward. 
The most prevalent privacy-enhancing technologies and strategies they introduce are personal information hubs, onion routing, and data aggregation via differential privacy or k-anonymity. \\ \addlinespace

2014 & \citep{P38} & Privacy threats and challenges in the IoT  & 
Classifies the threats and challenges that come along with privacy in IoT applications for individuals into seven categories: Re-identification of individuals through persistent pseudo-identifiers, localization and tracking, profiling for social engineering and price discrimination, information disclosure in life cycle transitions, information linkage of previously separated systems, inventory attacks, and the disclosure of private information to an uninvited audience. \\ \addlinespace

\bottomrule
\end{tabularx}
\end{centering}
%\end{landscape}
\end{table}

%TC:endignore

\clearpage
\section{Mappings of privacy- and authenticity-enhancing technologies}
\label{sec:PETs_tree_whole}

\begin{figure}[htpb]
\centering
\includegraphics[scale=0.6]{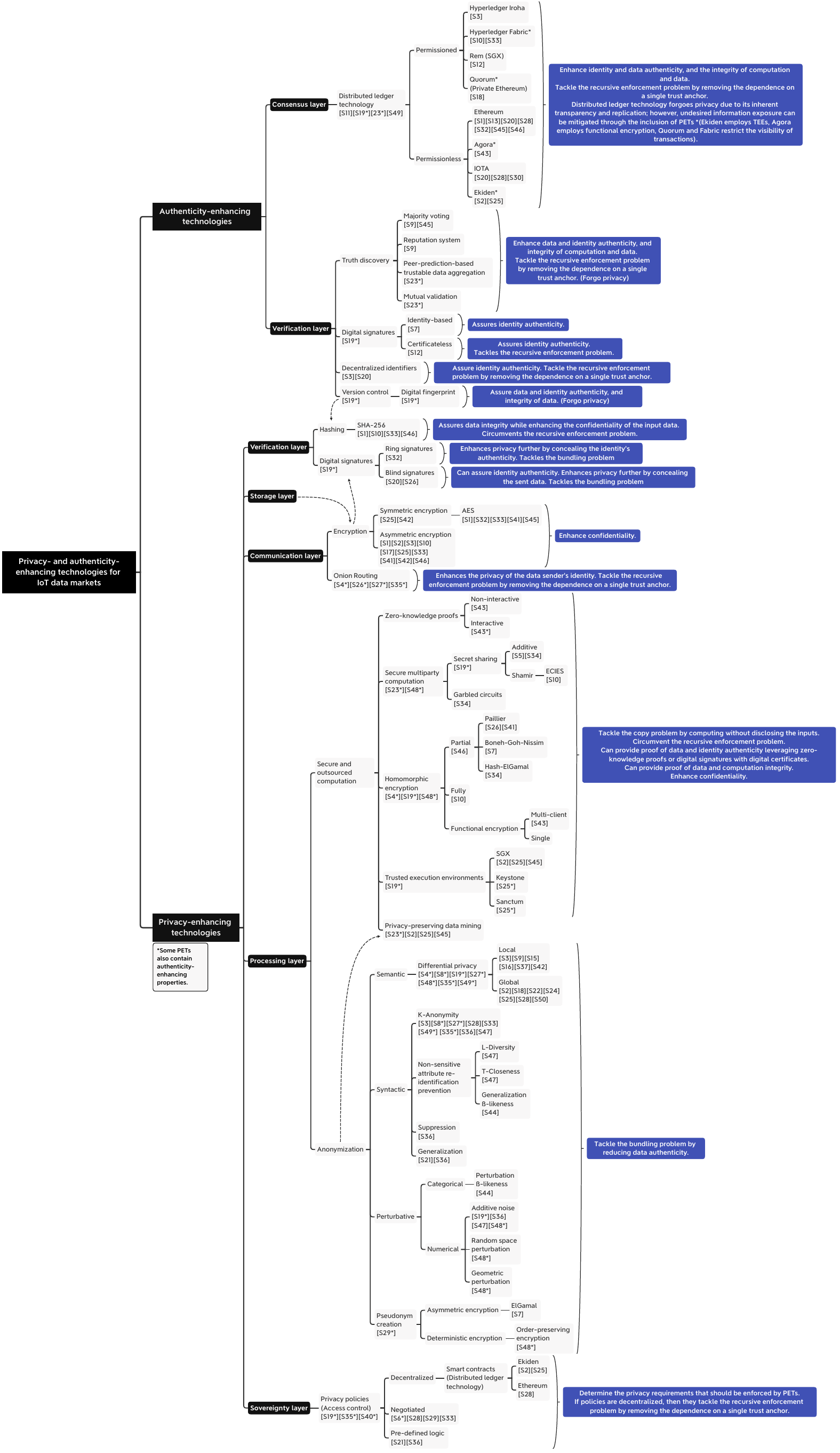}
\caption{Classification of the identified privacy- and authenticity-enhancing technologies in this SLR, together with the challenges they tackle.
Any other privacy approach encountered in the SLR without a succinct inclusion of the underlying technology was either not included in a leaf node but in a parent node or completely dismissed if too vague. 
*The publication reviews the technology without delving into it in-depth or using it as a building block of the architecture concept, e.g., the technology is only mentioned in the opportunities for future work.}
\label{fig:PETS_TREE_WHOLE}
\end{figure}

\onecolumn

\section{Summaries of the selected primary studies}
%\label{sec:summaries_architecture_papers}

% I would separate tables for architecture, and paradigms/proposals, and mathematical algorithms
%TC:ignore
% Table: Listing of experience reports, and mapping to case organizations
\begin{table*}[htpb]
%\begin{landscape}
\begin{centering}
\footnotesize
\begin{tabularx}{\linewidth}{%
>{\RaggedRight}p{0.5cm}%
>{\RaggedRight}p{0.5cm}%
>{\RaggedRight}p{3.5cm}%
>{}p{11.0cm}}

\caption{Set of examples from the selected studies describing different data marketplace architectures and auctions based on PETs and AETs with a focus on secure computation  technologies.}
%\jsnote{Maybe referring to the limitations / novelties would help a reader appreciate the summary more?} \gm{If we are going to put these summaries in the appaendix, is it worth the effort?} \jsnote{Yet another thing: In the table, it becomes pretty apparent that essentially, the auction with the help of metadata and the actual exchange of data can be separated (particularly as fair exchange is impossible). So what should we focus on? I guess the secure auction is less relevant to businesses. I am also not sure whether the anonymization of bidders in an auction through onion routing has to do with the copy problem for data.} \gm{Let us focus on the info exchange. Onion routing does not do anything to prevent the copy problem, yet people have used. We can diminish its importance in the paper.}  
\label{tab:homomorphic_SMC_Table} \\
\toprule
\textbf{Year}  & \textbf{Study} & \textbf{Privacy-enhancing approaches} & \textbf{Description}  \\
\midrule
\endfirsthead
\caption{Secondary studies \emph{(contd.)}} \\
\toprule
\textbf{Year}  & \textbf{Study} & \textbf{Privacy-enhancing approaches} & \textbf{Description} \\
\midrule
\endhead
\midrule
\endfoot
\bottomrule
\endlastfoot

2020 & \citep{P41} & Partially homomorphic encryption, symmetric encryption, and digital signatures & 
Develops a privacy-enhancing auction for big data trading using Paillier's cryptosystem~\cite{P41R17} and a one-time pad. 
They consider four entities: sellers, buyers, an auctioneer, and an intermediary platform.
A data auction is carried out without any entity seeing the data (except the auction winner) or the bid values, which are ordered obliviously by the auctioneer thanks to the homomorphic properties of the ciphertext. 
Furthermore, to efficiently encrypt the data, the authors use symmetric encryption (AES). 
Lastly, digital signatures are created with the same homomorphic cryptosystem, which the authors use to encrypt the symmetric keys. \\ 
\addlinespace
 
2018 & \citep{P7} & Partially homomorphic encryption, digital signatures, and pseudonym creation & 
Implements a platform for data markets that facilitates data processing and outcome verification while enhancing the privacy of identities and their data. 
The authors consider four entities: data contributor, service provider, data consumer, and a registration center; in a two-layer system model: data acquisition and trading. 
Furthermore, the platform synchronizes data processing and signature verification into the same homomorphic ciphertext space (encrypt and then sign). 
Additionally, they tightly integrate data processing with outcome verification via a set of homomorphic properties. 
To achieve a trade-off between functionality and performance, they selected a partially homomorphic scheme called Boneh-Goh-Nissim cryptosystem~\cite{boneh_evaluating_2005}. \\ \addlinespace

2018 & \citep{P10} & Fully homomorphic encryption, secure multiparty computation, distributed ledger technology, and hashing  &
Provides a distributed outsourcing computation architecture, whereby data owners may request fully homomorphic computations with a schema called fully homomorphic non-interactive verifiable secret sharing~\cite{FHE}. 
Moreover, the proposed architecture allows transactions to be verified by the participants of the permissioned blockchain thanks to the immutability properties of the blockchain; Hyperledger Fabric was the selected blockchain architecture.
Moreover, the hash value of the shared data is stored in the blockchain, for data recipients to verify the truthfulness of the received data.
Furthermore, for secure multiparty computation, the authors implement Shamir's secret sharing~\cite{Shamir_secret_sharing} with ECIES by leveraging the distributed nature of the blockchain. 
In this manner, the data owner may share verifiable pieces of information with a set of servers.
Then, the servers execute the necessary computations, and when several verified responses are received by the agreed data consumer, the true result is recovered. \\ \addlinespace
%\jsnote{I did not find MPC stated in the paper -- do you include it because of secret sharing?} \gm{Yes, should we not tag this as SMC then? it would change some plots and some conclusion numbers perhaps. If it is not a crime we could let it be.} 

2016 & \citep{P34} & Partially homomorphic encryption, and secure multiparty computation & 
Develops an auction cloud-based framework that cryptographically hides the bids from all auction participants until a winner is determined. 
It achieves this by combining PHE based on the hashed scheme~\cite{HashElGamal_HE} of ElGamal~\cite{elgamal}, and secure two-party computation through garble circuits and additive secret sharing. \\ \addlinespace

2015 & \citep{P26} & Partially homomorphic encryption, digital signatures, and onion routing & 
Proposes a combinatorial auction~\cite{combinatorial_auction} mechanism that ensures the privacy of the bidders.
The bidders bids are blindly signed through the third party~\cite{blind_signatures} so that the third party does not learn the contents. 
Later, these signatures are used by the bidder to prove the authenticity of the bids. 
The winner is determined by the third party through a partially homomorphically encrypted computation using the Paillier cryptosystem~\cite{P41R17}.
Lastly, the identities of the bidders are enhanced by using onion routing~\cite{Tor}. \\ \addlinespace

2012 & \citep{P5} & Secure multiparty computation & 
Implements a privacy-enhancing data mining service market, whereby data donors distribute confidential data among a set of participants employing additive secret sharing.
The miners collectively perform secure multiparty computation based on the author's algorithm~\cite{bogdanov_sharemind_nodate}.
Finally, the results are in turn sent to the previously agreed analyst, who combines them to obtain the intelligible output. \\ \addlinespace

% mathematical model, not an architecture
% 2012 & \citep{P44} & $\beta$-Likeness & Proposes a new privacy model ($\beta$-Likeness) to anonymize an individual's data so that the likelihood of an sensitive value to be true does not surpass a threshold, $\beta$. \\ \addlinespace

\bottomrule

\end{tabularx}
\end{centering}
%\end{landscape}
\end{table*}

%TC:endignore
\clearpage

% I would separate tables for architecture, and paradigms/proposals, and mathematical algorithms
%TC:ignore
% Table: Listing of experience reports, and mapping to case organizations
\footnotesize
\begin{longtable}{p{0.5cm}p{0.5cm}p{3.5cm}p{11cm}}
\caption{Set of examples from the selected studies describing different data marketplace architectures and trading mechanisms based on PETs and AETs with a focus on anonymization technologies.}
\label{tab:AnonymousTable}
\\\toprule
\textbf{Year}  & \textbf{Study} & \textbf{Privacy-enhancing\newline approaches} & \textbf{Description} \\
\midrule
\endfirsthead
\endhead

\caption{PETs and AETs in the context of anonymization \emph{(continued).}} \\
\toprule
\textbf{Year}  & \textbf{Study} & \textbf{Privacy-enhancing\newline approaches} & \textbf{Description} \\
\midrule
\endhead
\addlinespace
\bottomrule
\endfoot

2020 & \citep{P24}  & Differential privacy & 
Designs a privacy-enhancing crowd-sensed data trading mechanism.
First, the data broker orchestrates an auction whereby data consumers bid in a differentially private manner for a data asset.
Secondly, to form a data asset, the data broker creates a set of data generation tasks, some of which are fake to protect the privacy of the auction winner.
Lastly, the data broker selects data owner outputs in a differentially private manner.
More specifically, both the auction-based data pricing and the data collection are based on the differentially private exponential mechanism. \\ \addlinespace

2019 & \citep{P37} & Differential privacy & 
Proposes a differentially private data market auction framework with a fair negotiation method to set the price and noise; this study is extended in~\citep{P42}.
The entities involved are a data provider, a consumer, and a trusted market manager that matches providers with consumers and enforces Rubinstein bargaining.
Firstly, the data provider and consumer enter a negotiation phase that involves the data query, the $\varepsilon$ values, and the unit price for $\varepsilon$.
Once the negotiation is over, the data provider answers the query with the agreed $\varepsilon$ with local differential privacy.
\\ \addlinespace

2019 & \citep{P42}  & Differential privacy and digital signatures & 
Proposes a differentially private data market framework.
This study extends~\citep{P37} by specifying the type of differential privacy algorithm, and the digital signature schemata followed to deploy the framework in practice.  
The authors use local differential privacy for numeric~\cite{numeric_LDP} and for categorical~\cite{categorical_LDP} data types.
% Lastly, the provider encrypts the answer with a symmetric key shared with the consumer and, in turn, the provider encrypts that key and the encrypted data with the consumer's public key from the consumer. 
\\ \addlinespace

2019 & \citep{P15}  & Differential privacy & 
Designs contracts for a data marketplace whereby a data broker matches the required accuracy from a data consumer with the degree of privacy that data owners desire.
Furthermore, by handpicking the data sources, the differentially private algorithm incurs a bias that makes the output more accurate while maintaining the desired privacy from the data owners. 
Lastly, the authors derive an optimal data contract to minimize payment while satisfying accuracy and privacy.
\\ \addlinespace

2019 & \citep{P50}  & Differential privacy & 
Proposes a framework for counting trading range query results, and designs a pricing approach for the traded results.
Firstly, the framework calculates the range counts approximately, and secondly, it protects the results further by using differential privacy, while satisfying the accuracy demands of data consumers. 
The authors also design the pricing scheme in a way that prevents arbitrage. 
\\ \addlinespace

2019 & \citep{P22}  & Differential privacy & 
Designs an online auction with two stages, whereby a trusted auctioneer aggregates data from data owners and applies differential privacy before selling the data to consumers. 
In the first stage, the auctioneer selects data owners based on their privacy requirements to maximize profit. 
In the second stage, the auctioneer applies differential privacy to the aggregated data and subsequently sells the data to a consumer in an auction.
\\ \addlinespace

2018 & \citep{P2} & Differential privacy, distributed ledger technology, smart contracts, privacy policies, asymmetric encryption, and trusted execution environments & 
Implements an end-to-end privacy-enhancing decentralized data marketplace for data consumers to train machine learning algorithms. 
The authors achieve end-to-end privacy by protecting inputs with asymmetric encryption and differential privacy, and the execution with trusted execution environments.
More specifically, differential privacy prevents the weights of machine learning algorithms from overfitting to the inputs.
Because of the privacy limitations of current distributed ledger technology applications, the authors of this study and of the subsequent publication~\citep{P25} created a novel concept unlike any other blockchain-based system.  
For example, in principle, the smart contracts of their architecture may contain machine learning algorithms which may be executed in trusted execution environments, hold privacy policies and payment logic, and point to where encrypted data and decryption keys are stored privately.
On the other hand, data consumers also deploy smart contracts that may interact with the data owners. \\ \addlinespace

2018 &\citep{P16} & Differential privacy & 
Develops an auction framework for privacy-enhancing data aggregation for mobile crowdsensing.
The auctioneer chooses data owners based on their sensing capabilities, and the data owners apply differential privacy to their inputs sampling from a noise distribution tailored by the auctioneer for each data owner based on its qualities.
The goal of the platform is to optimize task allocation to a set of data owners while minimizing their payment, taking into account accuracy and privacy constraints. 
\\ \addlinespace

2018 & \citep{P9} & Differential privacy and truth discovery & 
Designs two locally differentially private mechanisms for truth discovery in crowd sensing, so that the answers from edge devices are protected while being useful in aggregate.
The second mechanism provides more utility for an equal degree of privacy, and consists on the users randomly selecting a probability distribution, and in turn, adding noise sampled thereof to their truthful answer.
\\ \addlinespace

2018 & \citep{P21} & Privacy policies and generalization & 
Proposes a data market framework that models and enforces privacy policies dynamically for data-intensive applications. 
More specifically, the authors implement a data-flow-focused system with a policy enforcement algorithm defined by users and a context.
In data-flow computing, directed graphs embody the application, where edges represent data streams and nodes represent functional operators and data sources or sinks. 
The data is anonymized based on policies and enforced by generalization, e.g., substituting Munich with Germany. 
To formalize a language to model the privacy policies, the authors use metric first-order temporal logic. \\ \addlinespace

2017 & \citep{P47} & $K$-anonymity and additive noise & 
Models a data marketplace in which groups of users may actively monetize their data through a mediator and a set of mobile crowd sensing service providers.
The authors use a reverse auction, where users bid for performing sensing tasks.
Individual users may set their own privacy preferences, and if they are a coalition of users, they are protected by k-anonymity, t-closeness, l-diversity and local noise addition approaches. 
The total coalition payoff is divided among the cooperative users based on their marginal contributions to the total data quality at the end of the sensing service.\\ \addlinespace

2016 & \citep{P36} & $K$-anonymity, additive noise, and privacy policies & 
Designs a one-to-one privacy enhancing paradigm for a data market place in which privacy policies and data requirements are defined based on the publication record of the data owner.
Because published records of a user aggregate over time and thus accrue privacy risk, the paradigm relies on privacy risk management, which is enforced by evaluating the risk associated with revealing yet another piece of information with regard to the privacy requirements. 
This evaluation is based on the preferences of the user, or if unfeasible, based on current regulation; furthermore, it is based on an assessment of the background information, achieved by semantically analyzing attributes that if released could be linked to externally available information.
Ultimately, to privatise the data, the authors propose syntactic technologies such as k-anonymity, suppression, and generalization, and semantic ones like additive noise.

\end{longtable}

%TC:endignore
\clearpage

% I would separate tables for architecture, and paradigms/proposals, and mathematical algorithms
%TC:ignore
% Table: Listing of experience reports, and mapping to case organizations
\footnotesize
\begin{longtable}{p{0.5cm}p{0.5cm}p{3.5cm}p{11cm}}
\caption{Set of examples from the selected studies describing different data marketplace architectures based on PETs and AETs with an underlying distributed ledger technology.}
\label{tab:DLTTable} 
\\\toprule
\textbf{Year}  & \textbf{Study} & \textbf{Privacy-enhancing\newline approaches} & \textbf{Description} \\
\midrule
\endfirsthead
\endhead

\caption{PETs and AETs in the context of DLT \emph{(continued).}} \\
\toprule
\textbf{Year}  & \textbf{Study} & \textbf{Privacy-enhancing\newline approaches} & \textbf{Description} \\
\midrule
\endhead
\addlinespace
\midrule
\endfoot
\bottomrule
\endlastfoot

2020 & \citep{P3} & Distributed ledger technology, smart contracts, decentralized identifiers, digital signatures, $k$-anonymity, and differential privacy & 
Implements a framework for mobility data markets with six layers, each with a purpose and a technology to execute.
Furthermore, the framework focuses on location-based services.
The identity layer uses asymmetric identity keys, i.e. a key issued only to a real person, to verify that an entity is a real individual.
The privacy layer leverages $k$-anonymity for Geomasking (low utility and high privacy), and when the service needs an exact location, differential privacy for Geo-Indistinguishability (high utility and low privacy).
Moreover, the contract layer is based on smart contracts that enforce fair trade and the resolve disputes automatically.
% Additionally in this layer, based on publicly available metadata stored in the blockchain, data brokers match potential data sellers with data consumers. 
For the private communication layer, the authors use decentralized identifiers (DID)~\cite{DIDs} issued by the device of a person itself. 
When devices communicate, the communication has a unique ID based on both of the DIDs, thus, even though the communication data is persisted in a blockchain, it is nontrivial to track the locations of a user.
Consecutively, the incentive layer uses smart contracts and data brokers to promote data exchange for a profit; however, this architecture does not tackle the copy problem.  
% \jsnote{might be interesting to look how that works. It should be difficult for a SC to check whether data has been provided bilaterally; and if it is provided on-chain, it is provided to everyone, which might lead to strange incentives and behaviour.}
The consensus layer is based on a consortium blockchain for distributed governance among non-anonymous honest entities. 
The blockchain selected was Hyperledger Iroha, chiefly because of its lightweight quality that couples with deployments in IoT devices. \\ \addlinespace

2020 & \citep{P43} & Distributed ledger technology, smart contracts, functional encryption, and zero-knowledge proofs & 
Proposes a privacy-enhancing decentralized data marketplace employing the Agora blockchain with verification technology that enable data prosumers to monetize their data.
The privacy-enhancing aspect is achieved by sending encrypted data to brokers employing a primitive called \textit{multi-client functional encryption}~\cite{boneh_functional_2011}\cite{chotard_decentralized_2018}, which ensures that the receiver may only decrypt the output of a formerly agreed-on function.
Moreover, consumers may purchase these outputs, together with a proof of correctness from the broker by using non-interactive zero-knowledge proofs.
For the decentralized architecture, the authors employ the Agora blockchain, and atomic payments are performed via smart contracts.
\\ \addlinespace

2020 & \citep{P45} & Distributed ledger technology, smart contracts, trusted execution environments, truth discovery, digital signatures & 
Proposes a data processing-as-a-service model based on a blockchain-based data trading ecosystem, whereby neither data brokers nor consumers have access to the raw data, only to the analysis.
The use of a blockchain (Ethereum) prevents a single point of failure and allows for immutability and transparency in transactions. 
% The authors chose the Ethereum blockchain for this purpose.
Furthermore, to protect the data, the analysis results, and the processing itself, the authors use Intel's SGX trusted execution environment~\cite{sgx}, in addition to the symmetric encryption algorithm AES-256 to provide encryption and decryption within and outside the secure environment.
The architecture uses the conventional Ethereum Virtual Machine (EVM) for traditional smart contracts, while the data analysis contracts are executed in a SGX-protected EVM where an initial key exchange is needed.
Lastly, the nodes in the network form a compute market, i.e. multiple nodes execute the analysis and only the most frequent result is delivered to the data consumer, and the corresponding nodes are rewarded.
\\ \addlinespace

2020 & \citep{P28} & Distributed ledger technology, privacy policies, differential privacy, $k$-anonymity, and digital signatures & 
Proposes an architecture for a personal data marketplace in which personal data is stored decentrally in a allegedly GDPR compliant manner.
To accomplish this, transactions and pointers to the data are encrypted and stored using a distributed ledger technology, namely IOTA.
The data is stored either in an interplanetary file system, or in an IOTA-based storage format.
In order to access such data, a data aggregator must request permissions through Ethereum-based smart contracts (whitelists) owned by data consumers.
Once the permission has been granted, the trusted data aggregator, whose mutually agreed privacy policies are persisted in another Ethereum-based smart contract, waits until enough data owners exist to fulfill a particular analysis, so that the aggregator may perform $k$-anonymity.
The data aggregator sells the anonymized data to consumers and remunerates data owners accordingly.
However, the presence of a trusted aggregator defeats to some extent the purpose of a decentralized platform. 
Furthermore, the link between Ethereum and IOTA is carried out by trusted authentication services, which allow data aggregators to decrypt the data.
Lastly, in order for data owners to grant access to their data, the authors recommend dynamic threshold encryption~\cite{dynamic_threshold_encryption} over centralized forms of authentication services. \\ \addlinespace

2020 & \citep{P18}  & Distributed ledger technology, smart contracts, and differential privacy & 
Proposes a data trading approach in which privacy loss is publicly auditable and data owners set their privacy requirements on publicly available contracts. 
To accomplish this, the author uses a private Ethereum blockchain called Quorum that supports a set of built-in privacy measures, such as private transactions, messaging, and contracts; however, this design also restricts the interactions that are possible with smart contracts outside the private subset.
Furthermore, the data owner applies differential privacy locally before sharing the data with the consumer. 
\\ \addlinespace

2019 & \citep{P25} & Distributed ledger technology, trusted execution environments, smart contracts, privacy policies, digital signatures, and differential privacy & 
Implements an end-to-end privacy-enhancing decentralized data marketplace for data consumers to train machine learning algorithms, among other Turing-complete tasks.
The architecture proposed is a mature version of~\citep{P2}.
Containing all the features of~\citep{P2}, R. Cheng, et al.~\citep{P25} improve the performance of a newly designed distributed ledger technology to allow for horizontal scaling, i.e. the more nodes are added to the network, the more performant the network is, unlike e.g. Bitcoin or Ethereum; furthermore, the authors tackle the problem of confidentiality by separating consensus from execution, whose computations are performed in a trusted execution environment.
Horizontal scaling is achieved by allowing for parallel transaction execution, which is, in turn, accomplished by a set of transaction schedulers, and by creating dedicated committees for computation, storage, merging outputs, key management, and consensus.
However, scalability through restricting the degree of redundancy entails a security/integrity tradeoff.
Key management committees are necessary for the use of trusted execution environments to enable confidential computations.
The architecture uses symmetric keys for state encryption and asymmetric encryption for concealing user inputs. 
The authors achieve end-to-end privacy by protecting inputs with asymmetric encryption and differential privacy, and the execution with trusted execution environments.
More specifically, differential privacy prevents the weights of machine learning algorithms from overfitting to the inputs.
Because of the privacy limitations of current distributed ledger technology applications, they created a new concept so that smart contracts allow for privacy-enhancing features; this concept was introduced by~\citep{P2}. \\
\addlinespace
2019  & \citep{P32} & Distributed ledger technology, smart contracts, and digital signatures & 
Proposes a privacy-enhancing fair data trading protocol.
The protocol relies on the Ethereum blockchain to achieve a decentralized nature, however, the authors claim the protocol is blockchain agnostic.
Nonetheless, despite using a decentralized network, the market manager holds non-negligible authority, as it may trace the identity of sellers so that they can be punished monetarily in case they misbehave.
Furthermore, once the buyers have decided which data asset to purchase, the sellers use symmetric keys to encrypt data in chunks before sending it to the buyers.
Upon receiving the data chunks, the buyer (i) challenges a set of data chunks, and upon verification of truthful data, (ii) employs similarity learning, a machine learning technology~\cite{similarity_learning}, to decide whether to finally purchase the data.
Consequently, once the buyer decides to purchase the data, the seller and buyer interact via a payment smart contract and double-authentication-preventing signatures~\cite{poettering_double_authentication_preventing_2017} to ensure payment and data decryption.
Lastly, in order to enhance the anonymity of the actors, the protocol uses ring signatures~\cite{P32R26}\cite{GroupSignaturesPaper}.
\\ \addlinespace

2019  & \citep{P20} & Distributed ledger technology, smart contracts, digital signatures, and decentralized identifiers & 
Implements a decentralized data market architecture with secure data processing for the IoT.
To achieve decentralization, the authors rely on the distributed ledger technology IOTA, and Ethereum-based smart contracts for subscribing to data streams.
The constellation of actors consists of three entities: a data provider, a consumer, and a broker; the former two entities are included in a registry via decentralized identifiers~\cite{DIDs}.
The product that the consumers purchase is a key to a data stream for a predetermined period of time, created but not accessible by the data broker.
For the consumer to attain data access in a private manner, blind signatures~\cite{blind_signatures} are employed, which enable a data broker to verify stream access keys from the data provider without ever accessing these keys.
More specifically, the data provider "blinds" the session key with the broker’s public key and sends the blinded key to the broker, consequently, the broker certifies the key with its signature and returns the signature to the provider who removes the blinding factor to access the stream.
Lastly, to exchange stream data, an inter-planetary file system is employed.
\\ \addlinespace

2019 & \citep{P12}  & Distributed ledger technology, trusted execution environments, and digital signatures &
Proposes a distributed IoT data storage system and a data trading scheme.
The authors use the blockchain for its distributed nature, immutability, and requester authentication; moreover, their solution is blockchain agnostic.
However, for the consensus algorithm, the authors rely on Intel's Software Guard Extension (SGX)~\cite{sgx} to deploy a trusted execution environment, to perform "Proof of Useful Work".
The blockchain only contains pointers (addresses) to a distributed hash table, where the data is stored off-chain by peers of the network.
Only certified data consumers, e.g. other IoT devices, would be able to query addresses in the blockchain.
Furthermore, the authors employ certificateless cryptography so that the key generation center of conventional identity-based encryption does not need to be trusted~\cite{certificateless_crypto}.
To perform the cryptographic operations, edge devices are deployed.
Lastly, to share data with purchasers, the authors propose to use either asymmetric encryption or re-encryption~\cite{P12R27}.
\\ \addlinespace

2019 & \citep{P1} & Distributed ledger technology, digital signatures, and hashing & 
Prototypes a decentralized fair data trading platform.
The authors rely on the Ethereum blockchain to avoid third-party data brokers and to leverage the ledger's immutability properties.
Moreover, data sellers utilize smart contracts to propose their data offers and to interact with sellers.
Sellers include the hash of the data in the ledger so that the buyer may initiate a rebuttal if there is an expectation mismatch.
Additionally, to ensure accountability the authors rely on digital signatures to verify that the data was encrypted using a specific key that belongs to the seller, and encrypt the data efficiently using symmetric encryption.
The asset traded are decryption keys, and buyers may retrieve data as ciphertexts from untrusted storage. \\ \addlinespace

2018 & \citep{P46} & Distributed ledger technology, smart contracts, partially homomorphic encryption, hashing, and digital signatures & 
Proposes two secure, and fair data trading decentralized schemata built on the Ethereum blockchain. 
One scheme enables entities to trade raw data, while the other scheme enables them to exchange statistics. 
The authors chose blockchain in both schemata for its immutability, smart contracts, P2P payment, and disintermediation. 
Furthermore, for the second scheme, the authors use partially homomorphic encryption to perform confidential statistics. 
For the data structure to compute the statistics, the authors chose a Merkle Accumulative Tree, where the leaf nodes hold the encrypted data and the non-leaf nodes contain the hash values and a cumulative sum of homomorphic ciphertexts. The data exchanged is verifiable through digital signatures based on asymmetric encryption. \\ \addlinespace

2017  & \citep{P33} & Distributed ledger technology, privacy policies, digital signatures, hashing, and $k$-anonymity & 
Prototypes a decentralized data market platform for anonymized data.
The underlying distributed ledger technology is Hyperledger Fabric, whose peers act as data brokers.
The data brokers may only handle datasets based on a set of privacy policies in the interest of the data owner and dictated by a data domain-specific privacy policy manager.  
The blockchain acts as an auditable ledger for transactions between data brokers and consumers, while the exchange of data is handled off-chain.
Furthermore, the anonymization of data is suggested to be performed employing $k$-anonymity by the broker upon dataset reception from a secure channel, however, the solution remains anonymization-agnostic.
For every actor to verify that the correct anonymized dataset has been shared, the broker sends its hash value using SHA-256 to the blockchain before sending it to the consumer. Upon reception, both the policy manager and data receiver may verify the dataset.
Lastly, cryptography technologies are employed to encrypt the dataset symmetrically (128-bit AES) before sharing the dataset to the consumer, and the actors use ECDSA to sign confirmations and transactions.
% These smart contracts of data owners act as announcing boards, hold privacy policies and payment logic, privately store the decryption key of their corresponding data, and point to where such data is stored.
% On the other hand, data consumers also deploy smart contracts to interact with the data owners'. 

\end{longtable}
%TC:endignore

\newpage

\section{Distribution of selected studies by publication channels}
\label{sec:AppendixDistributionSelectedStudies}

\renewcommand{\arraystretch}{1.25}
\begin{table}[H]
\centering
\caption{Publication channels for the studies from our SLR.}
\label{tab:publication_sources}
\footnotesize\vspace{0.25em}
\begin{tabular}{rllcc}
\toprule
\textbf{\#\,} & \textbf{Publication source}                                                                & \textbf{Type}     & \textbf{No.} & \textbf{\%}  \\ \midrule
1	&	 ACM International Conference Proceeding Series     &	 Conference 	&	2	&	4	\\
2	&	 IEEE International Conference on Internet of Things     &	 Conference 	&	2	&	4	\\
3	&	 VLDB Endowment     &	 Journal 	&	2	&	4	\\
4	&	ACM Conference on Computer and Communications Security     &	 Conference 	&	1	&	2	\\
5	&	ACM International Workshop on Mobile Commerce     &	 Workshop 	&	1	&	2	\\
6	&	ACM SIGKDD International Conference on Knowledge Discovery and Data Mining     &	 Conference 	&	1	&	2	\\
7	&	ACM SIGMOD Record     &	 Journal 	&	1	&	2	\\
8	&	CEUR Workshop     &	 Workshop 	&	1	&	2	\\
9	&	Computer Law and Security Review     &	 Journal 	&	1	&	2	\\
10	&	Computer Networks     &	 Journal 	&	1	&	2	\\
11	&	Concurrency Computation: Practice and Experience    &	 Journal 	&	1	&	2	\\
12	&	Conference on Information and Knowledge Management    &	 Conference 	&	1	&	2	\\
13	&	Electronic Markets    &	 Journal 	&	1	&	2	\\
14	&	IACR Cryptology ePrint Archive   &	 Journal 	&	1	&	2	\\
15	&	IEEE Access   &	 Journal 	&	1	&	2	\\
16	&	IEEE Cloud Computing   &	 Journal 	&	1	&	2	\\
17	&	IEEE Communications Magazine &	 Journal 	&	1	&	2	\\
18	&	IEEE Computer &	 Journal	&	1	&	2	\\
19	&	IEEE Eurasia Conference on IOT, Communication and Engineering 	&	 Conference &	1	&	2	\\
20	&	IEEE European Symposium on Security and Privacy 	&	 Conference &	1	&	2	\\
21	&	IEEE International Conference on Big Data 	&	 Conference &	1	&	2	\\
22	&	IEEE International Conference on Blockchain and Cryptocurrency 	&	 Conference &	1	&	2	\\
23	&	IEEE International Conference on Collaboration and Internet Computing 	&	 Conference &	1	&	2	\\
24	&	IEEE International Conference on Pervasive Computing and Communications Workshops 	&	 Conference &	1	&	2	\\
25	&	IEEE International Conference on Software Engineering Research, Management and Applications 	&	 Conference &	1	&	2	\\
26	&	IEEE Internet Computing & Journal &	1	&	2	\\
27	&	IEEE Internet of Things Journal 	&	 Journal &	1	&	2	\\
28	&	IEEE International Conference on Parallel and Distributed Processing with Applications, Big Data and  & Conference &	1	&	2	\\
	&	Cloud Computing, Sustainable Computing and Communications, Social Computing and Networking 	&	  & &	\\
29	&	IEEE Symposium on Reliable Distributed Systems	&	 Conference &	1	&	2	\\
30	&	IEEE Transactions on Information Forensics and Security	&	 Journal &	1	&	2	\\
31	&	IEEE Transactions on Knowledge and Data Engineering	&	 Journal &	1	&	2	\\
32	&	IEEE Transactions on Network Science and Engineering	&	 Journal &	1	&	2	\\
33	&	IEEE Transactions on Services Computing	&	 Journal &	1	&	2	\\
34	&	IEEE Wireless Communications	&	 Journal &	1	&	2	\\
35	&	Information Sciences	&	 Journal &	1	&	2	\\
36	&	International Conference on Distributed Computing Systems	&	 Conference &	1	&	2	\\
37	&	International Conference on Smart Systems and Technologies	&	 Conference &	1	&	2	\\
38	&	International Conference on Software Engineering	&	 Conference &	1	&	2	\\
39	&	International Symposium on Mobile Ad Hoc Networking and Computing	&	 Conference &	1	&	2	\\
40	&	International Workshop on Security and Privacy in Big Data	&	 Workshop &	1	&	2	\\
41	&	International Workshop on Social Sensing	&	 Workshop &	1	&	2	\\
42	&	LNCS 7299 -- Intelligence and Security Informatics	&	 Journal &	1	&	2	\\
43	&	Online Information Review	&	 Journal &	1	&	2	\\
44	&	Security and Communication Networks	&	 Journal &	1	&	2	\\
45	&	Sensors	&	 Journal &	1	&	2	\\
46	&	Transportation Research Part C: Emerging Technologies	&	 Journal &	1	&	2	\\
47	&	Workshop on the Economics of Networks, Systems and Computation	&	 Workshop &	1	&	2	\\\midrule

   & \textbf{Total}                                                                             &    & \textbf{50} & \textbf{100} \\ \bottomrule

\end{tabular}
\end{table}

\newpage
\section{Electronic data sources and inclusion and exclusion criteria}

\renewcommand{\arraystretch}{1.5}
\begin{table}[htpb] % htpb --> Where to place the table? (here, top of page, top of next page or bottom of page). first letter has priority.
  \caption{Electronic data sources (SDS) used in automated search.}\label{tab:databases} %The first parentheses is the name in the List of tables, the second one is the caption in the document. 
  
  \centering
  \footnotesize\vspace{0.25cm}
  \begin{tabular}{l l l} % it has three rows, and l means that the content is shifted to the left. 
    \toprule
      \textbf{ID} & \textbf{Name (Acronym)} & \textbf{Website} \\
    \midrule
      EDS1 & IEEE Xplore (IEEE) & \url{https://ieeexplore.ieee.org/} \\
      EDS2 & ACM Digital Library (ACM) & \url{https://dl.acm.org/} \\
      EDS3 & ISI Web of Science (WoS) & \url{https://www.webofknowledge.com} \\
      EDS4 & ScienceDirect (SD) & \url{https://www.sciencedirect.com/} \\
      EDS5 & SpringerLink (SL) & \url{https://link.springer.com/} \\
      EDS6 & Wiley InterScience (WIS) & \url{https://onlinelibrary.wiley.com/} \\
      EDS7 & SCOPUS (SCOPUS) & \url{https://www.scopus.com/} \\
    \bottomrule
  \end{tabular}
\end{table}

\vspace{2cm}

\renewcommand{\arraystretch}{1.5}
\begin{table}[htpb] % htpb --> Where to place the table? (here, top of page, top of next page or bottom of page). first letter has priority.
  \caption[Selection criteria]{Selection criteria used to identify relevant papers. Fulfilling only one exclusion criterion discards the publication from being included. }\label{tab:selection criteria} %The first parentheses is the name in the List of tables, the second óne is the caption in the document. 
  \centering
  \footnotesize\vspace{0.25cm}
  \begin{tabular}{l l p{6cm} p{4.5cm}}
    \toprule
      \textbf{ID} & \textbf{Facet} & \textbf{Inclusion criterion} & \textbf{Exclusion criterion} \\
    \midrule
      F1 & Coarse focus & The privacy and data market topic must be within the field of computer science and technology & Any other privacy and data market sub-field \\
      F2 & Narrow focus & The paper must explicitly focus on privacy within data marketplaces within the defined applications & The paper does not explicitly address this research direction \\
      F3 & Publication channel type & Conference publication OR journal publication (full text) OR workshop publication & The paper is any other type of publication \\
      F4 & Language & English & Non-English \\
      F5 & Duplicates & Publications are new to the filtering process & Publication has already been processed \\
      F6 & Peer-review & The publication has been peer-reviewed & The publication is a grey publication \\
      F7 & Full-text access & TUM-Access granted & TUM-Access not granted \\
    \bottomrule
  \end{tabular}
\end{table}

\newpage
\section{Figures of the metadata analysis}

\begin{figure*}[htpb]
\centering
    \includegraphics[scale=0.7]{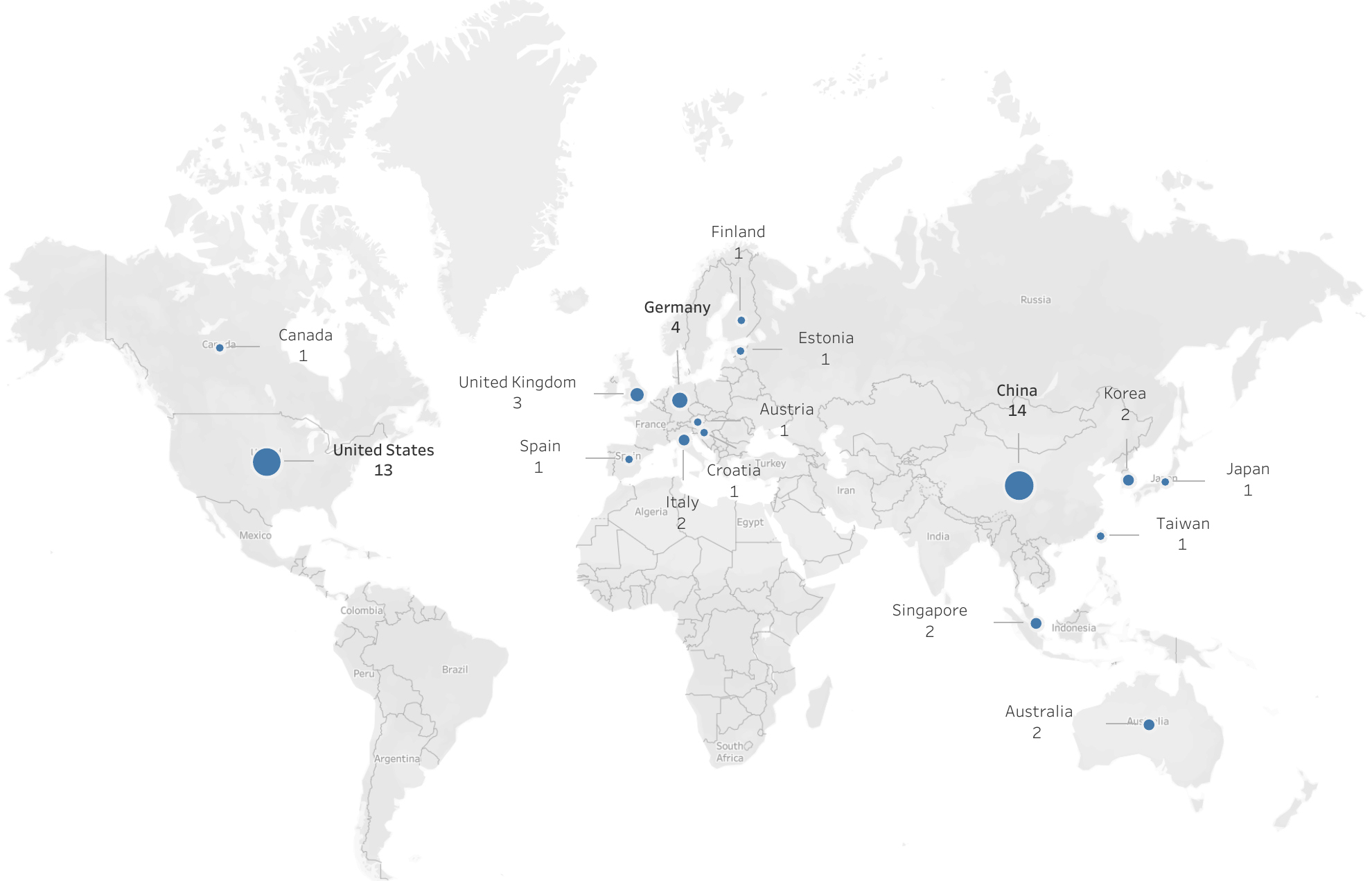}
\caption{Map of most active countries in the field of privacy-enhancing data markets for the IoT research.}
\label{fig:world_map}
\end{figure*}

\renewcommand{\arraystretch}{1.5}
\begin{figure*}[htpb]
\centering
\includegraphics[width=\textwidth, trim=0cm 1cm 0cm 0cm, clip]{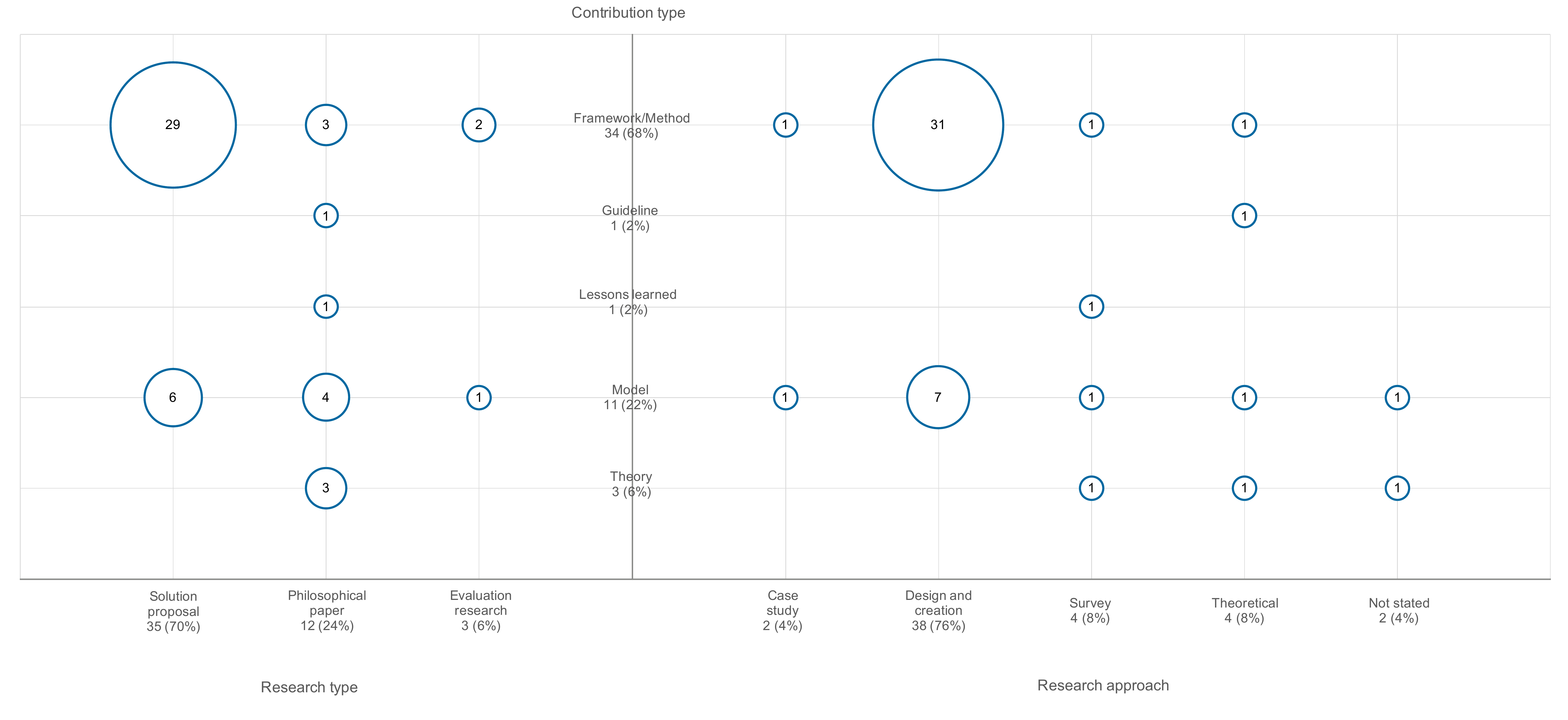}
\caption{Mapping of contribution types against research types and approaches.}
\label{fig:contribution_type_research_type_research_approach}
\end{figure*}

\vspace{2cm}
\begin{table}[htpb] % htpb --> Where to place the table? (here, top of page, top of next page or bottom of page). first letter has priority.
  \caption[Classification scheme of research types as described by \cite{Wieringa2006}]{Classification scheme of research types as described by \cite{Wieringa2006}.}\label{tab:research types} %The first parentheses is the name in the List of tables, the second óne is the caption in the document. 
  \footnotesize\vspace{0.25cm}
  \centering
  \begin{tabular}{p{3.5cm} p{11cm}} % it has three columns, and l means that the content is shifted to the left. p{6cm} defines the length of the second column.
    \toprule
      \textbf{Research type} & \textbf{Description}\\
    \midrule
    Evaluation research & The authors implement existing techniques, and the solutions are evaluated in practice. \\
    Philosophical papers & These studies present a new perspective on existing research by organizing the domain into a taxonomy or a conceptual framework.\\
    Solution proposal & The authors propose a solution to a problem. The solution can be either novel or a significantly enhanced version of an existent technique. A small example or argumentation demonstrates the benefit and applicability of the solution.\\
    \bottomrule
  \end{tabular}
\end{table}

\begin{figure*}[htpb]
\centering
\includegraphics[width=0.7\textwidth]{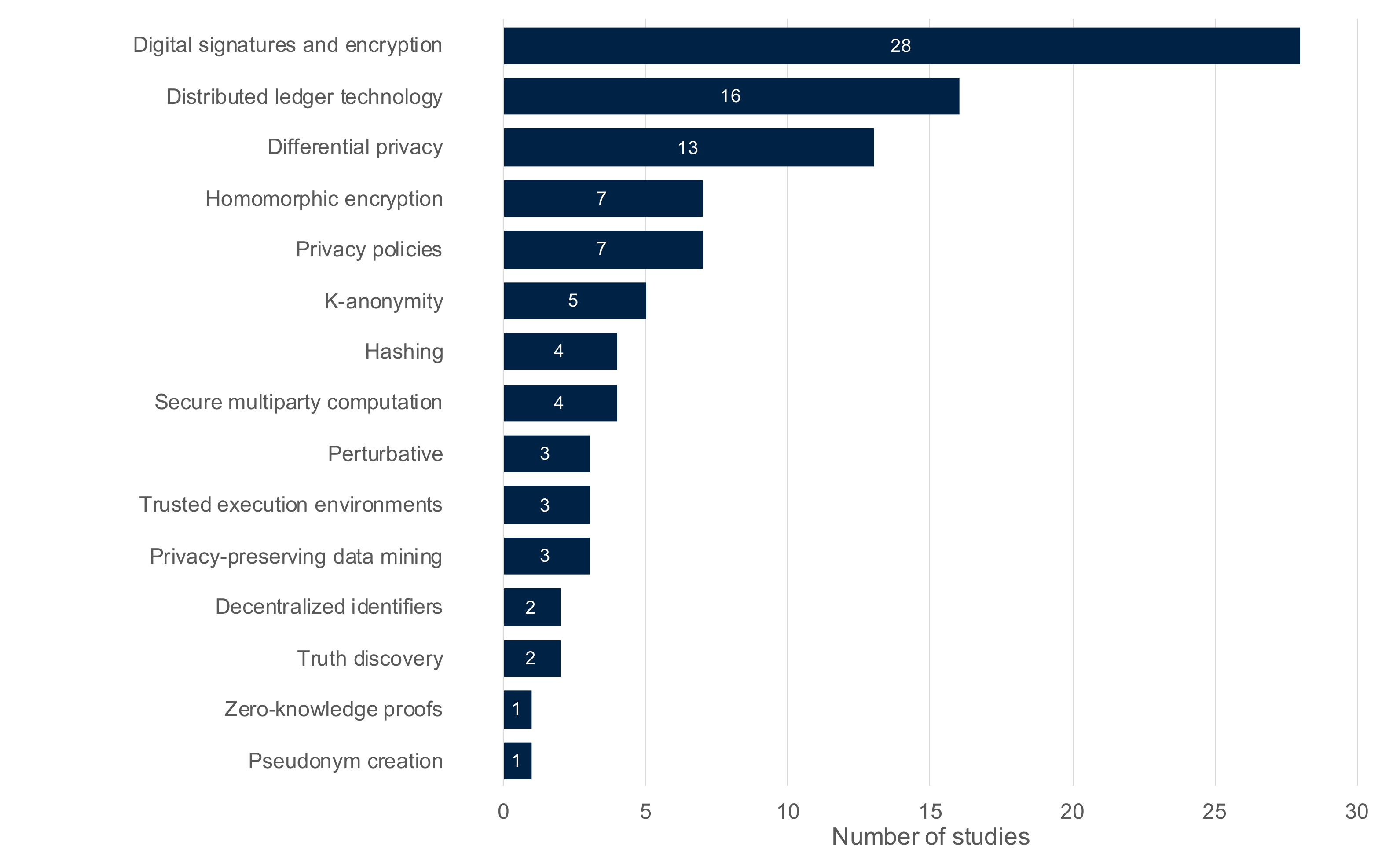}
\caption{Distribution of PETs and AETs explicitly employed in the corpus of selected studies.}
\label{fig:PETs_distributions}
\end{figure*}

\begin{figure*}[htpb]
\centering
\includegraphics[width=0.7\textwidth, trim=0cm 0cm 0cm 0cm, clip]{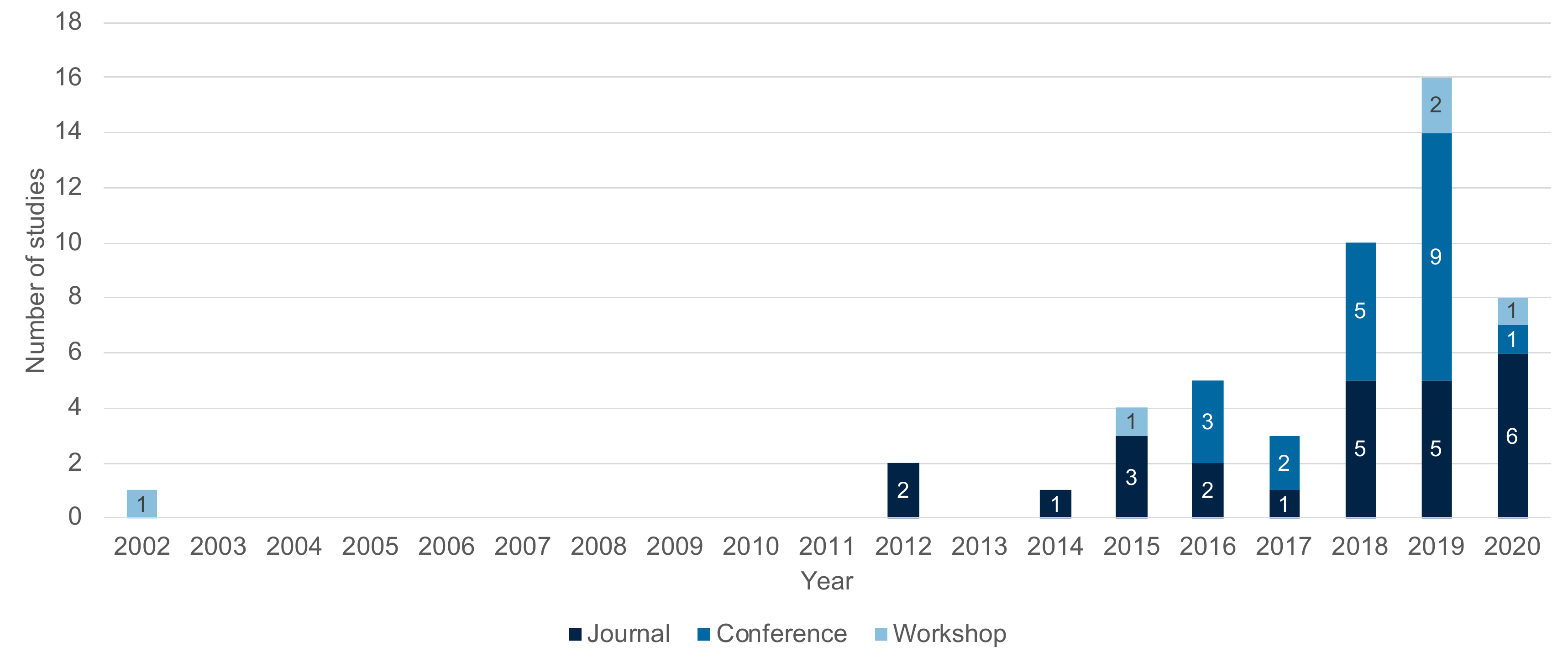}
\caption{Distribution of studies over publication domains.}
\label{fig:publicationdomain}
\end{figure*}

\begin{figure*}[htpb]
\centering
\includegraphics[width=0.7\textwidth]{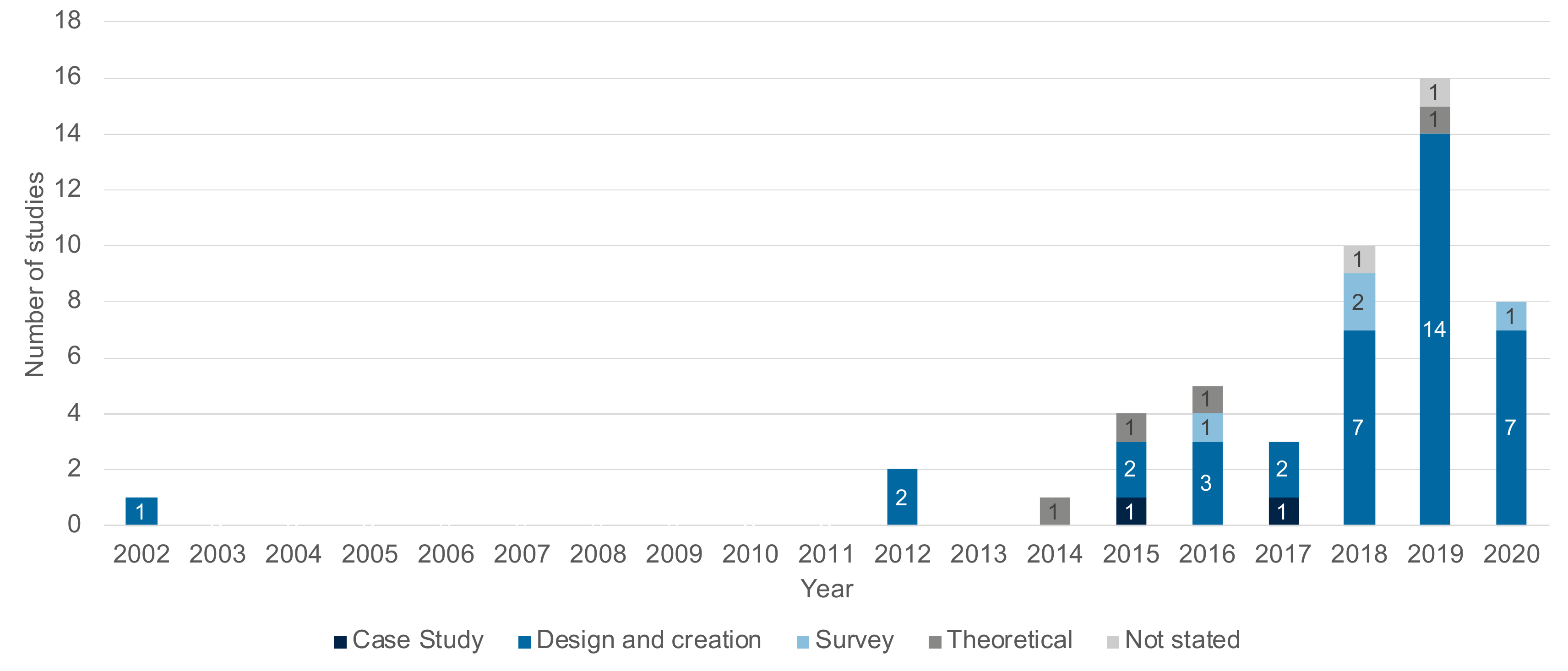}
\caption{Number of studies per research approach over time.}
\label{fig:researchapproachtrend}
\end{figure*}

\begin{figure*}[htpb]
\centering
\includegraphics[width=0.7\textwidth]{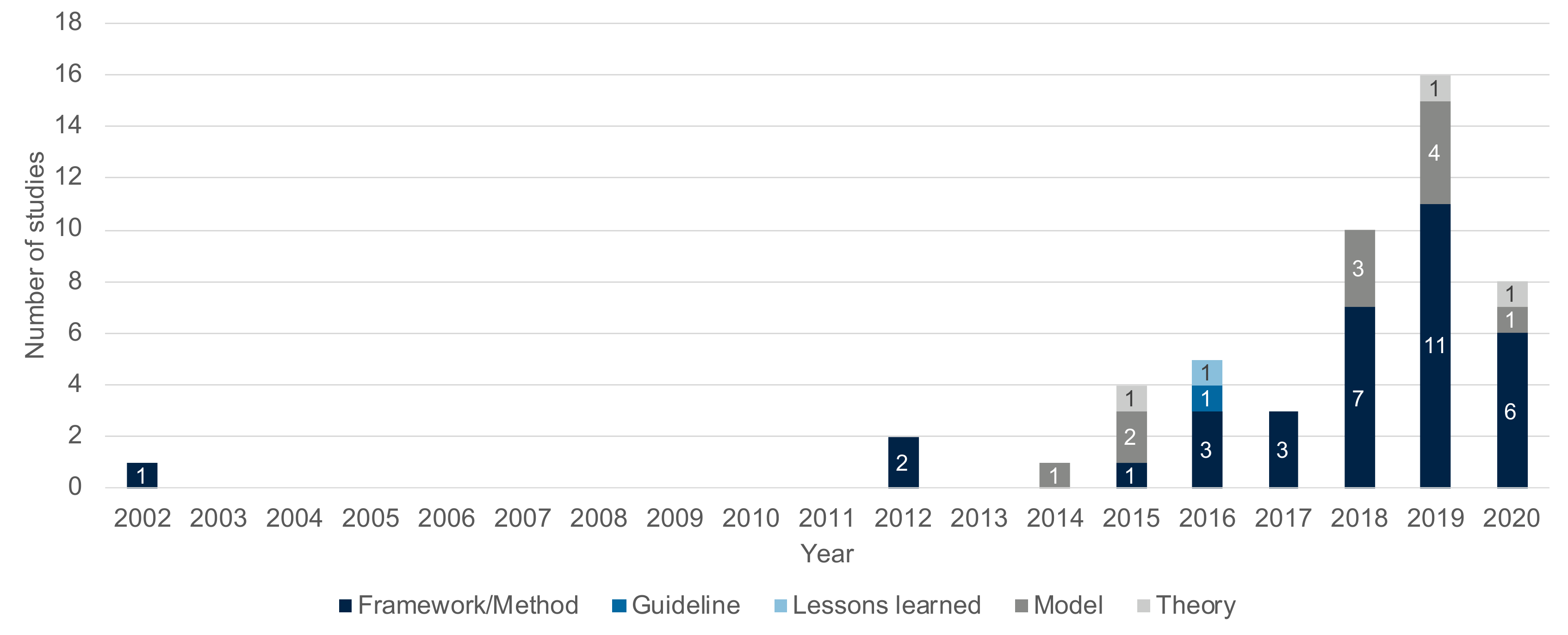}
\caption{Distribution of research outcomes over time.}
\label{fig:researchcontributiontrend}
\end{figure*}

% \section{PETs maturity}
% \include{challenges_tech}
% \include{challenges_tech2}

\else
%nothing
\fi
%TC:endignore

\end{document}